\documentclass[useAMS,usenatbib]{mn2e}

\title[]{OV] $\lambda\lambda$1213.8,1218.3 emission from extended
  nebulae around quasars: contamination of Ly$\alpha$ and a new
  diagnostic for AGN activity in Ly$\alpha$-emitters}
\author[A. Humphrey]{A. Humphrey$^{1}$\thanks{E-mail:
    andrew.humphrey@astro.up.pt}
\\
$^1$Instituto de Astrof\'{i}sica e Ci\^encias do Espa\c{c}o,
     Universidade do Porto, CAUP, Rua das Estrelas, PT4150-762 Porto, Portugal}

\begin{document}

\date{Accepted 2019 March 5.
      Received 2019 March 1;
      in original form 2018 October 25}

\pagerange{\pageref{firstpage}--\pageref{lastpage}}
\pubyear{2011}

\maketitle

\label{firstpage}

\begin{abstract}
We investigate the potential for the emission lines OV]
$\lambda\lambda$1213.8,1218.3 and HeII $\lambda$1215.1 to contaminate
flux measurements of Ly$\alpha$ $\lambda$1215.7 in the extended
nebulae of quasars. We have 
computed a grid of photoionization models with a
substantial range in the slope of the ionizing powerlaw (-1.5 $<$
$\alpha$ $<$ -0.5), gas metallicity (0.01 $<$ $Z/Z_{\odot}$ $<$ 3.0),
gas density (1 $<$ $n_H$ $<$ 10$^4$ cm$^{-3}$), and ionization parameter
(10$^{-5}$ $<$ U $<$ 1.0). We find the contribution from HeII $\lambda$1215.1 to be
negligible, i.e., $<$ 0.1 of Ly$\alpha$ flux, across our entire model
grid. The contribution from OV] $\lambda\lambda$1213.8,1218.3 is
generally negligible when U is low ($\la$10$^{-3}$) and/or when the
gas metallicity is low ($Z/Z_{\odot}$ $\la$ 0.1). However, at higher
values of U and Z we find that OV] can significantly contaminate
Ly$\alpha$, in some circumstances accounting for more than half the
total flux of the Ly$\alpha$+HeII+OV] blend. We also provide means to estimate the fluxes of
OV] $\lambda\lambda$1213.8,1218.3 and HeII $\lambda$1215.1 by
extrapolating from other lines. We estimate the
fluxes of OV] and HeII for a sample of 107 Type 2 active galaxies at
z$>$2, and find evidence for significant ($\ge$10\%) contamination of Ly$\alpha$
fluxes in the majority of cases (84\%). We also discuss
prospects for using OV] $\lambda\lambda$1213.8,1218.3 as a diagnostic
for the presence of AGN activity in high-z Ly$\alpha$ emitters, and
caution that the presence of significant OV] emission could impact 
the apparent kinematics of Ly$\alpha$, potentially mimicking the
presence of high-velocity gas outflows. 

\end{abstract}

\begin{keywords}
galaxies: active; quasars: emission lines; galaxies: ISM; ultraviolet:
ISM; line: formation
\end{keywords}

\begin{figure*}
\includegraphics{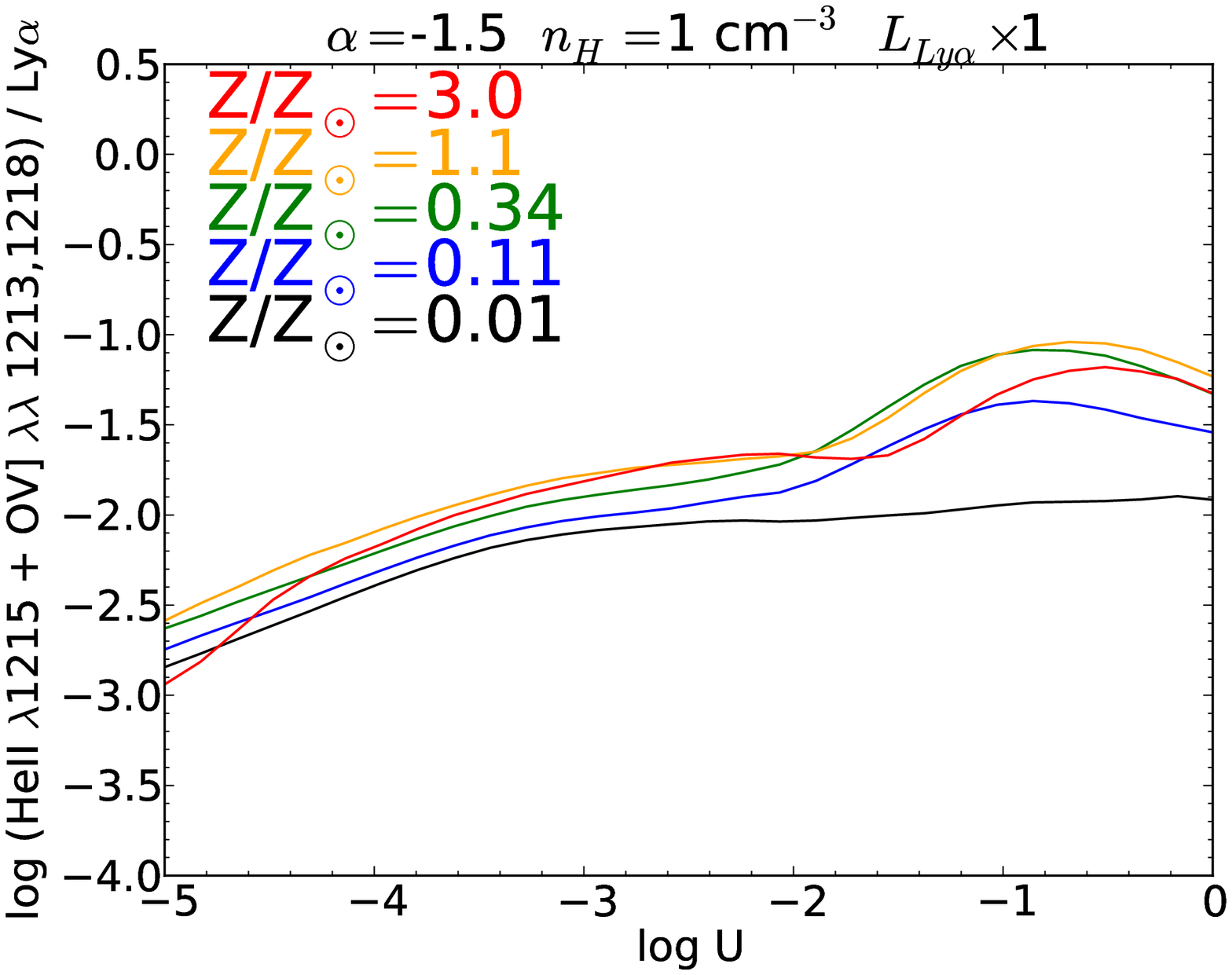}
\includegraphics{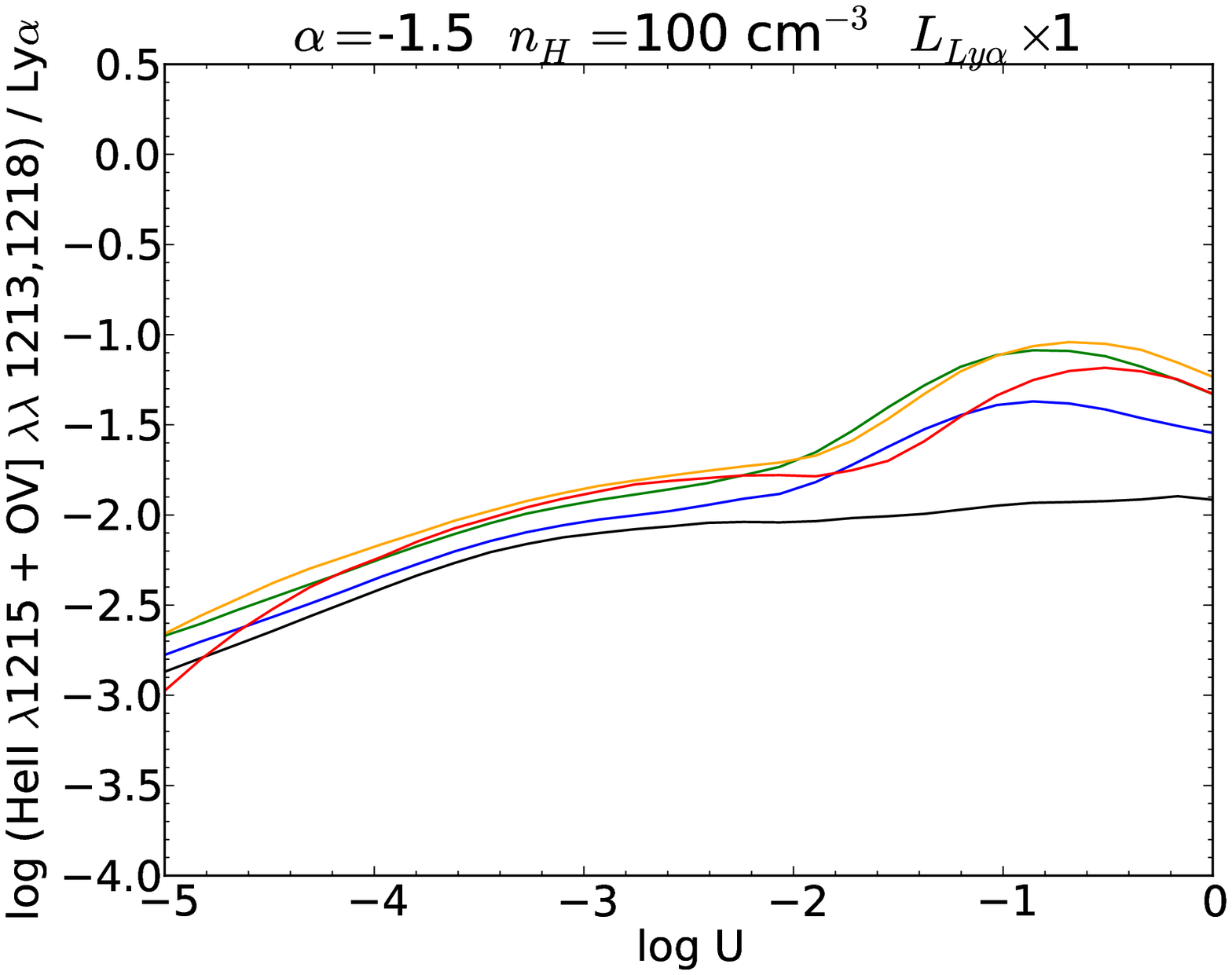}
\includegraphics{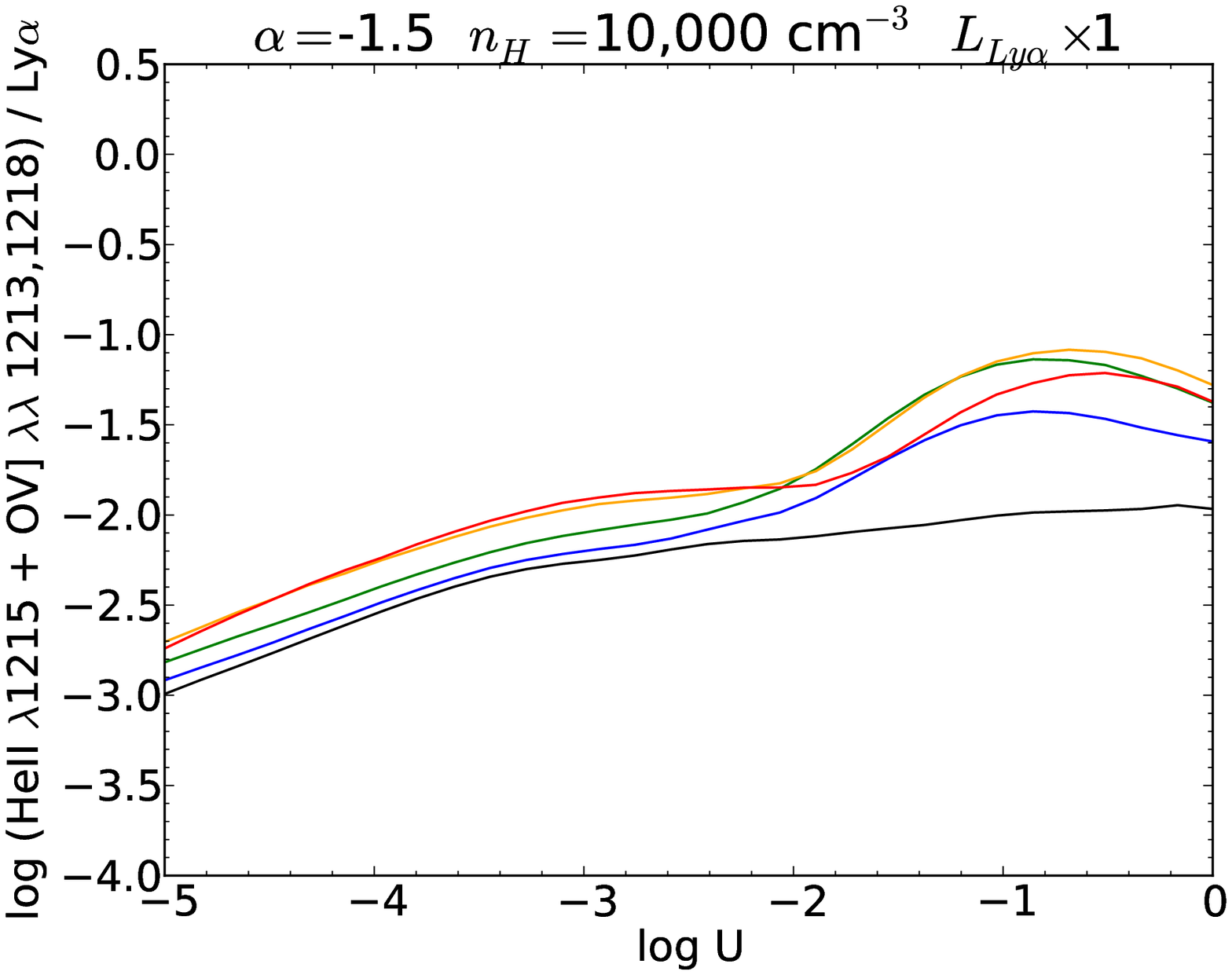}
\includegraphics{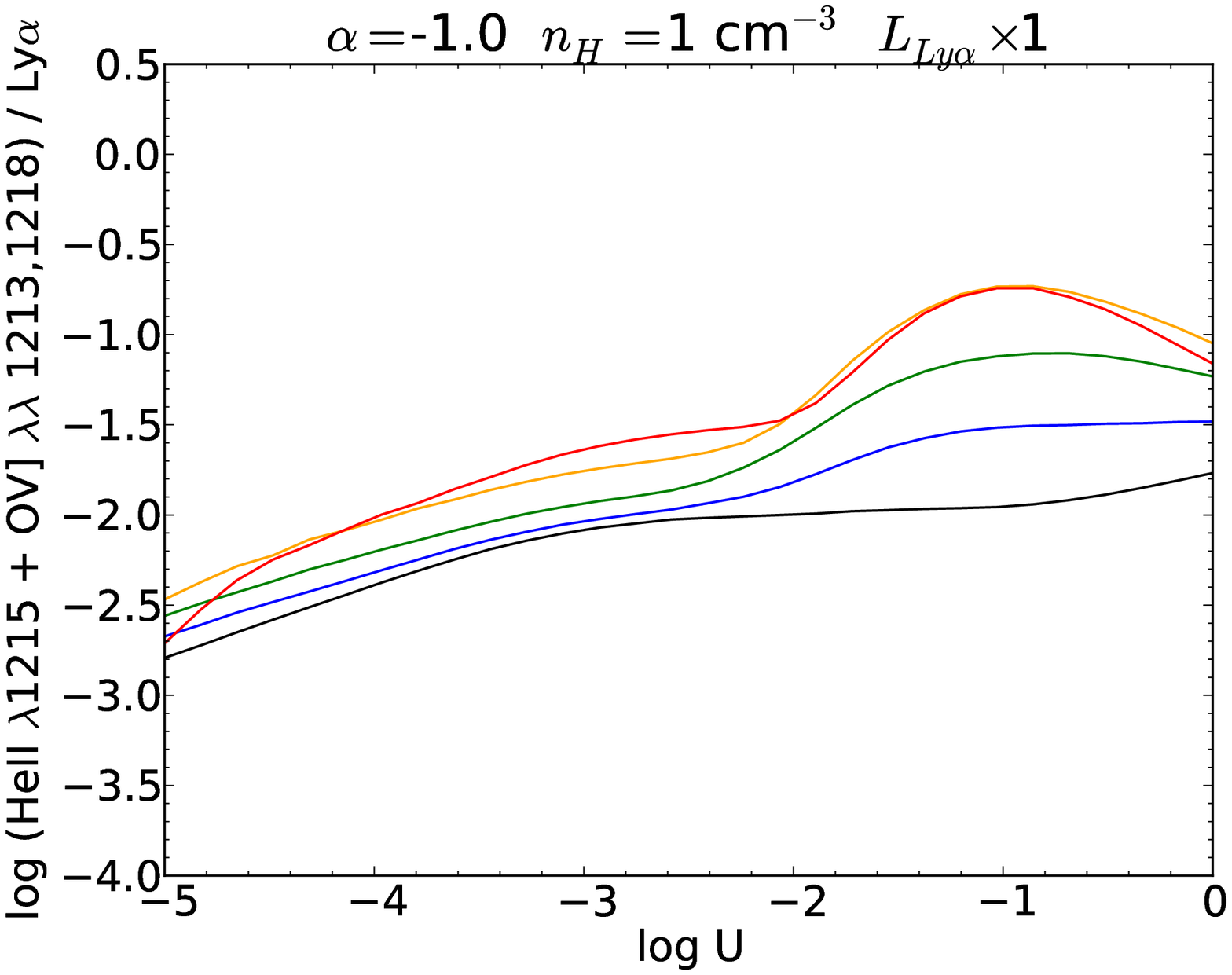}
\includegraphics{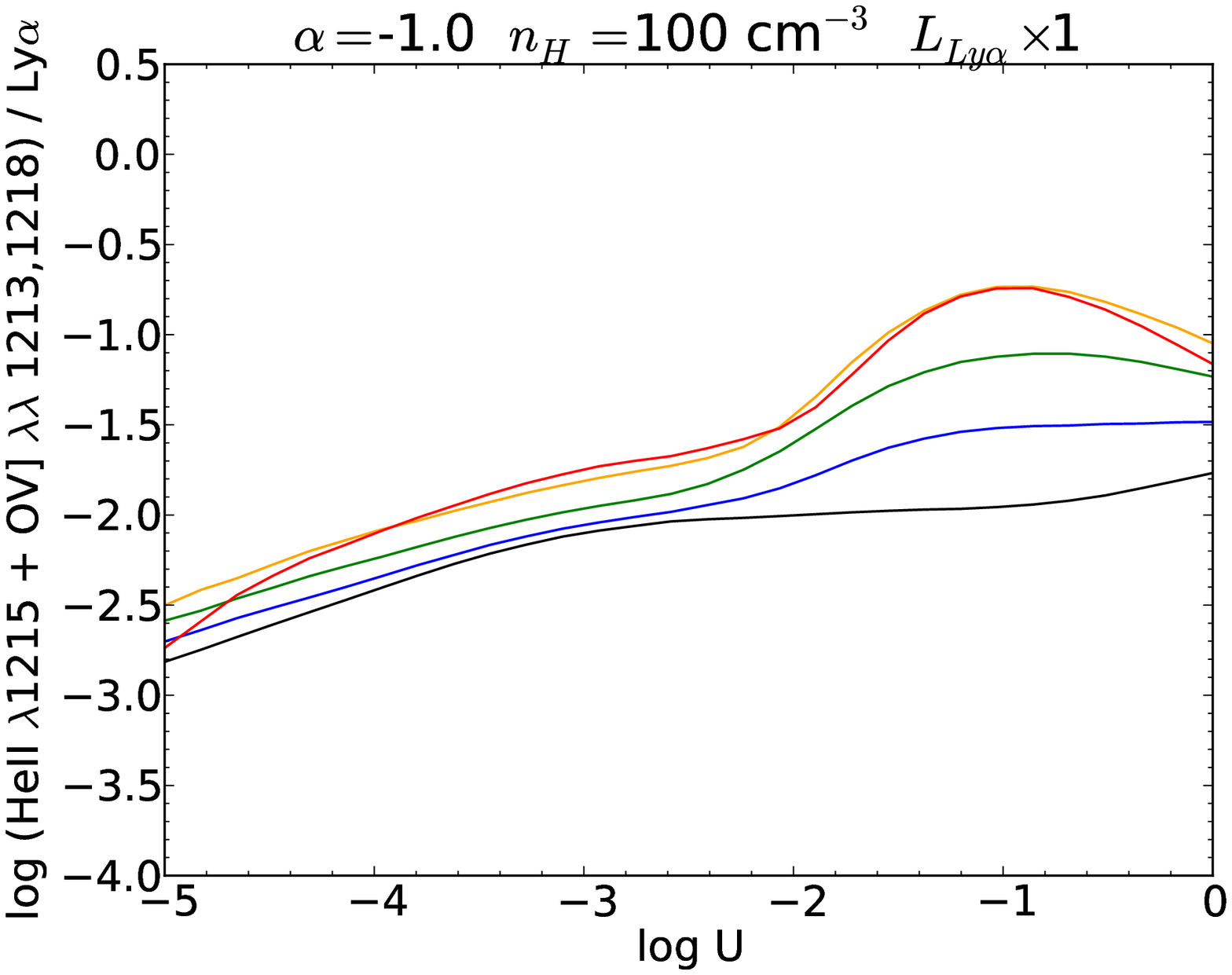}
\includegraphics{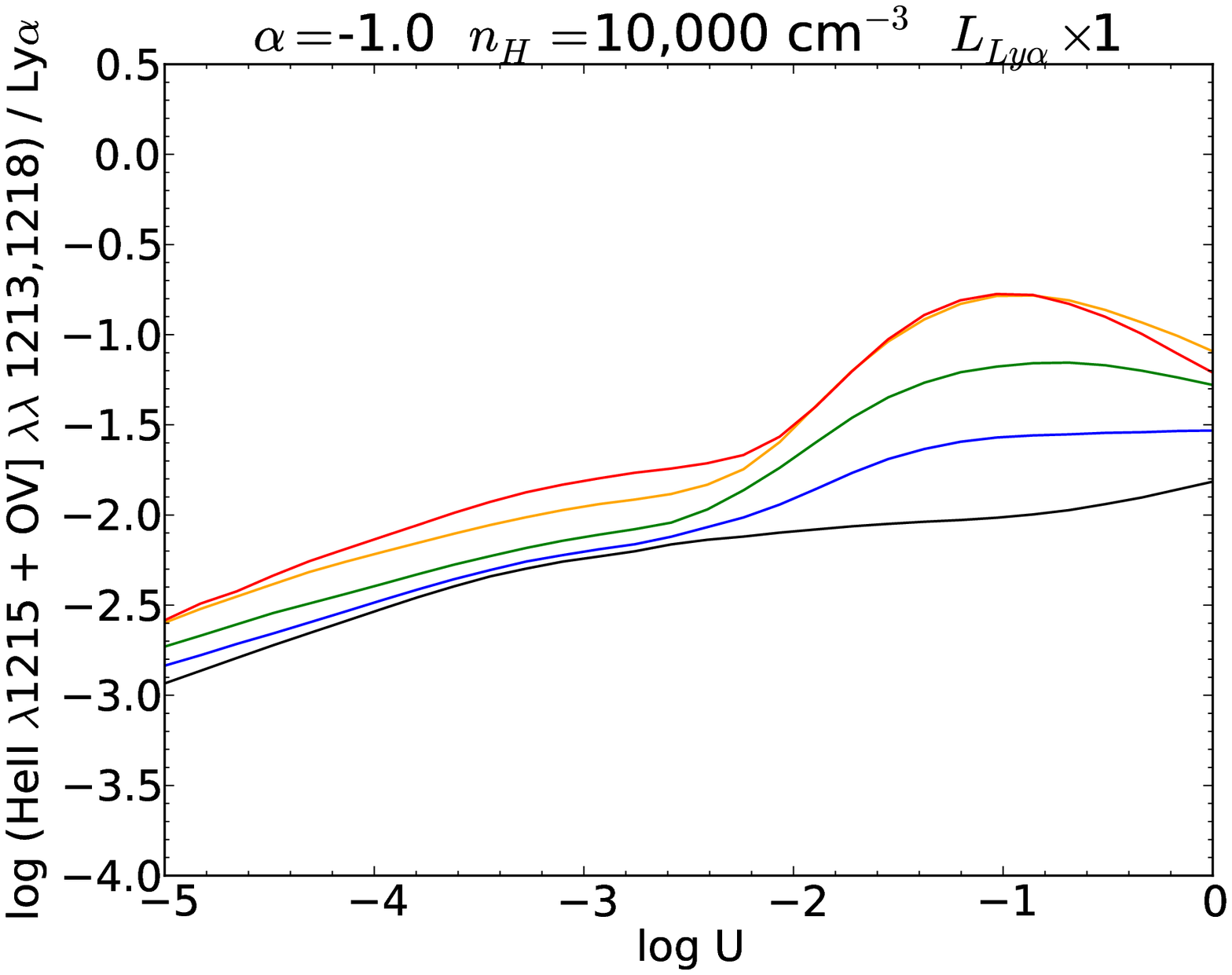}
\includegraphics{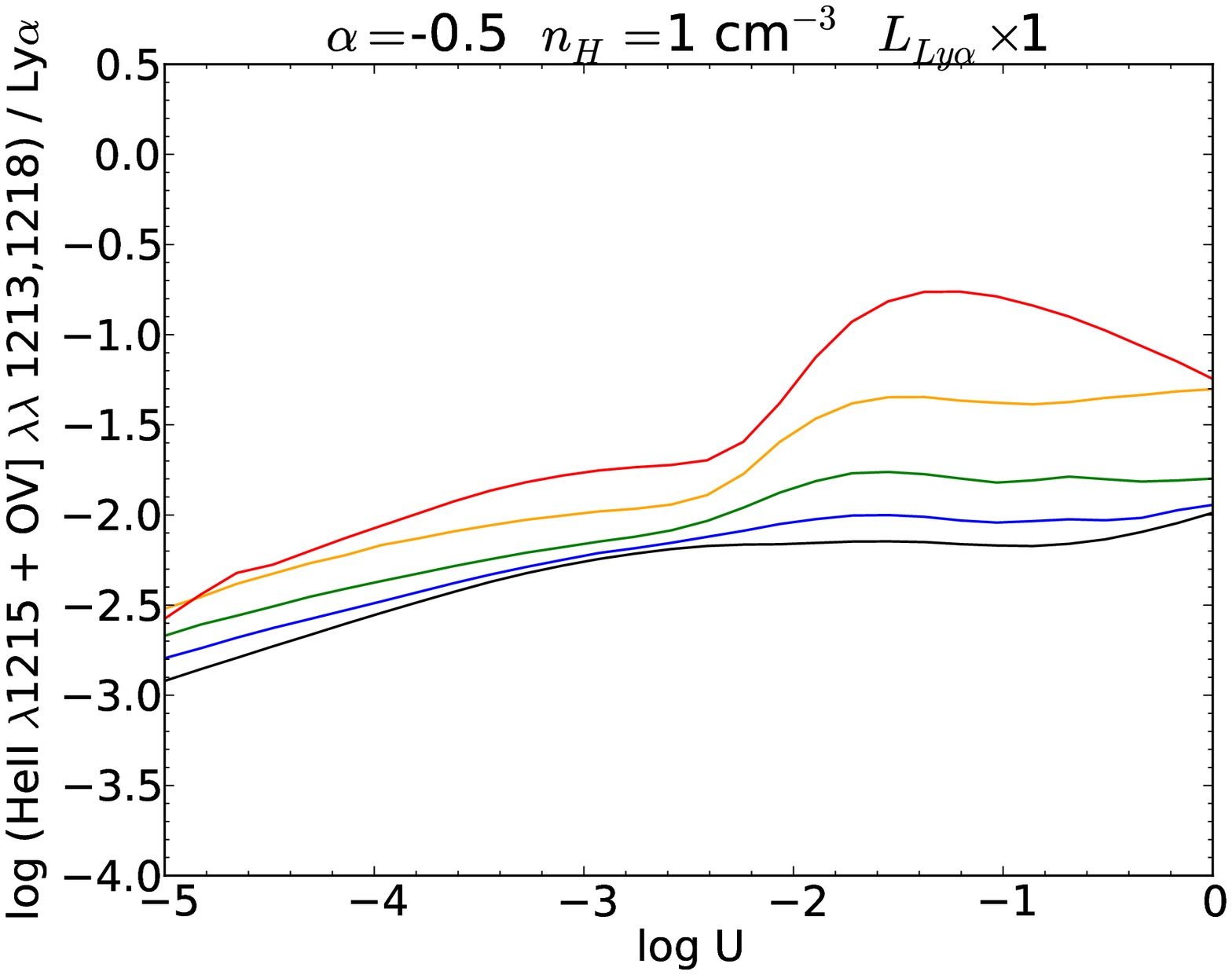}
\includegraphics{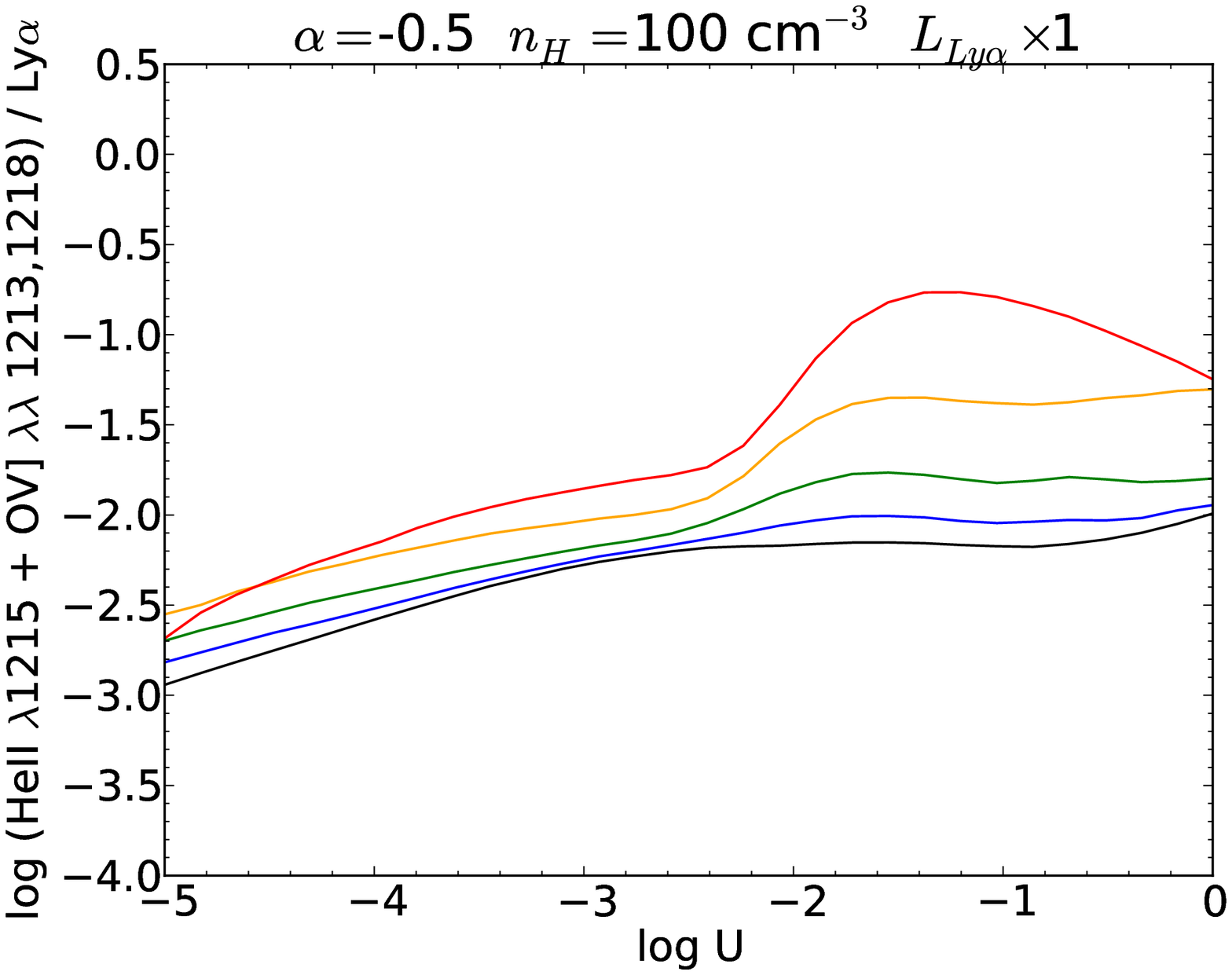}
\includegraphics{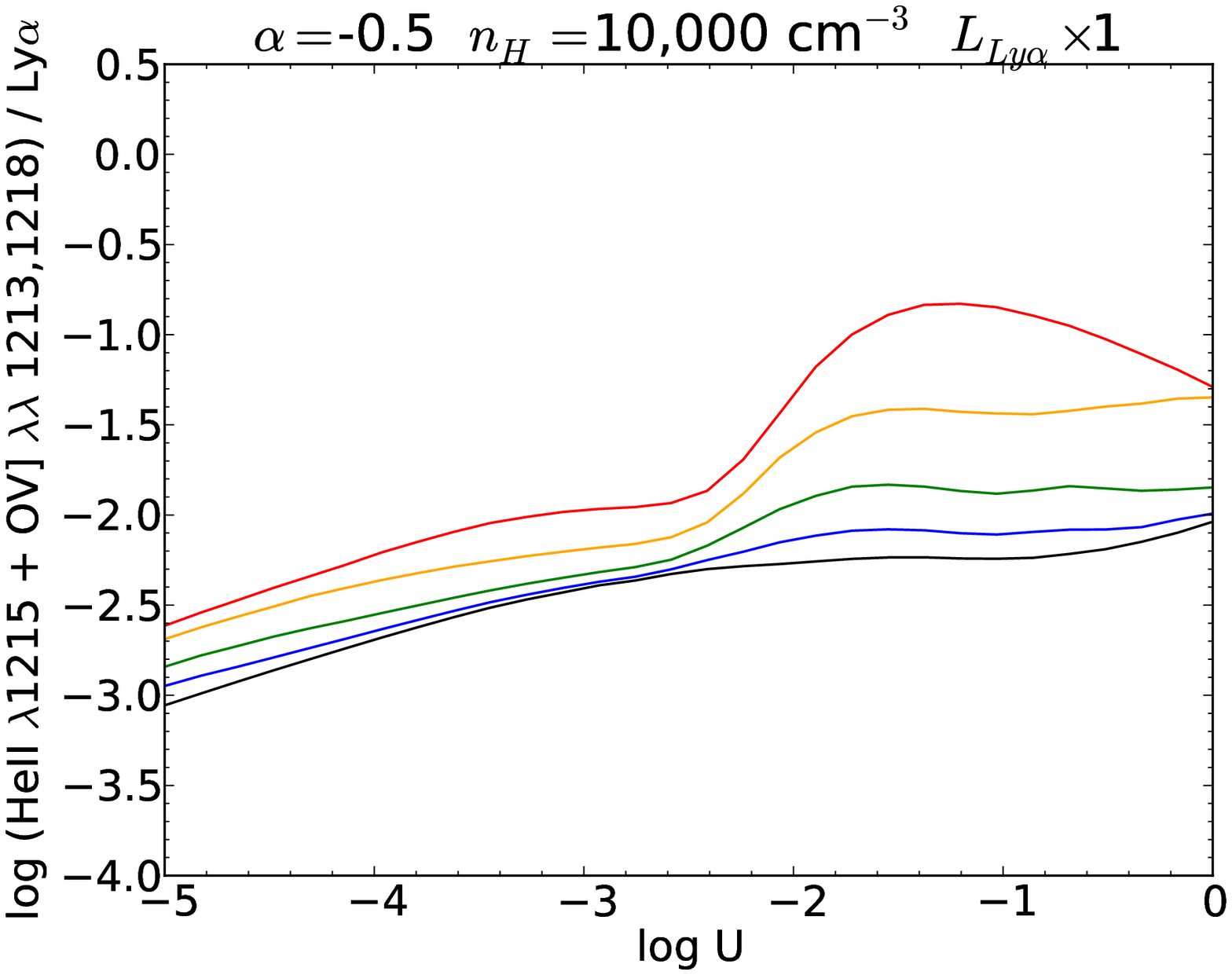}
\vspace{6.05in}
\caption{Cross-cuts through our grid of ionization-bounded
  (optically-thick) photoionization model grid, showing HeII+OV] /
  Ly$\alpha$ vs U curves for different fixed values of gas
  metallicity, $\alpha$ and $n_H$. In Figs. 1-6, the multiplicative
  transmission factor of Ly$\alpha$ ($\le$1) is shown above each panel.}
\label{fig1}
\end{figure*}

\begin{figure*}
\includegraphics{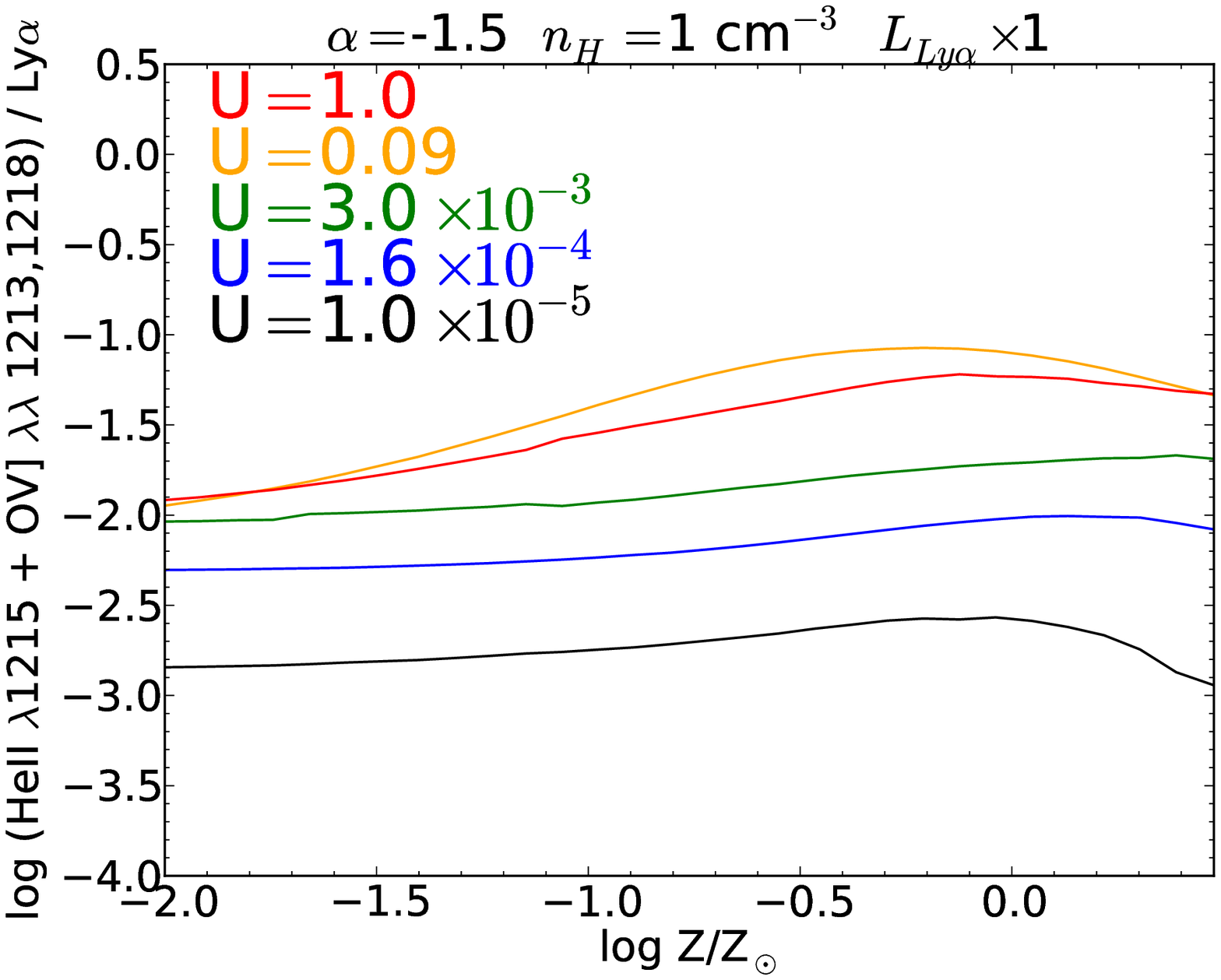}
\includegraphics{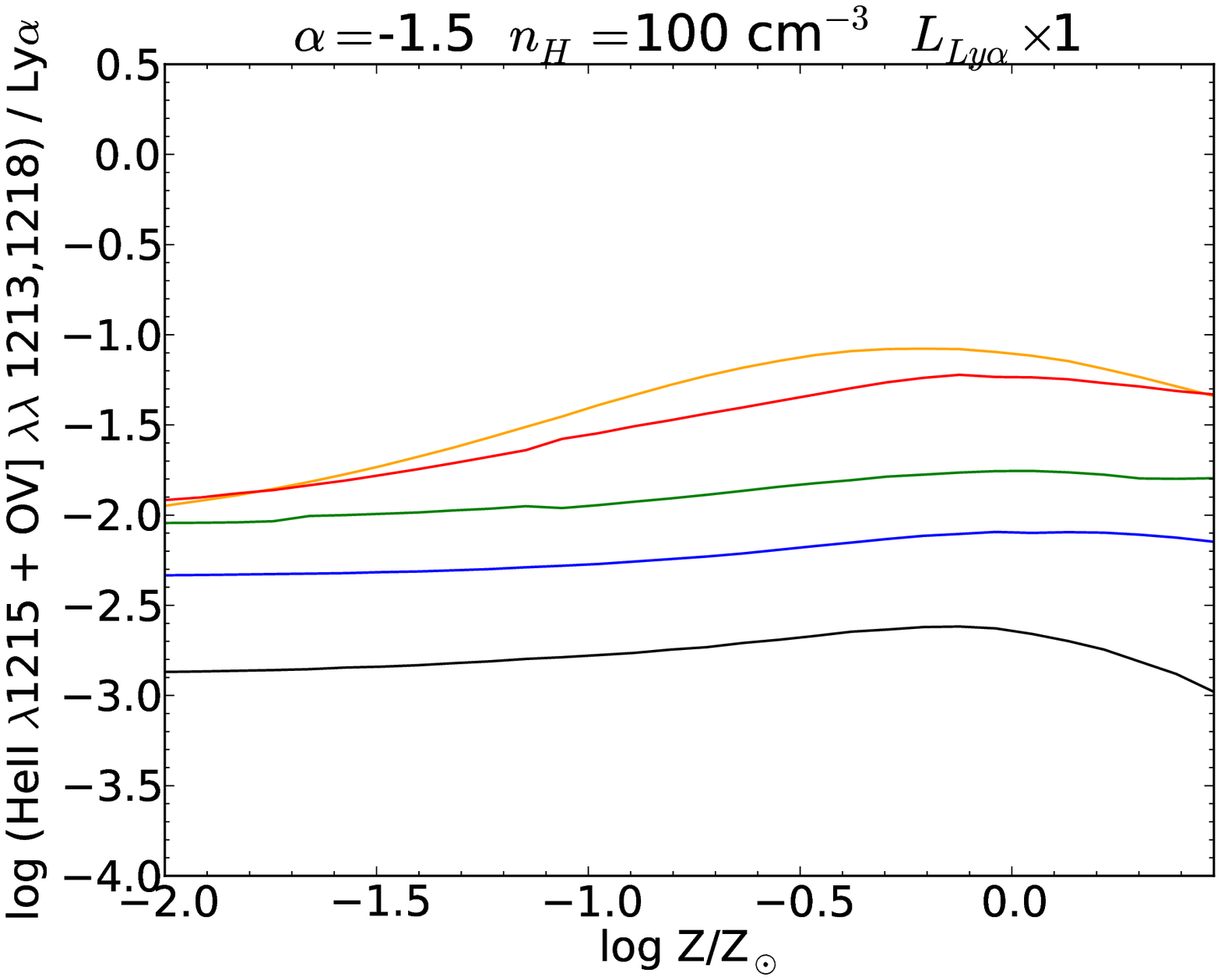}
\includegraphics{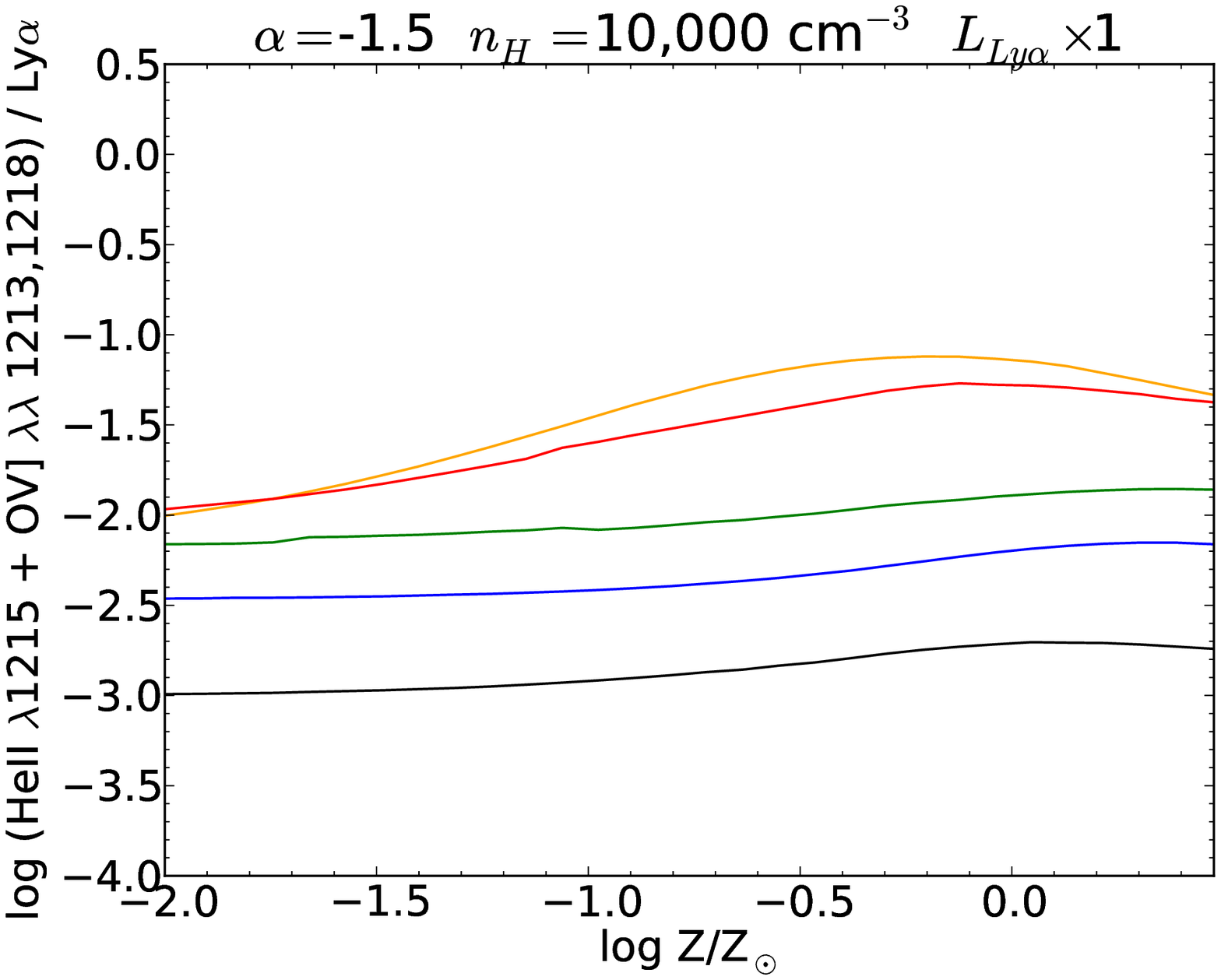}
\includegraphics{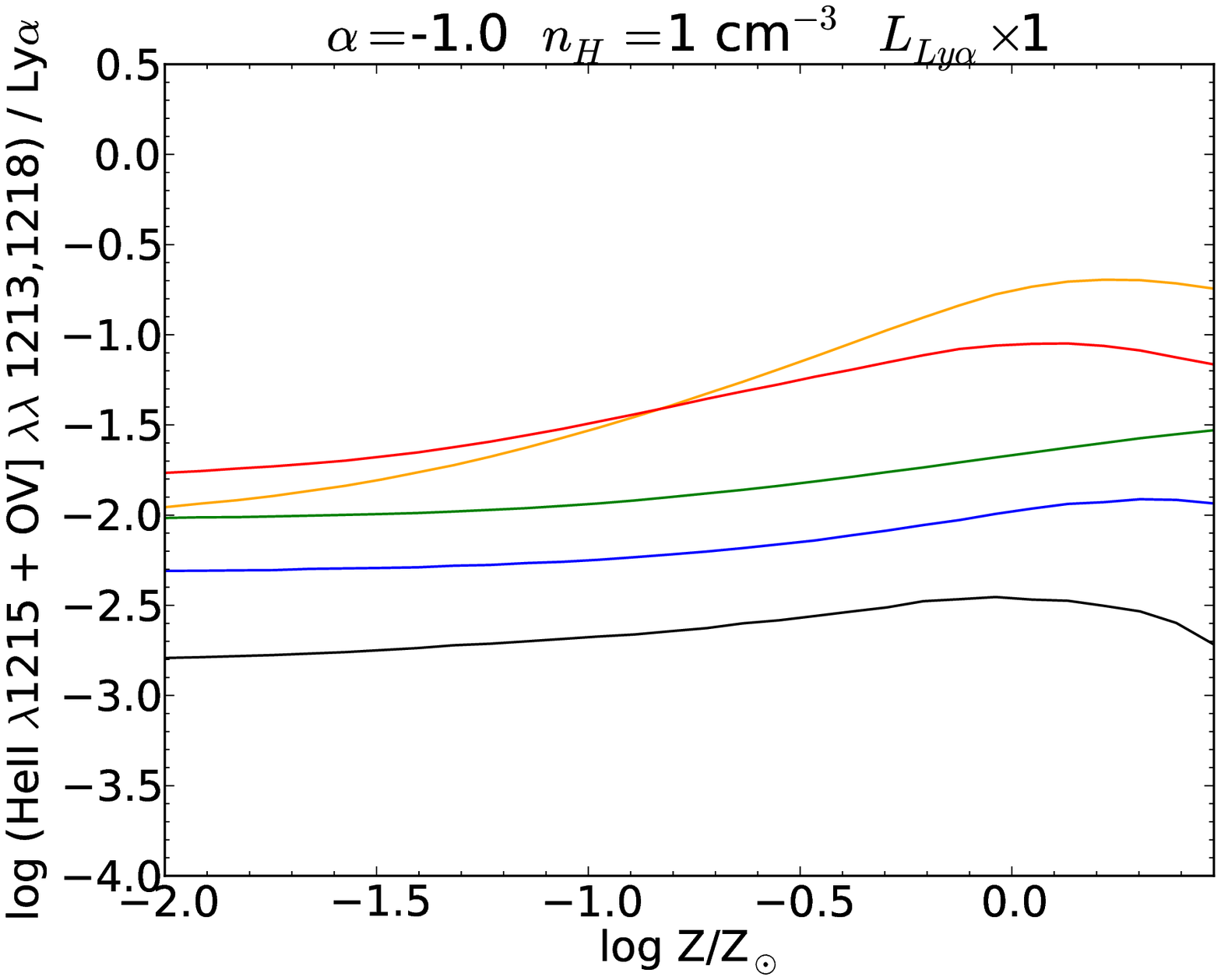}
\includegraphics{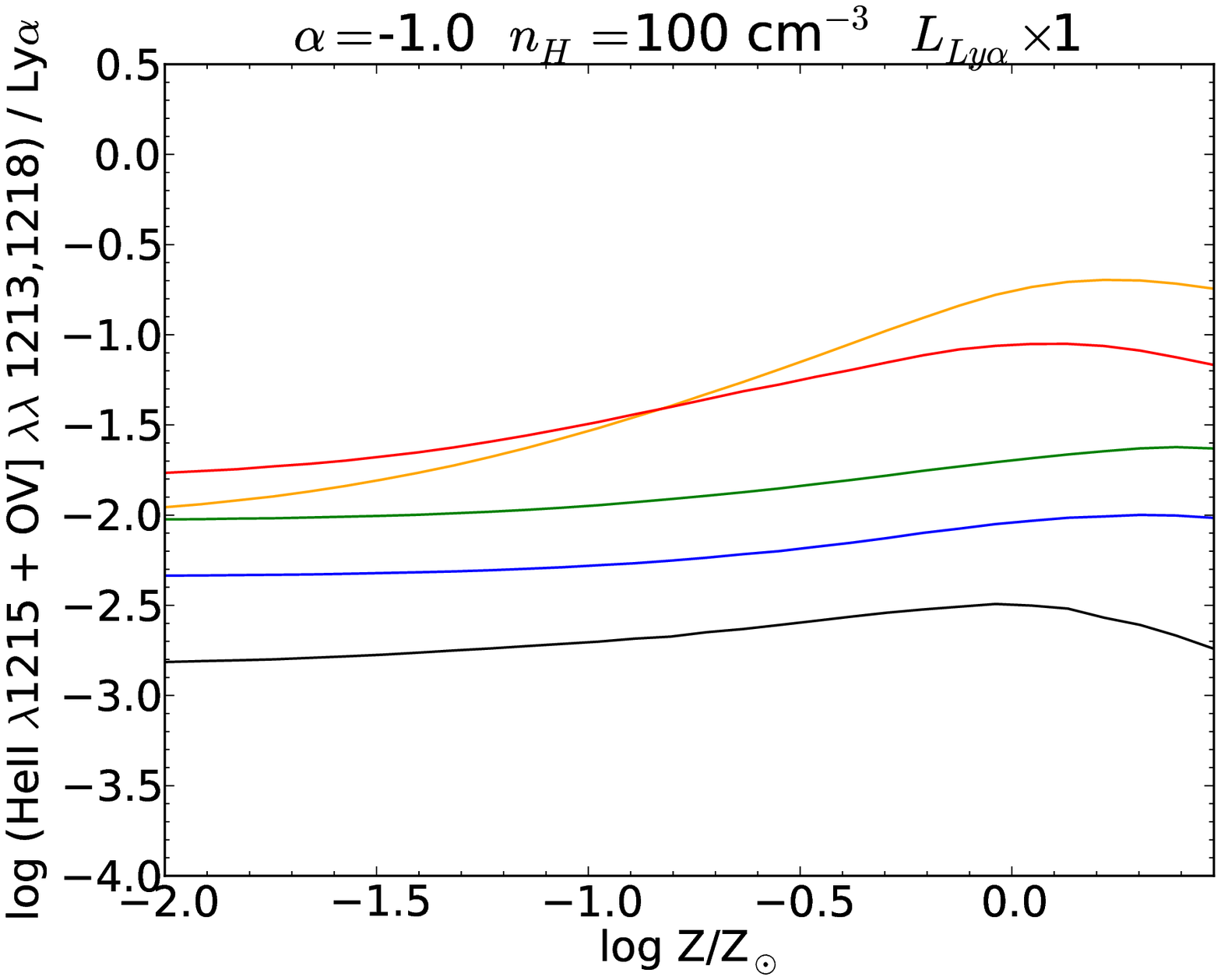}
\includegraphics{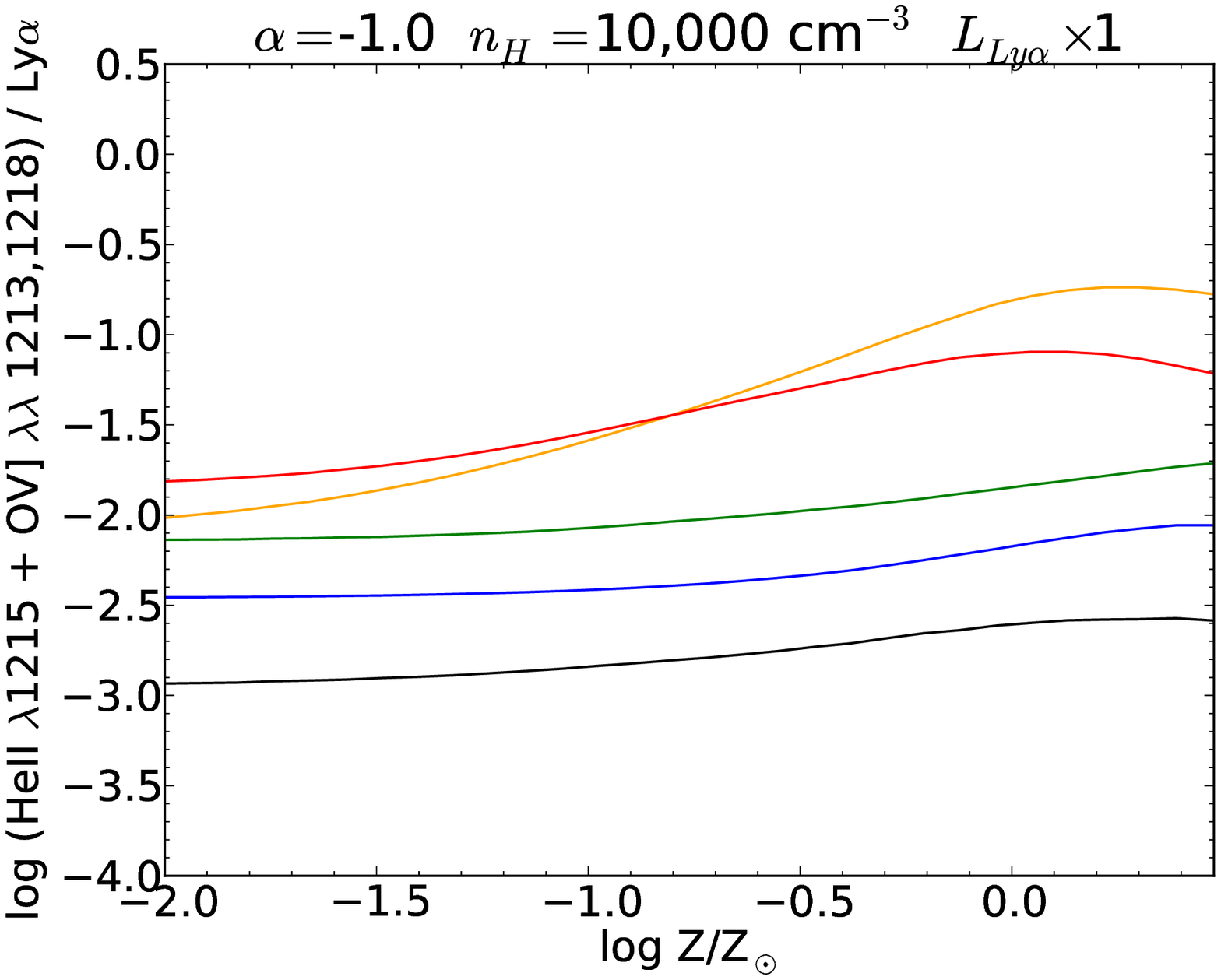}
\includegraphics{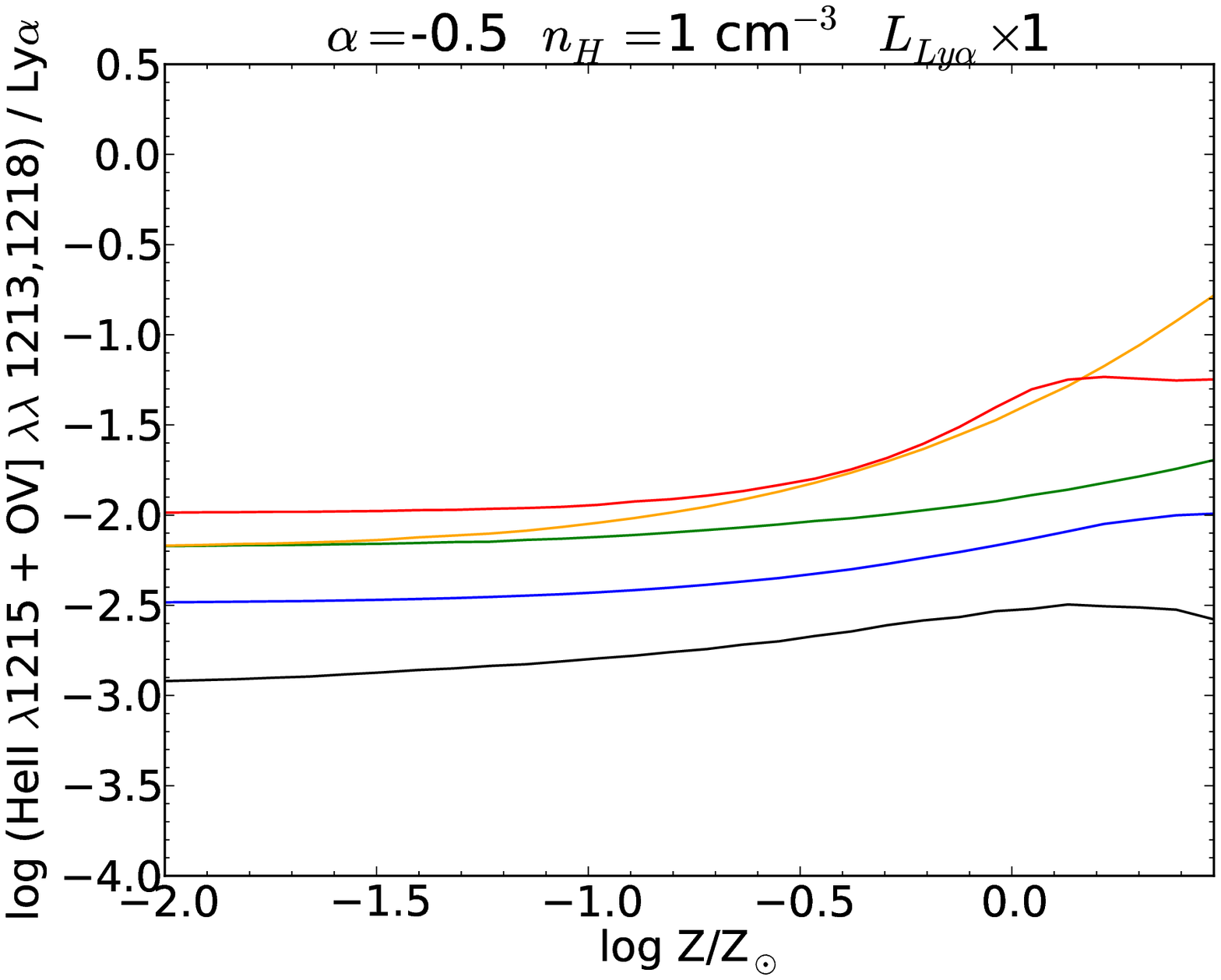}
\includegraphics{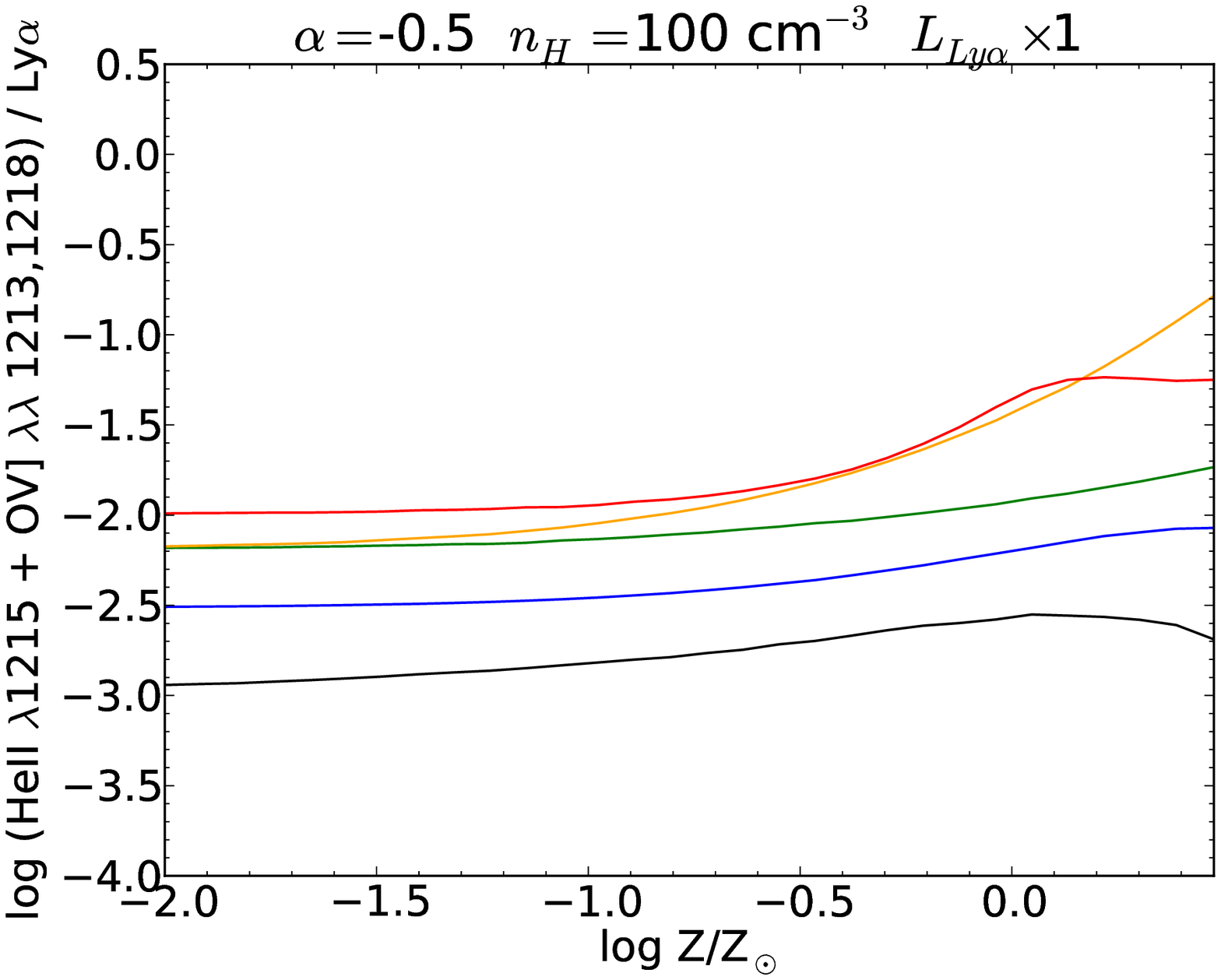}
\includegraphics{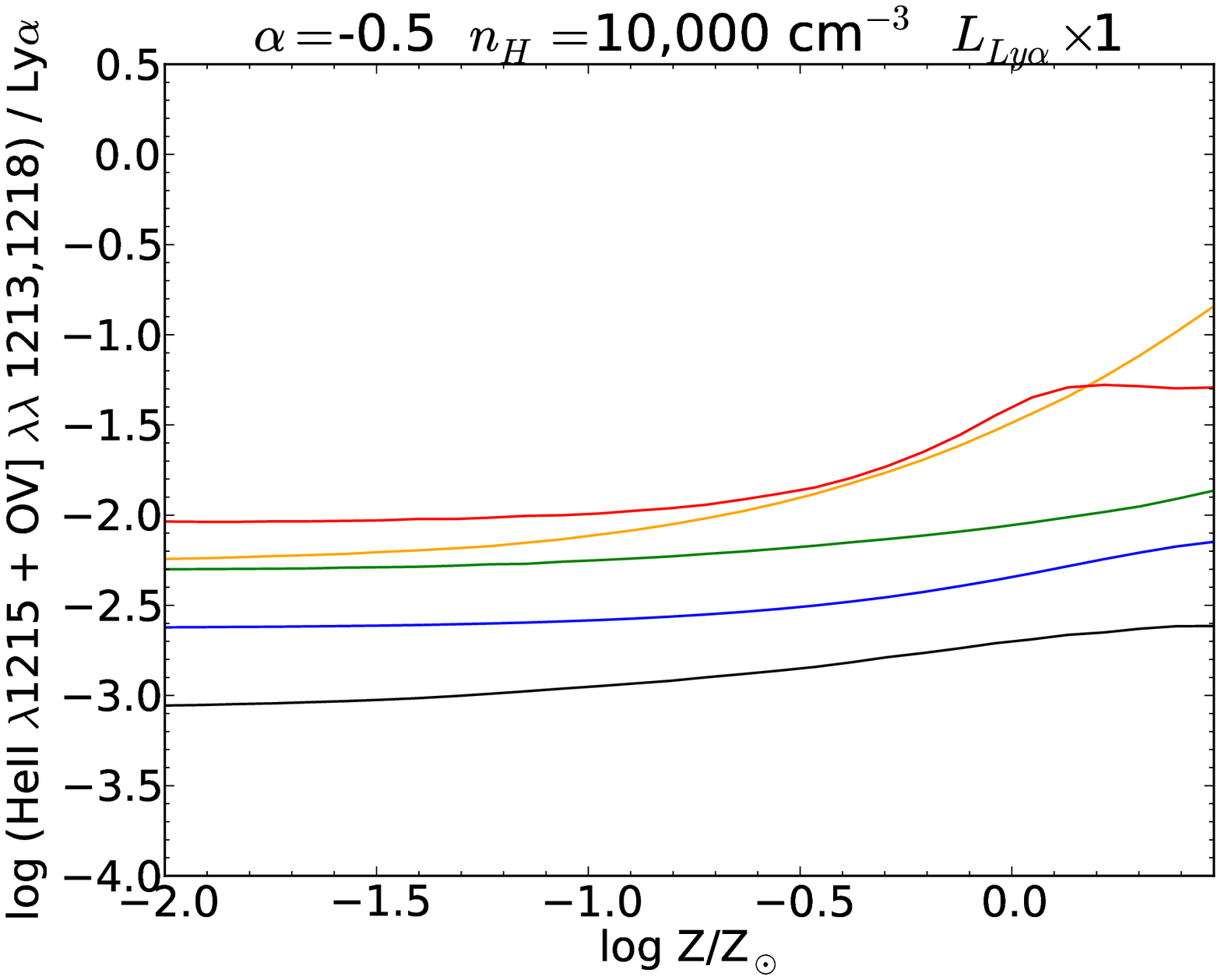}
\vspace{6.05in}
\caption{Similar to Fig. ~\ref{fig1}, but showing HeII+OV] / Ly$\alpha$ vs. metallicity curves for different fixed values of U, $\alpha$ and $n_H$.}
\label{fig2}
\end{figure*}

\begin{figure*}
\includegraphics{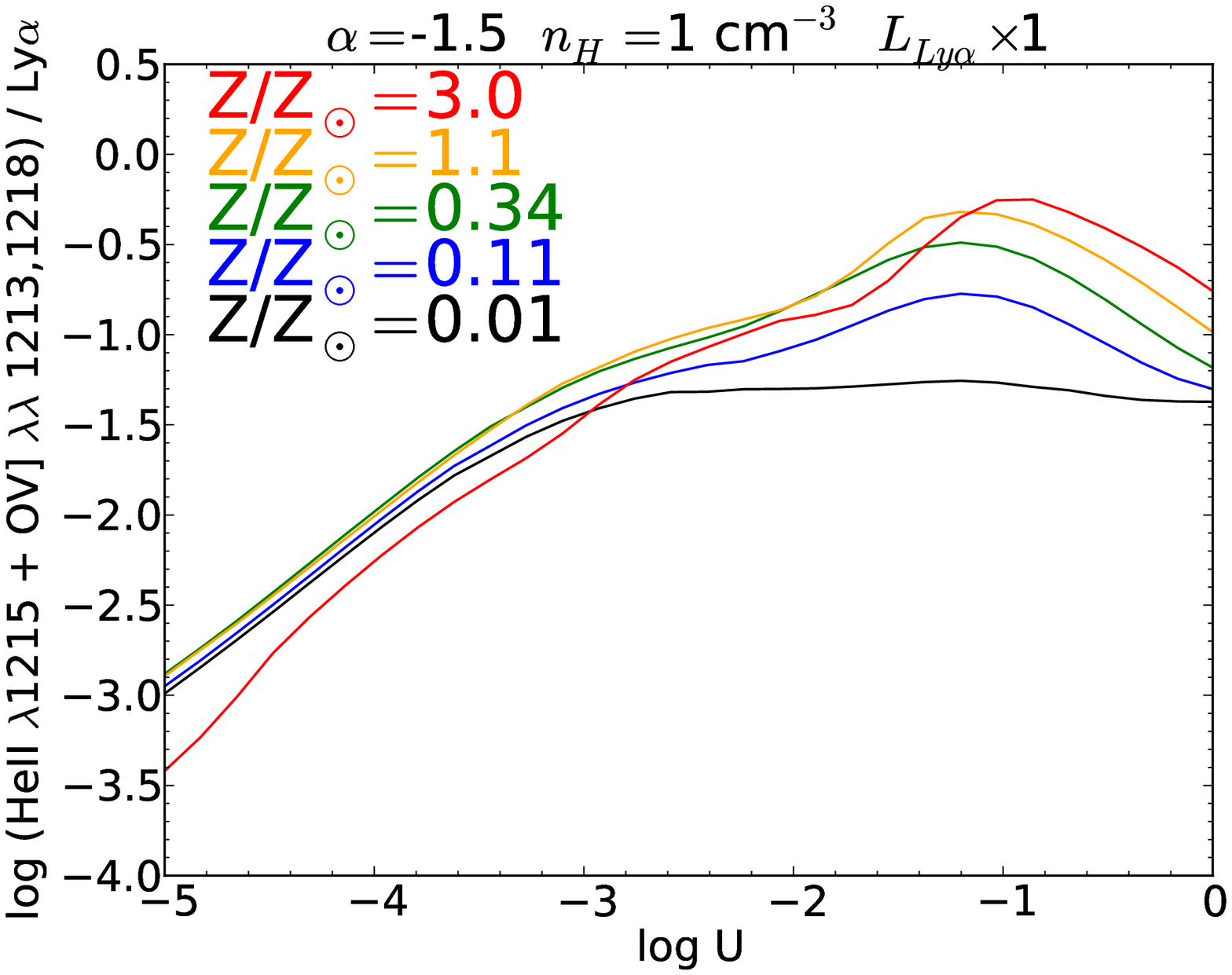}
\includegraphics{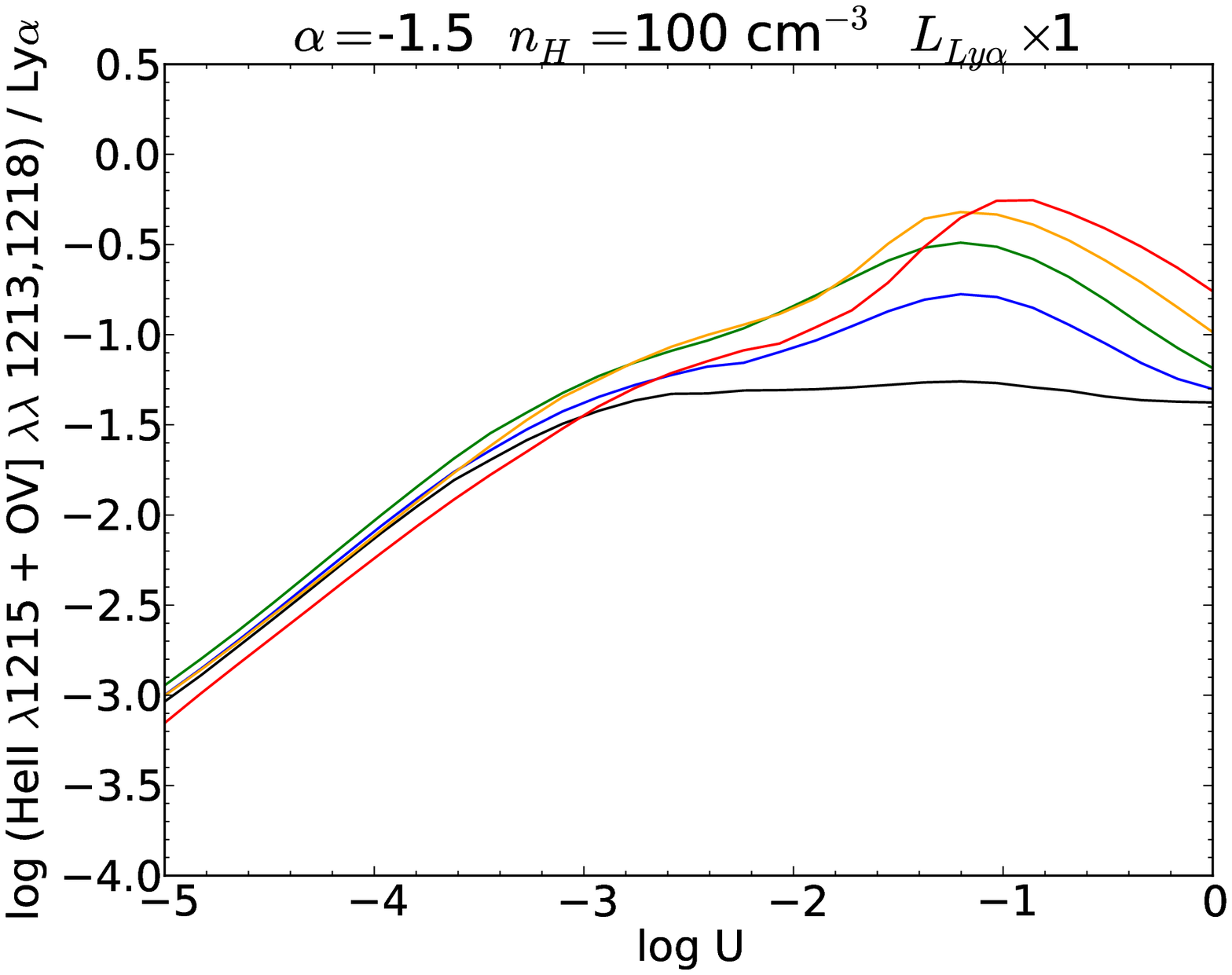}
\includegraphics{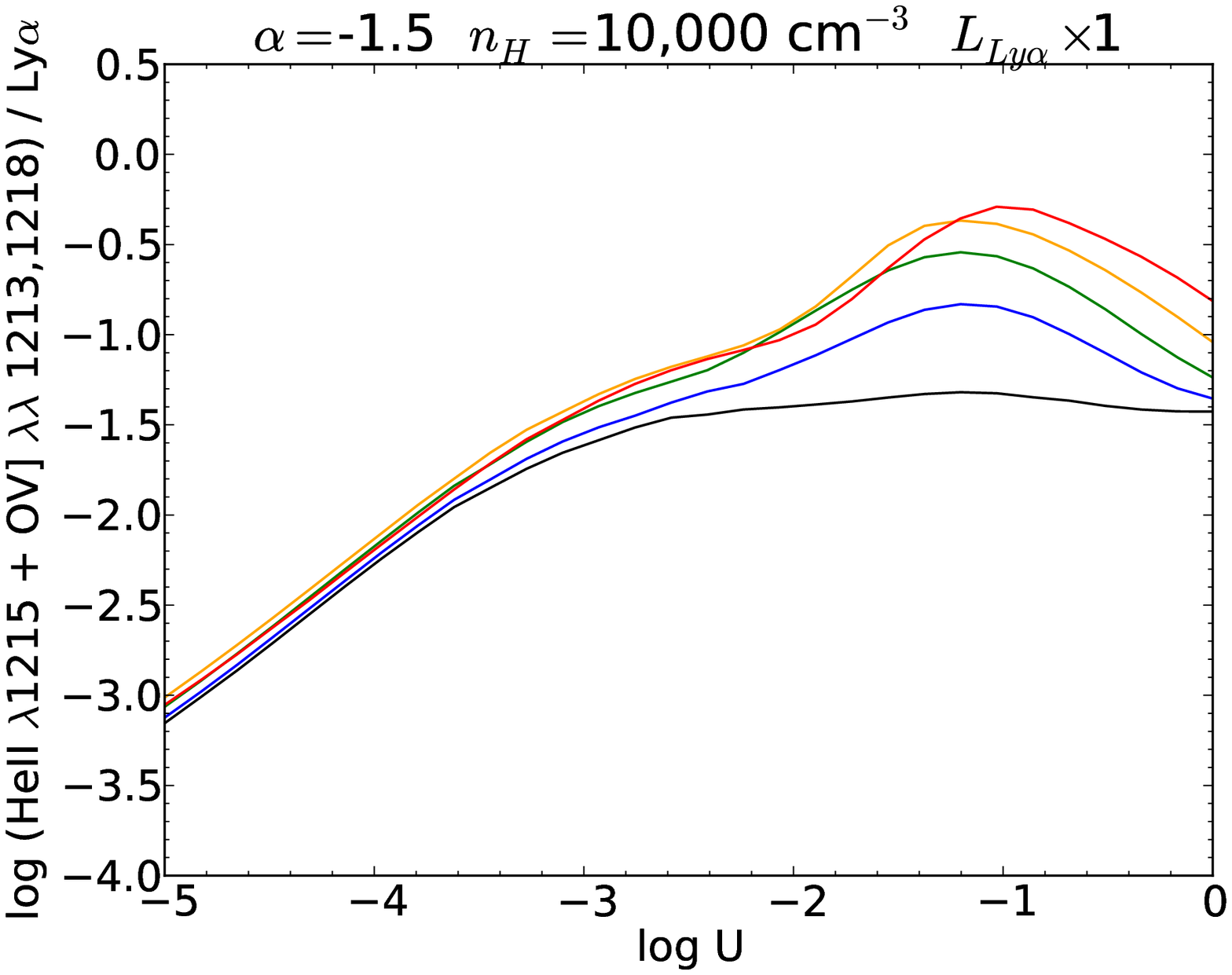}
\includegraphics{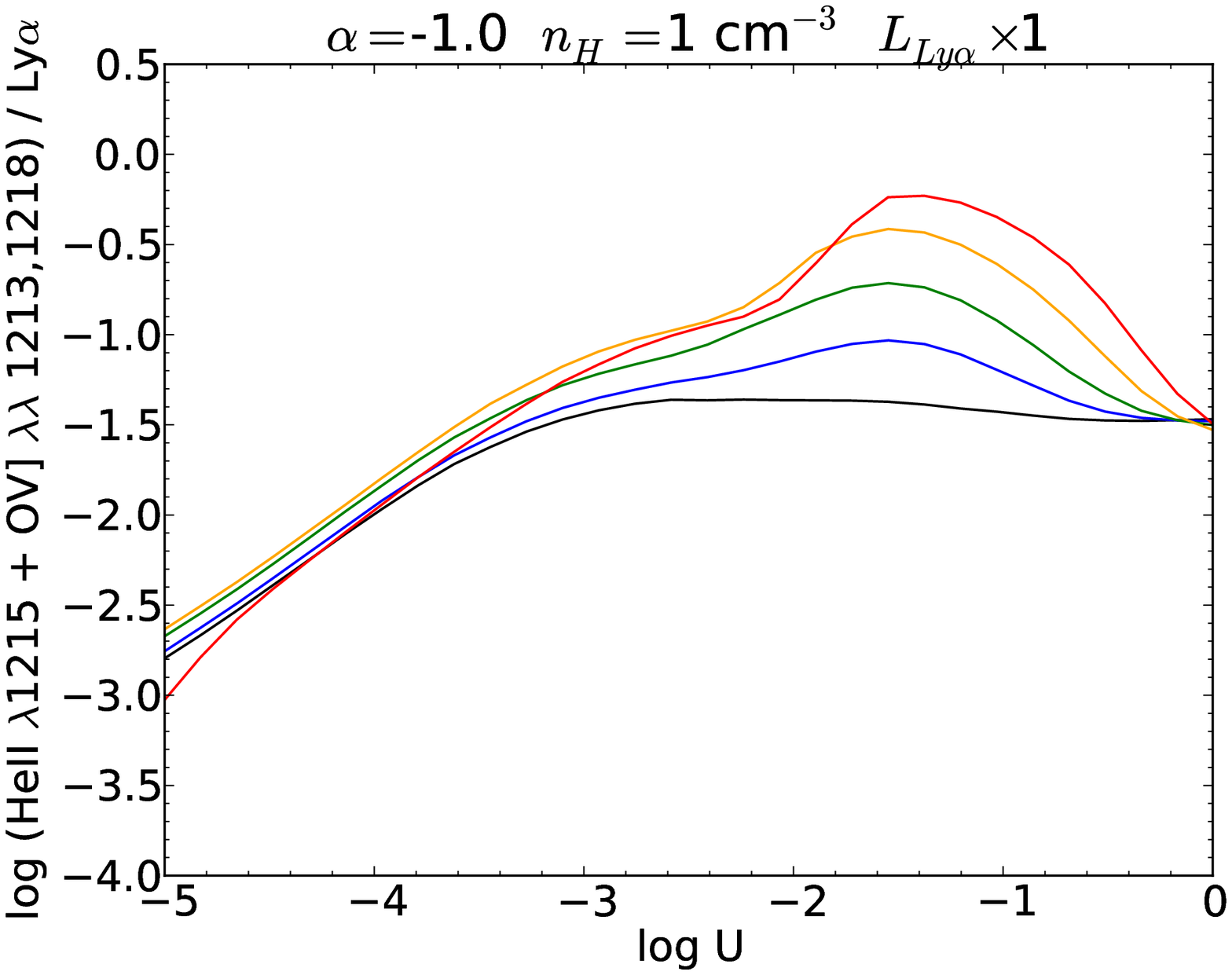}
\includegraphics{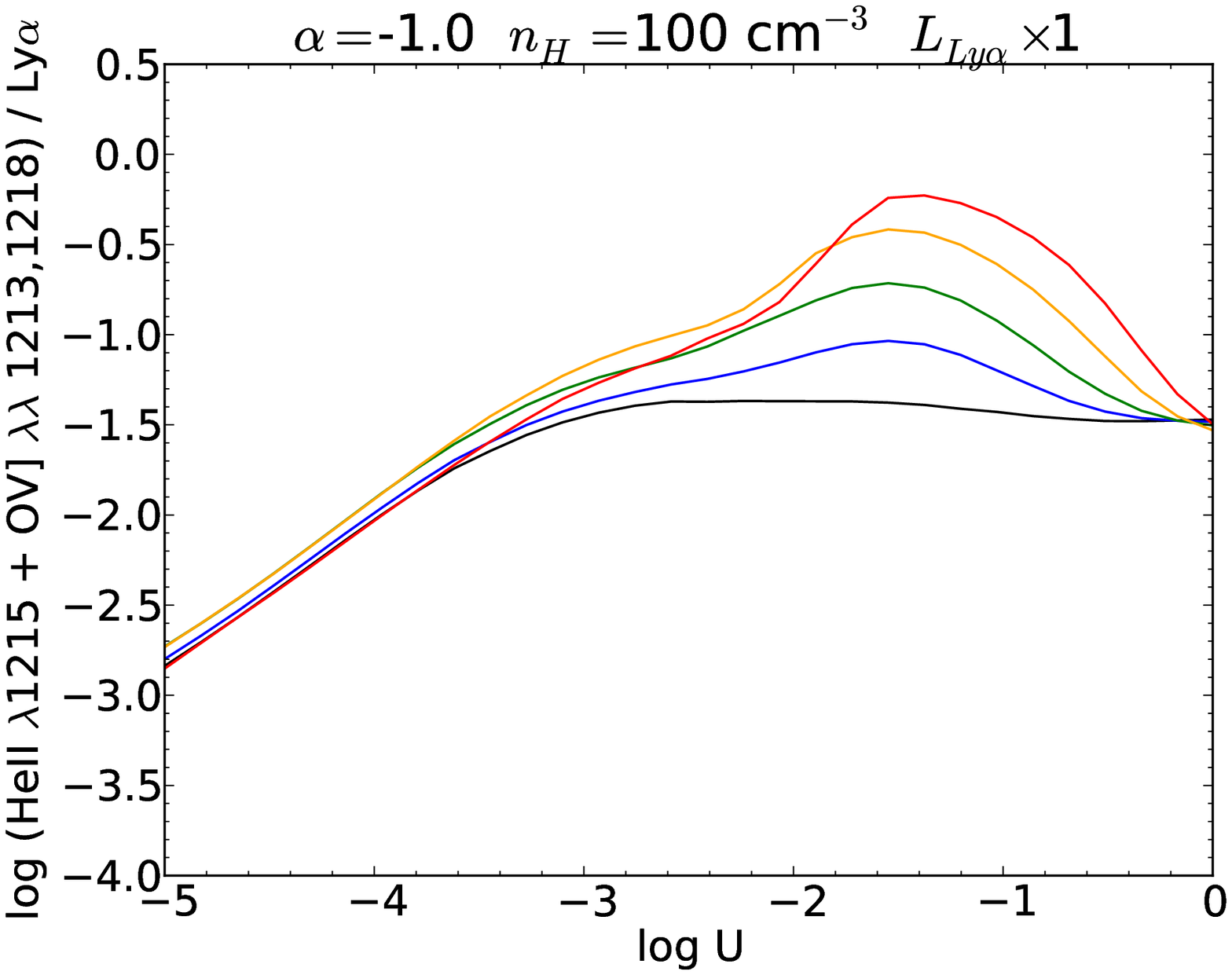}
\includegraphics{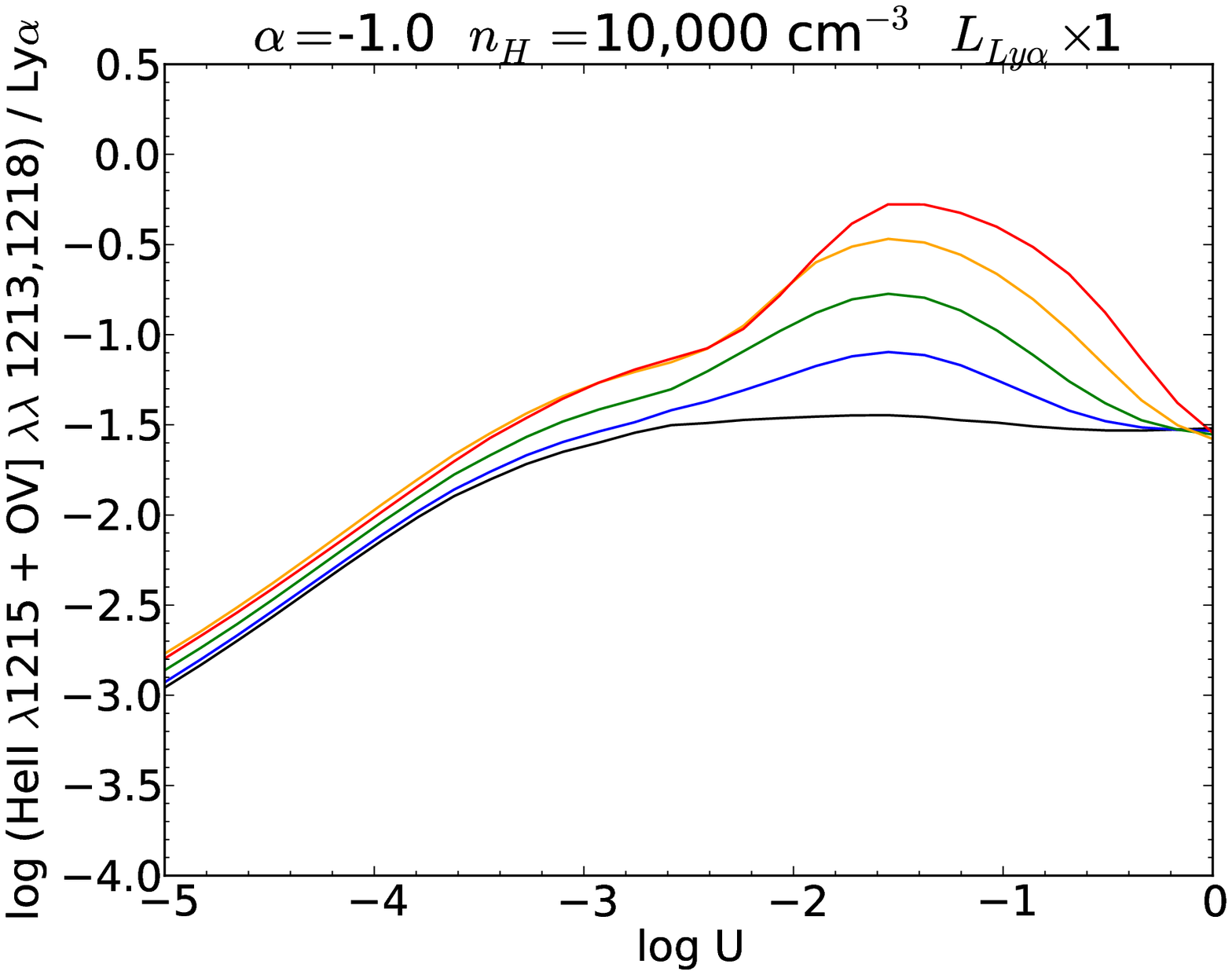}
\includegraphics{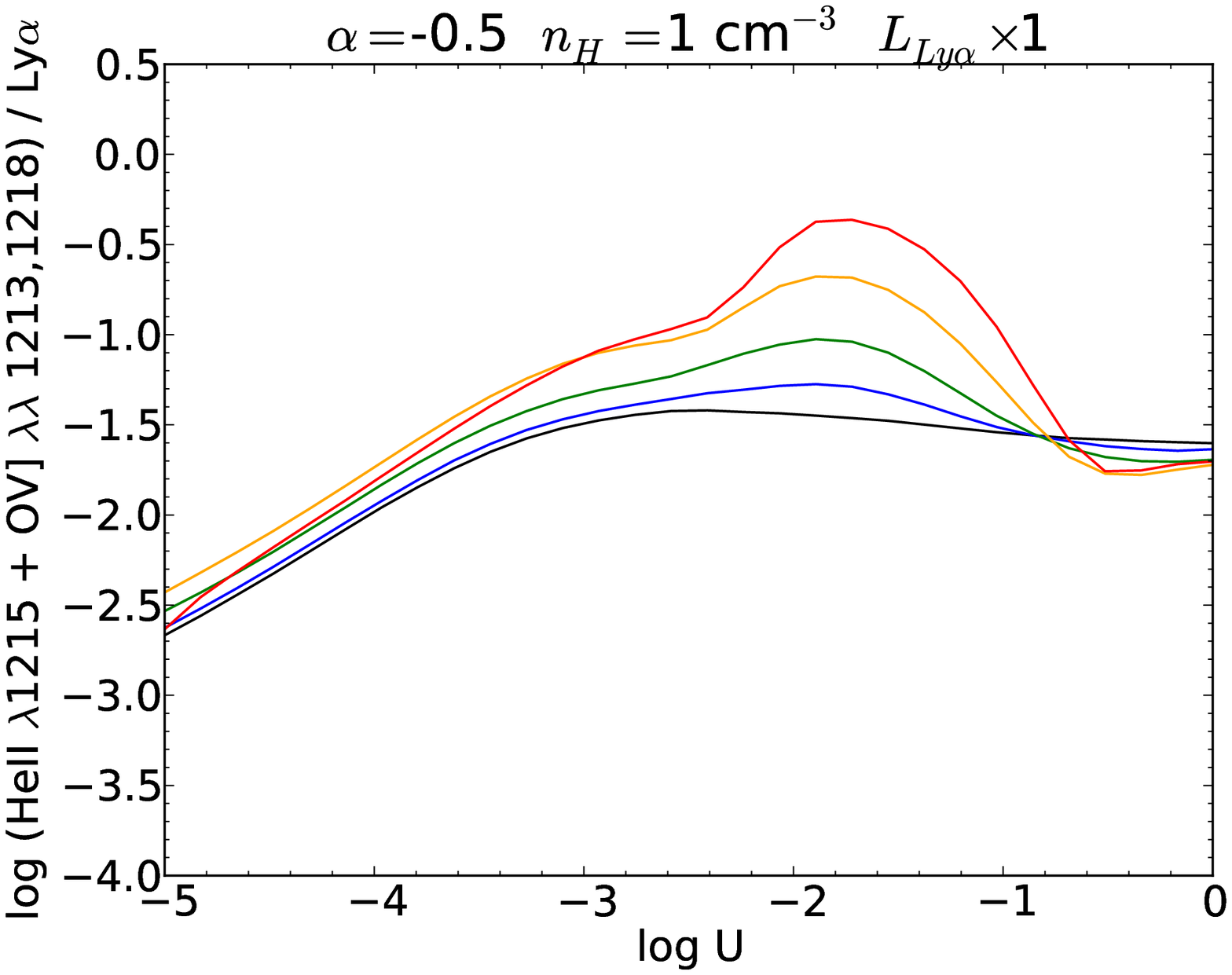}
\includegraphics{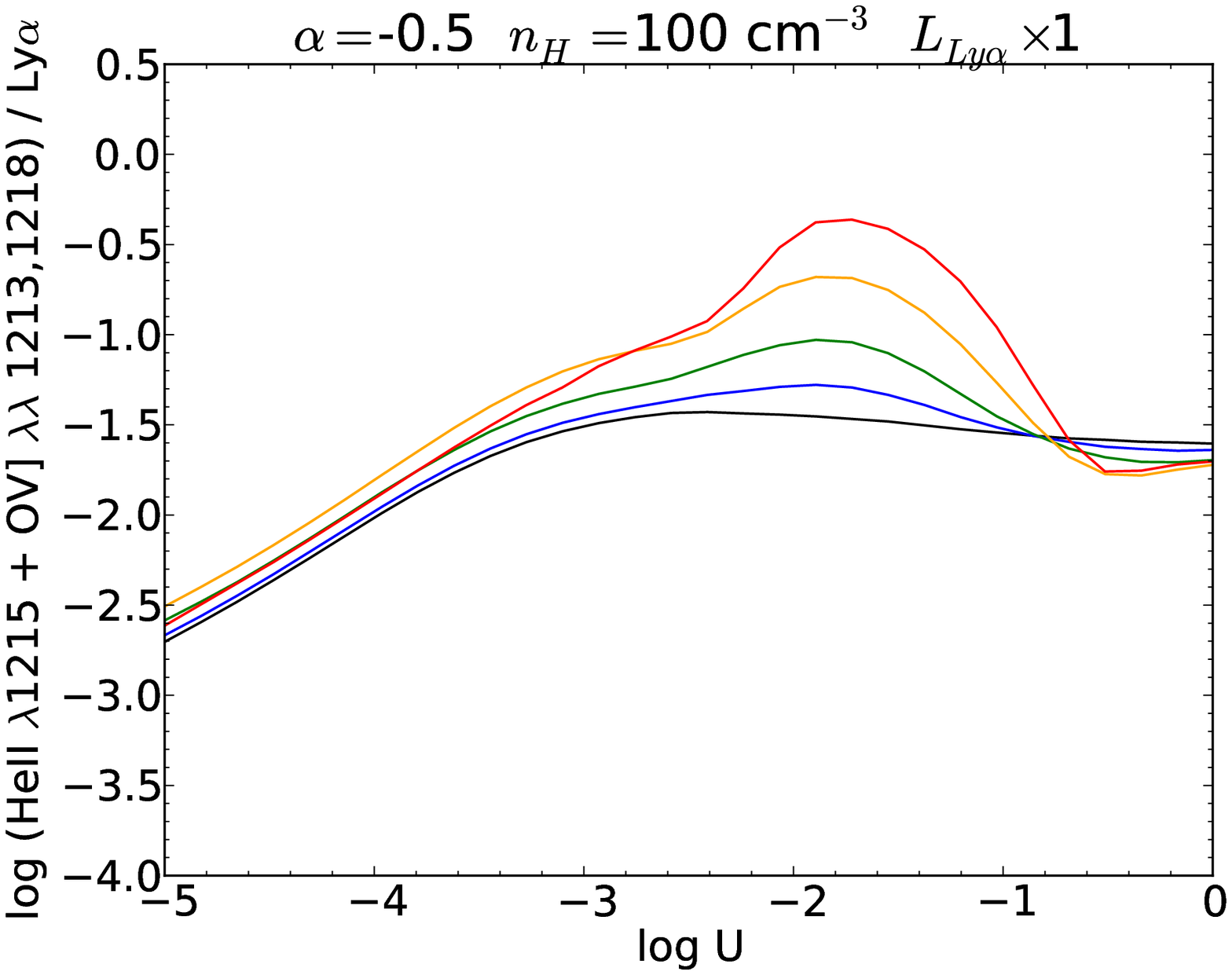}
\includegraphics{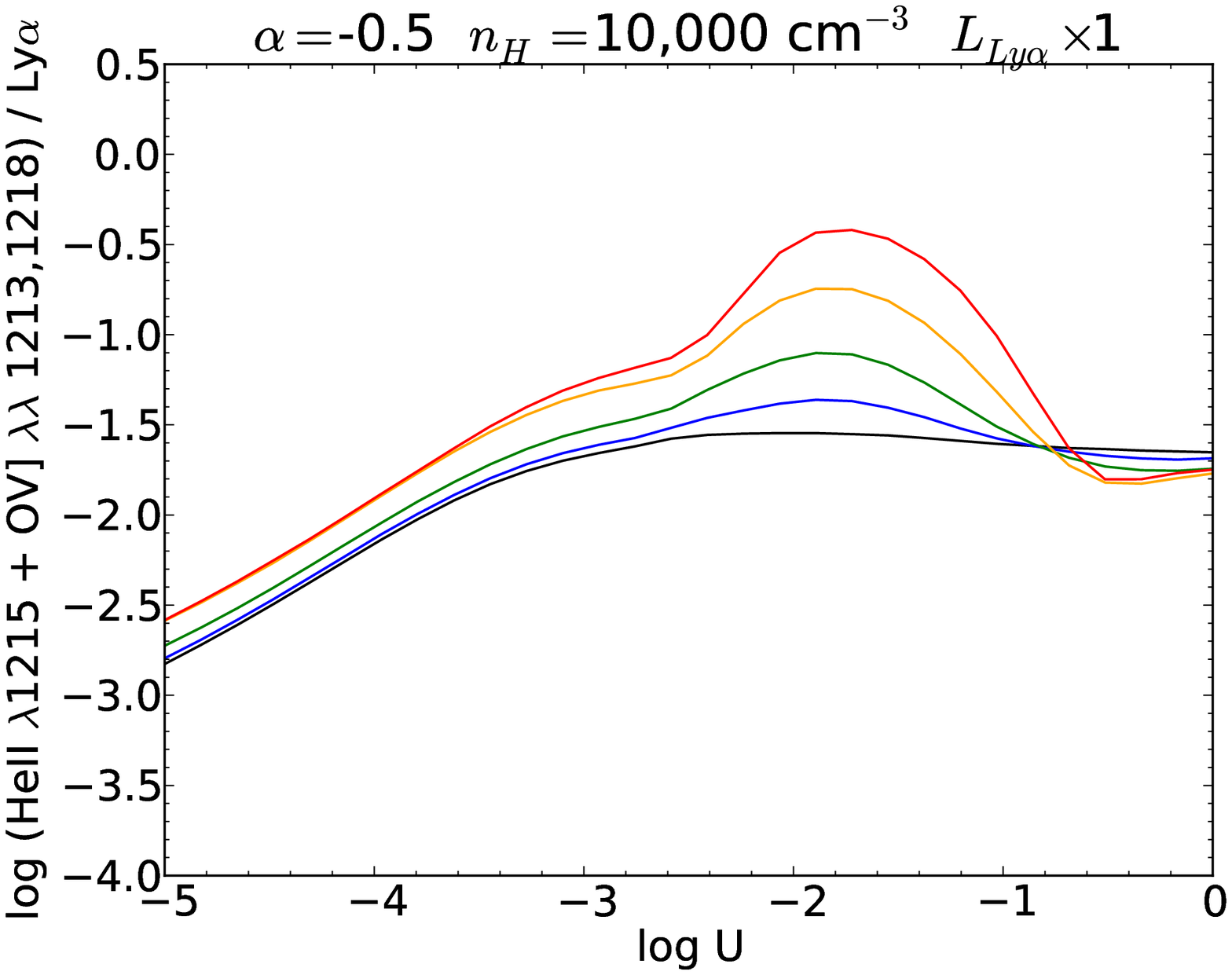}
\vspace{6.05in}
\caption{Cross-cuts through our grid of matter-bounded (optically-thin) photoionization model grid, showing HeII+OV] / Ly$\alpha$ vs U curves for different fixed values of gas metallicity, $\alpha$ and $n_H$.}
\label{fig3}
\end{figure*}

\begin{figure*}
\includegraphics{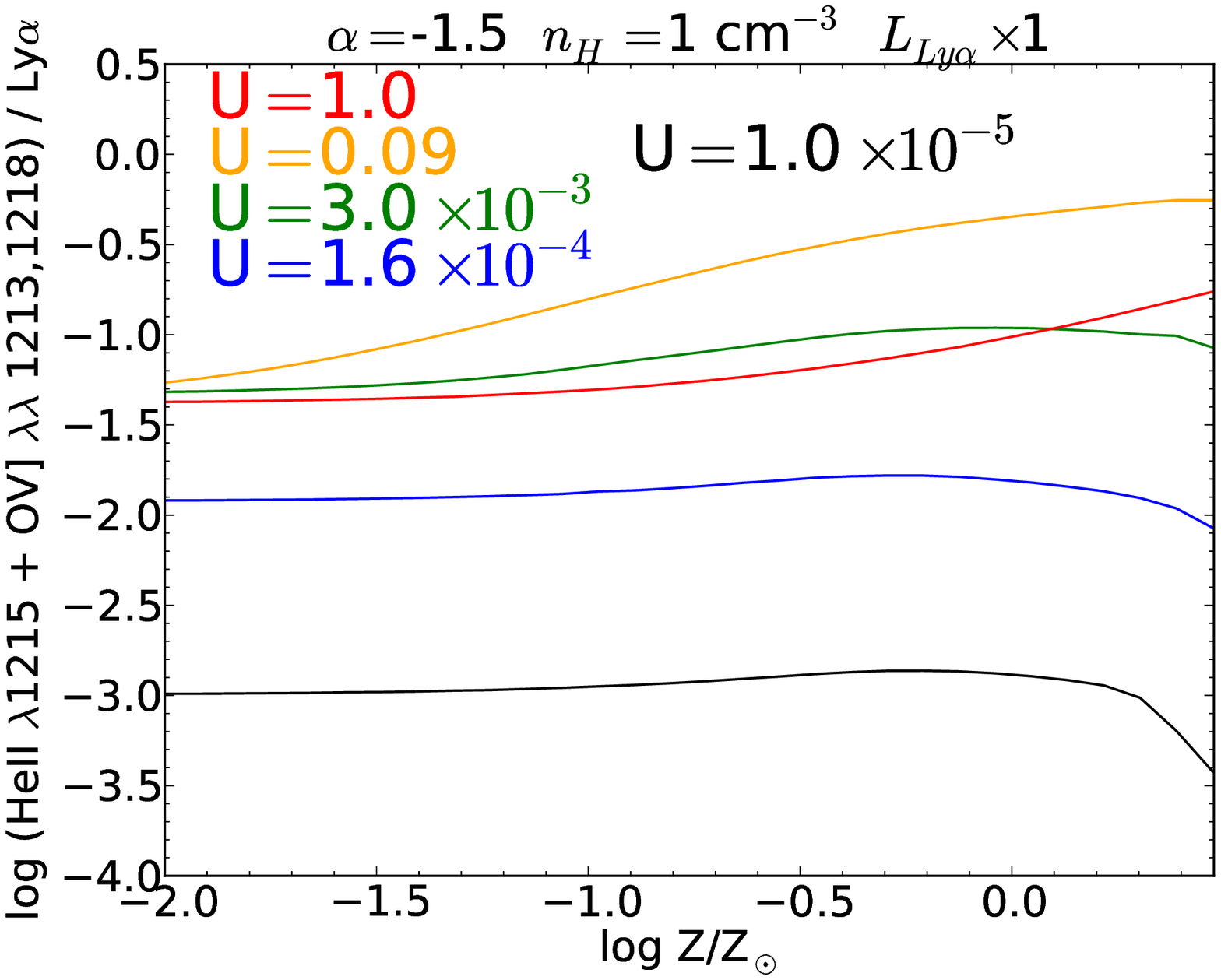}
\includegraphics{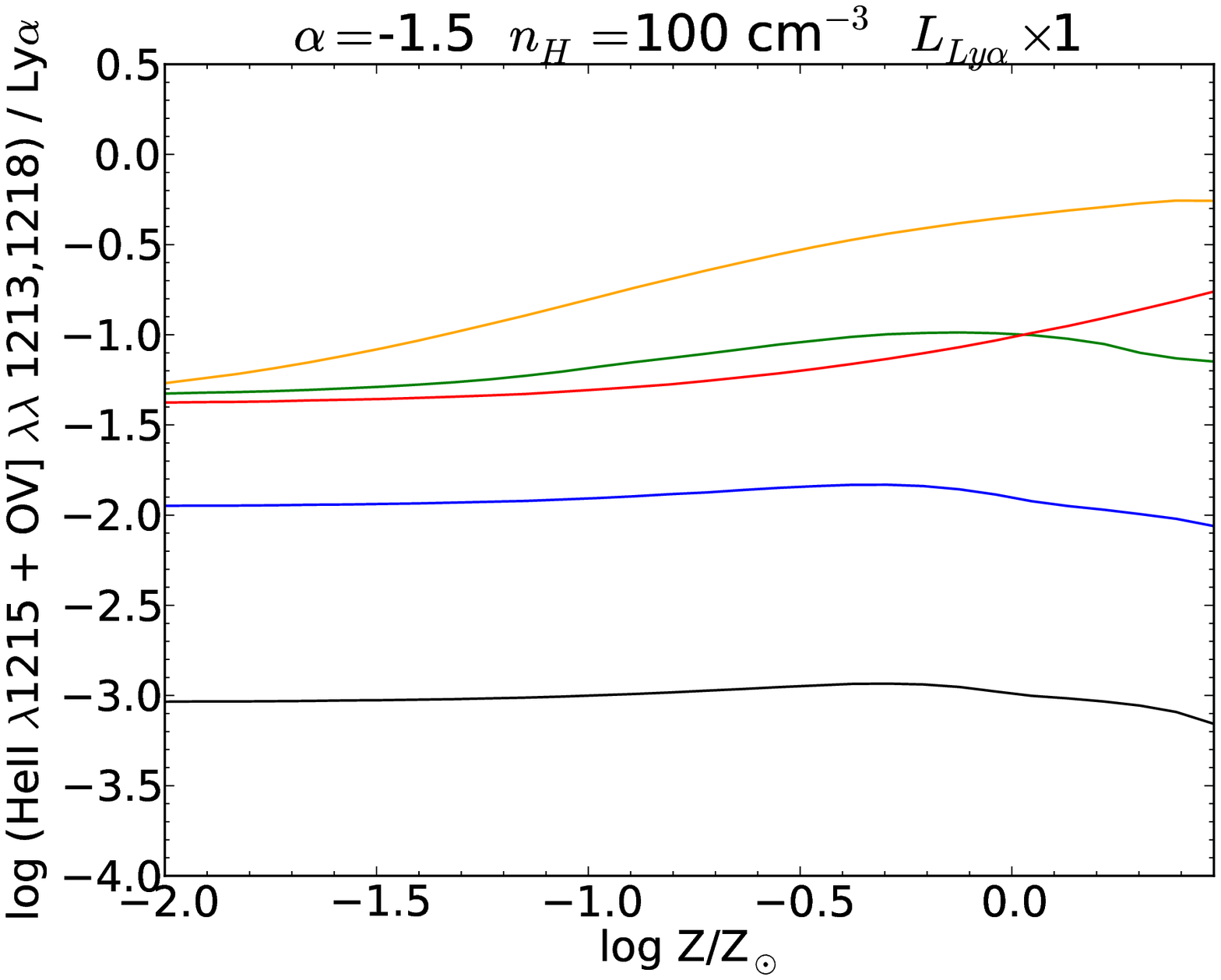}
\includegraphics{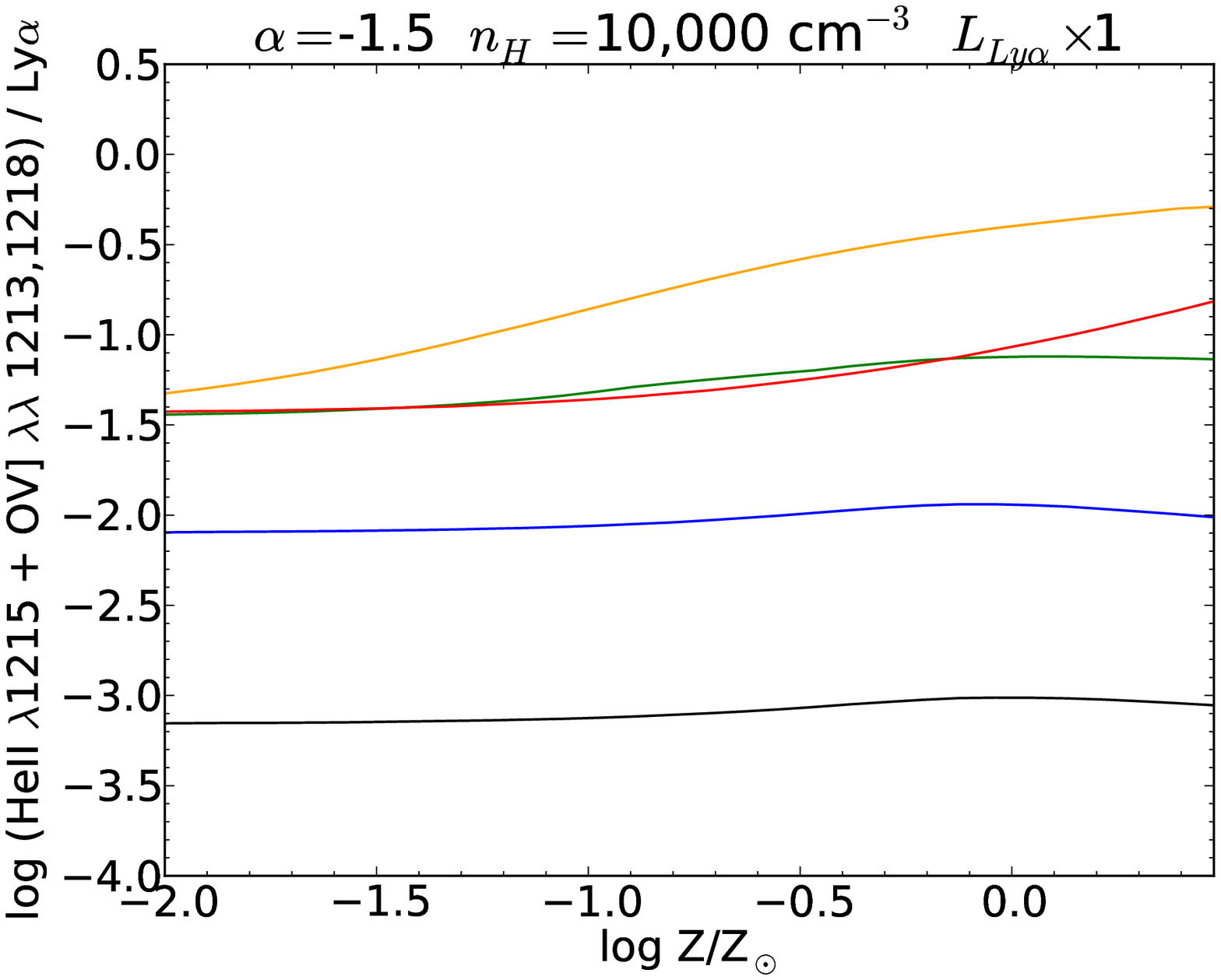}
\includegraphics{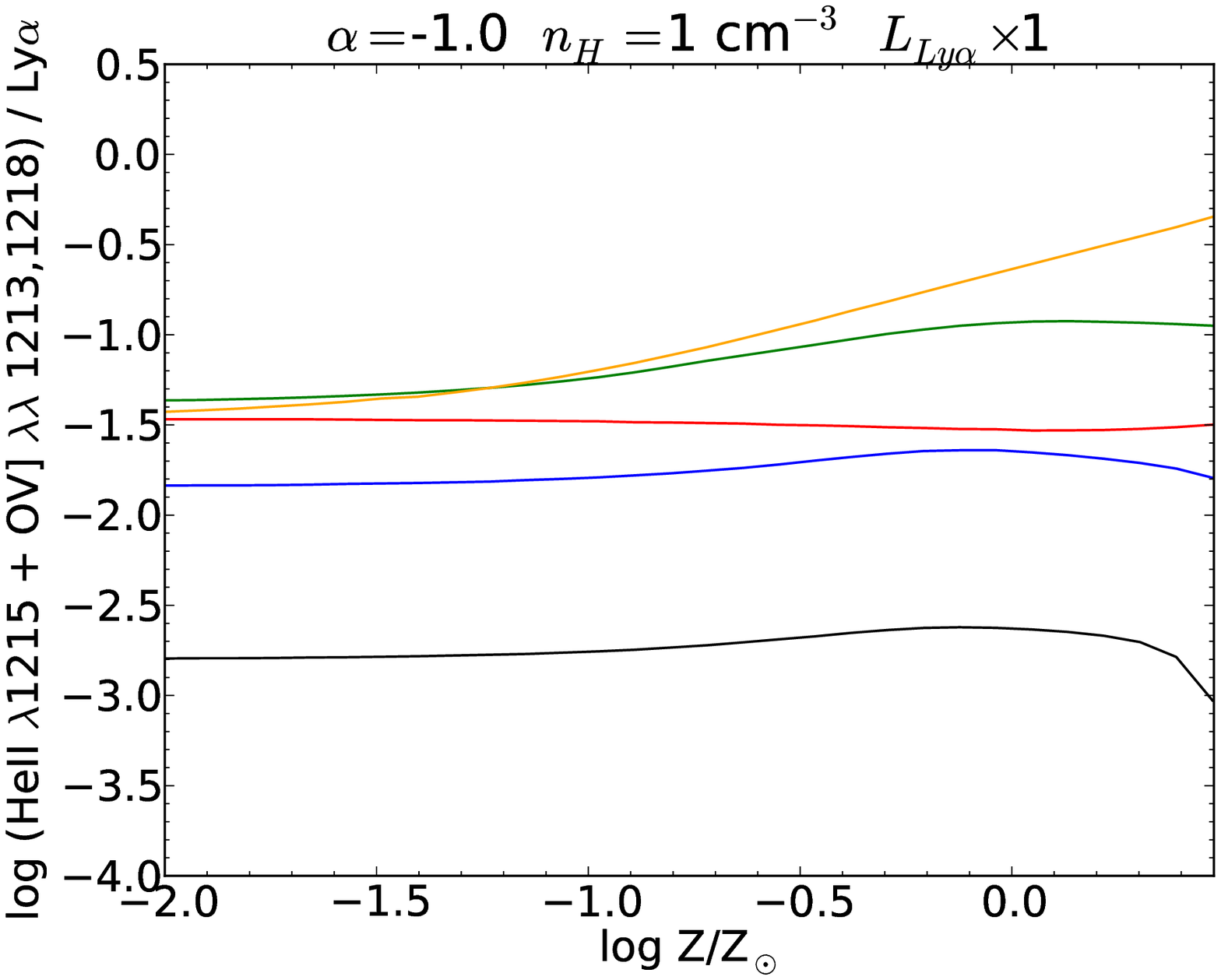}
\includegraphics{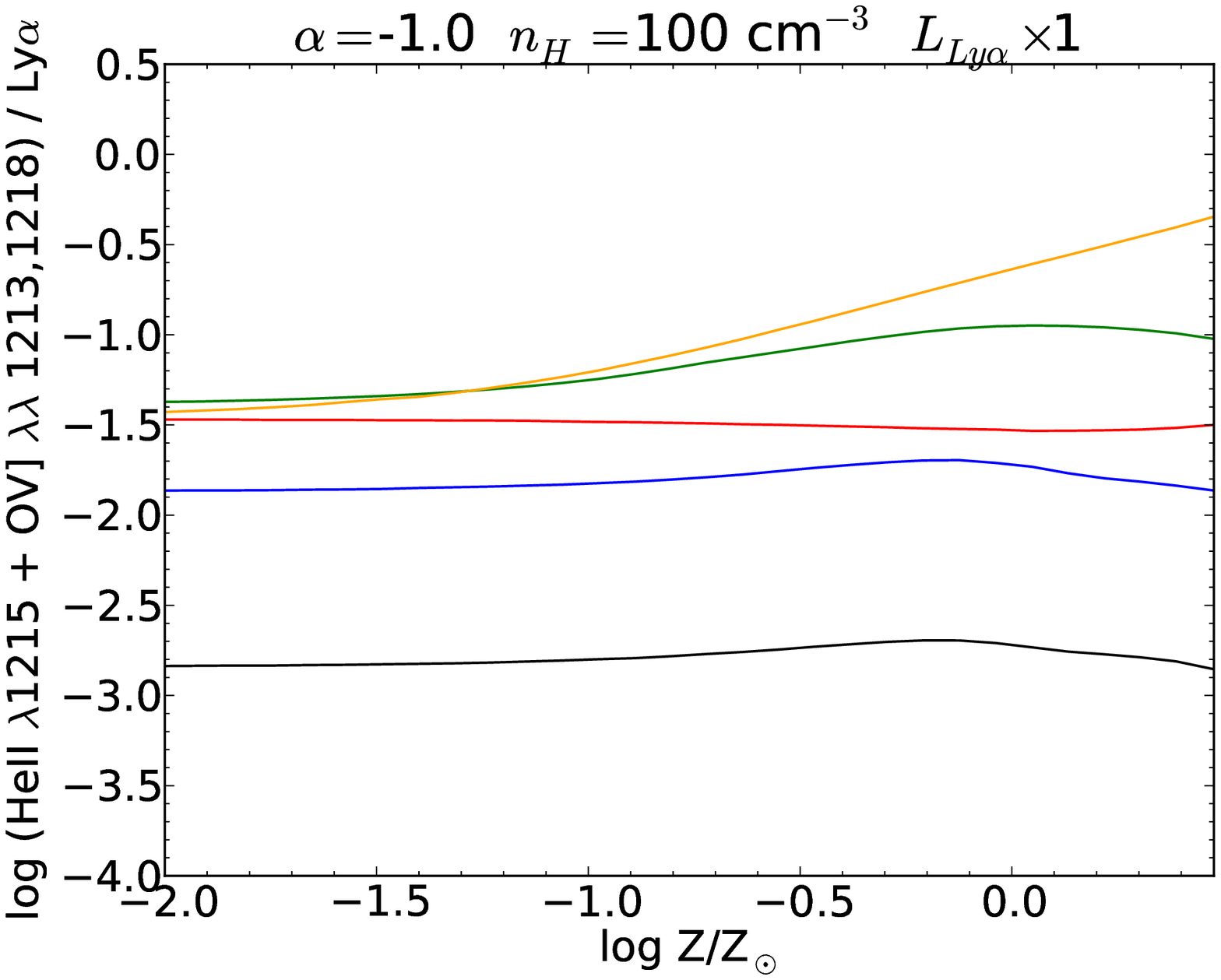}
\includegraphics{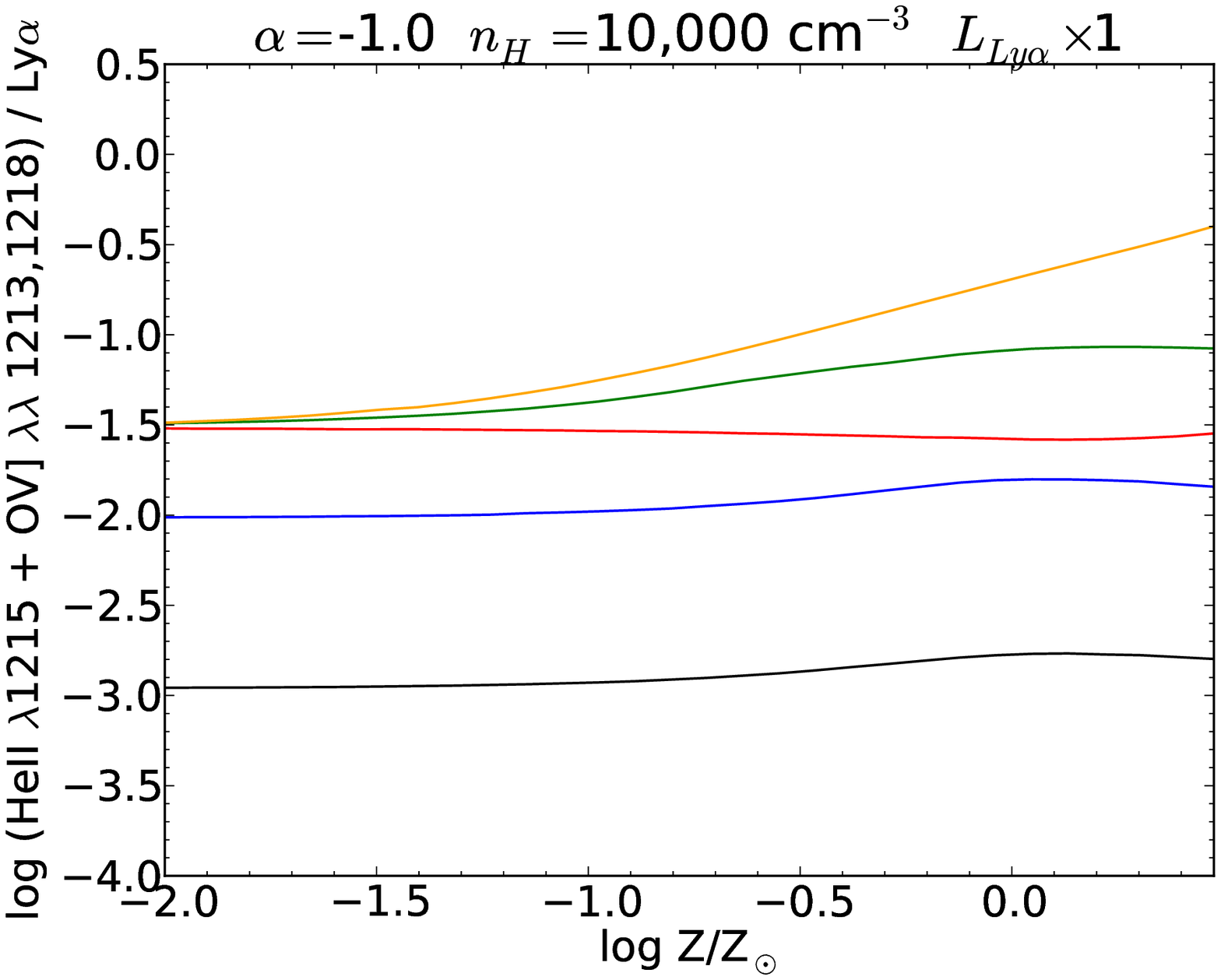}
\includegraphics{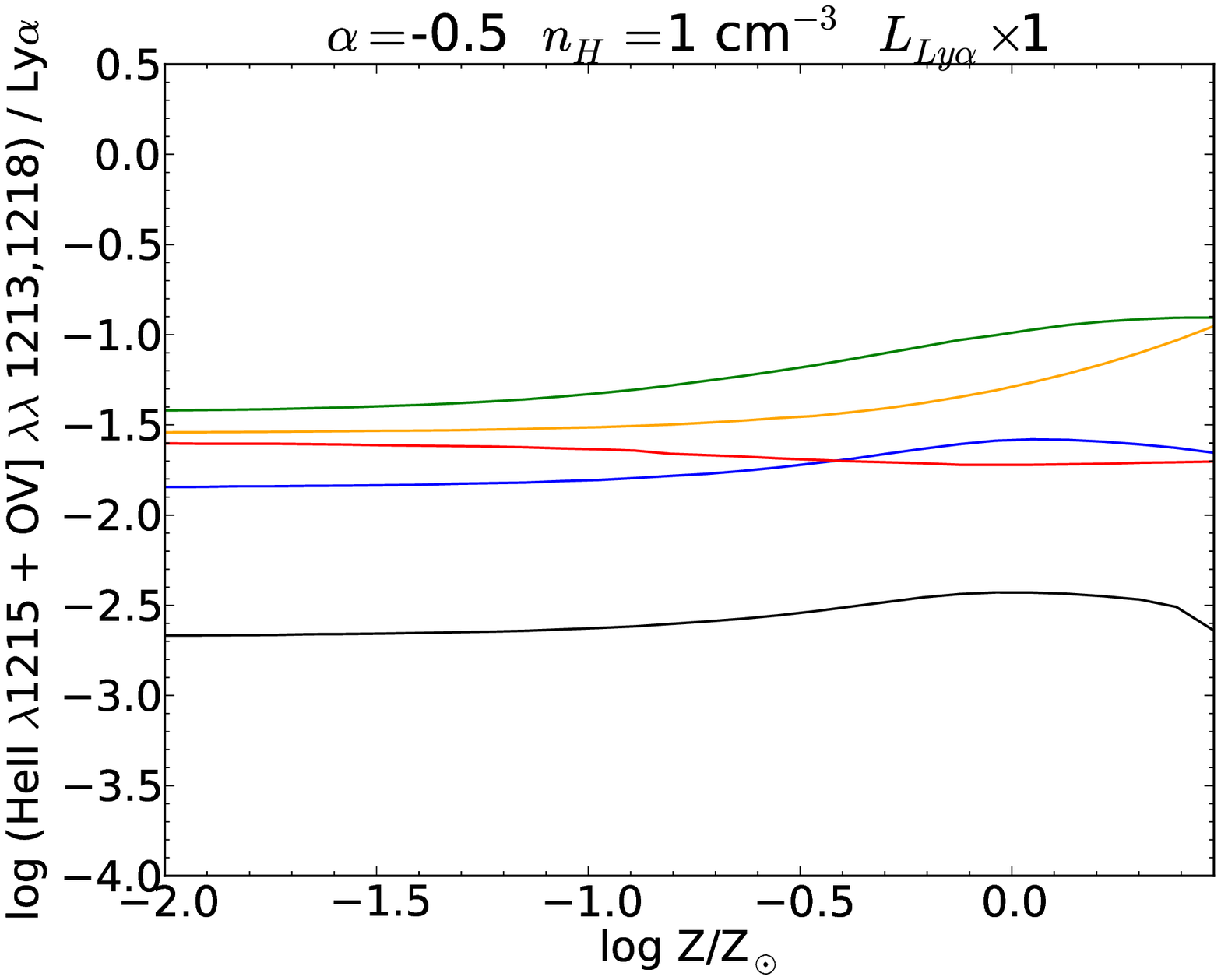}
\includegraphics{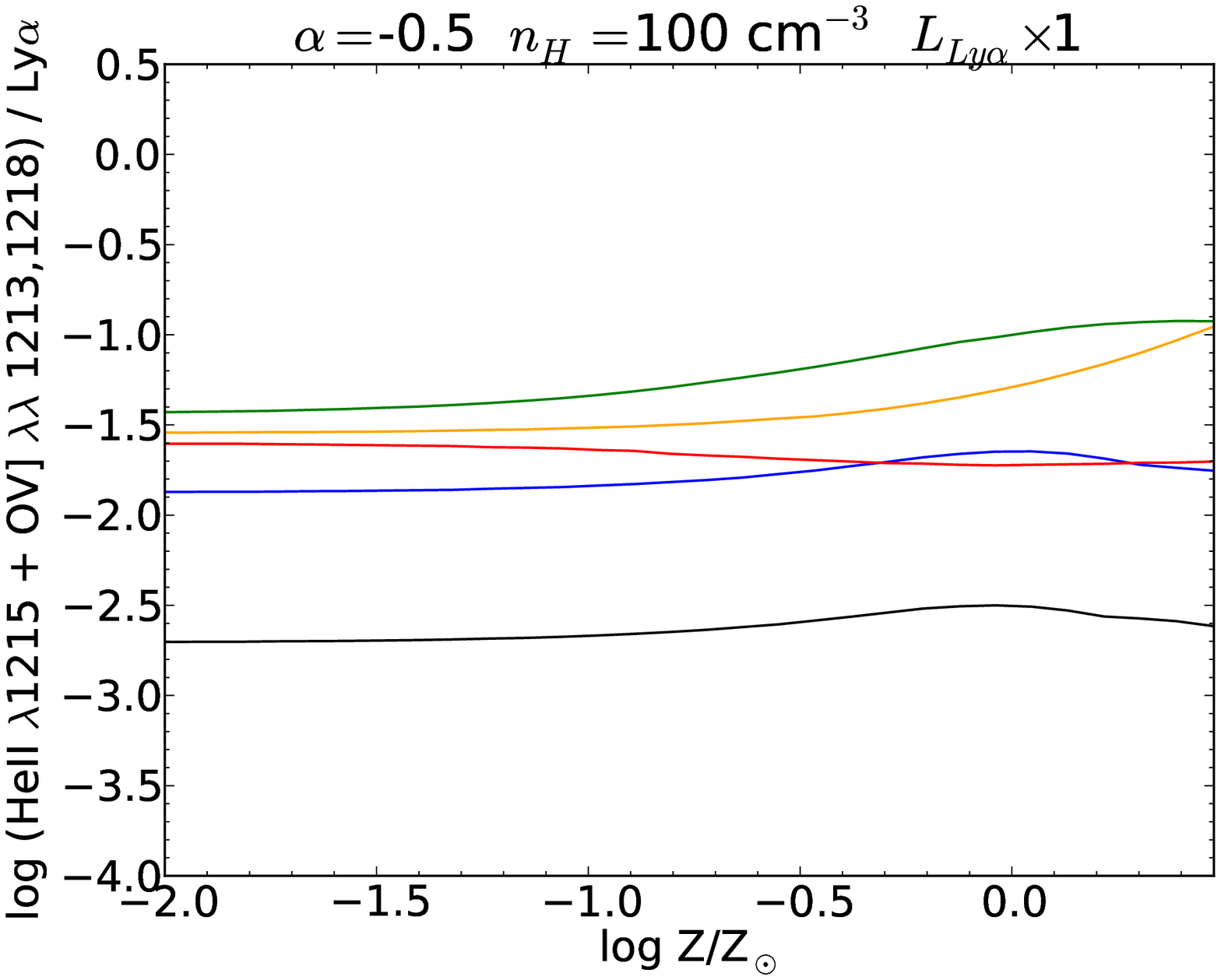}
\includegraphics{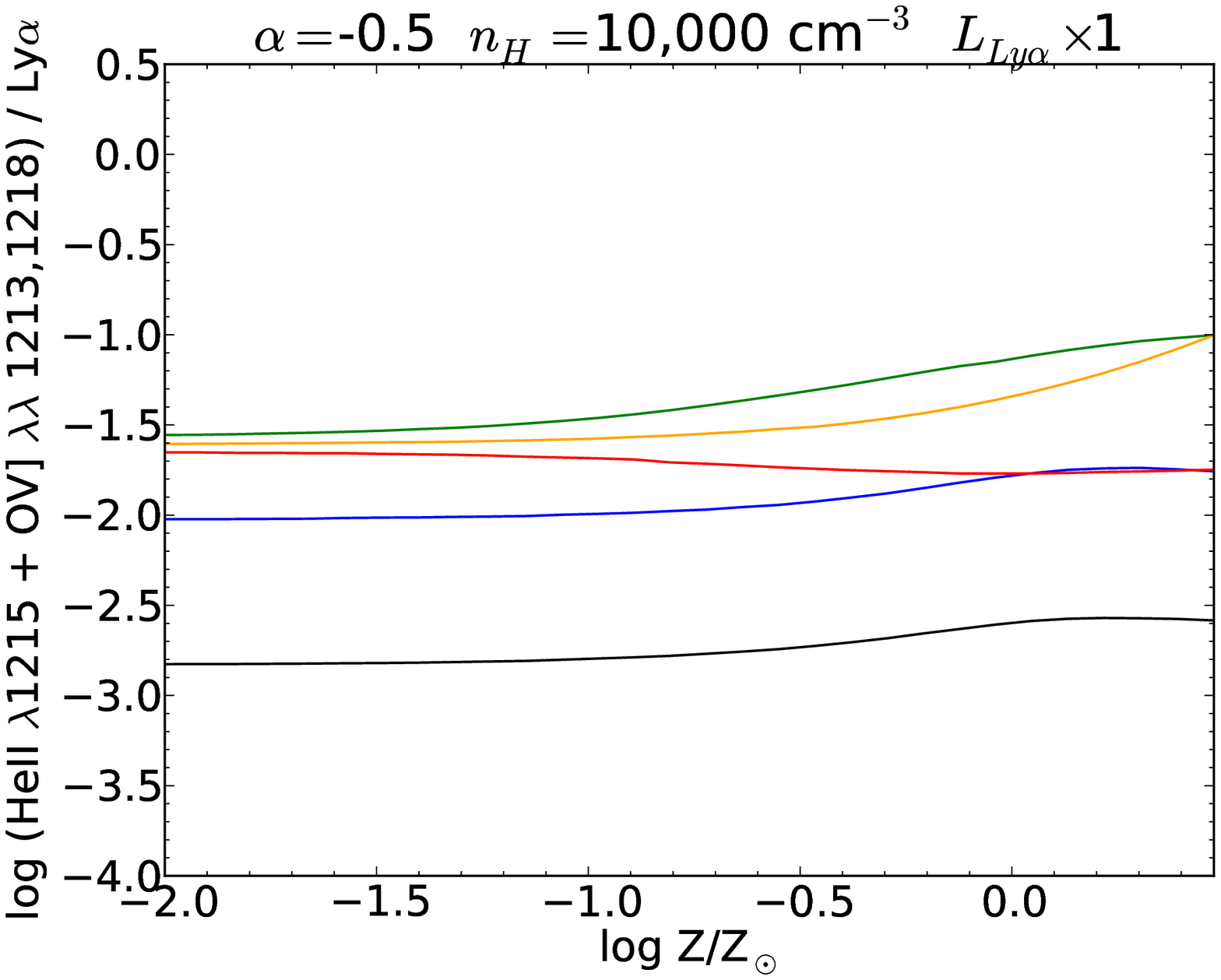}
\vspace{6.05in}
\caption{Similar to Fig. ~\ref{fig3}, but showing HeII+OV] / Ly$\alpha$ vs. metallicity curves for different fixed values of U, $\alpha$ and $n_H$.}
\label{fig4}
\end{figure*}

\begin{figure*}
\includegraphics{./OV_fig4_U.ps}
\includegraphics{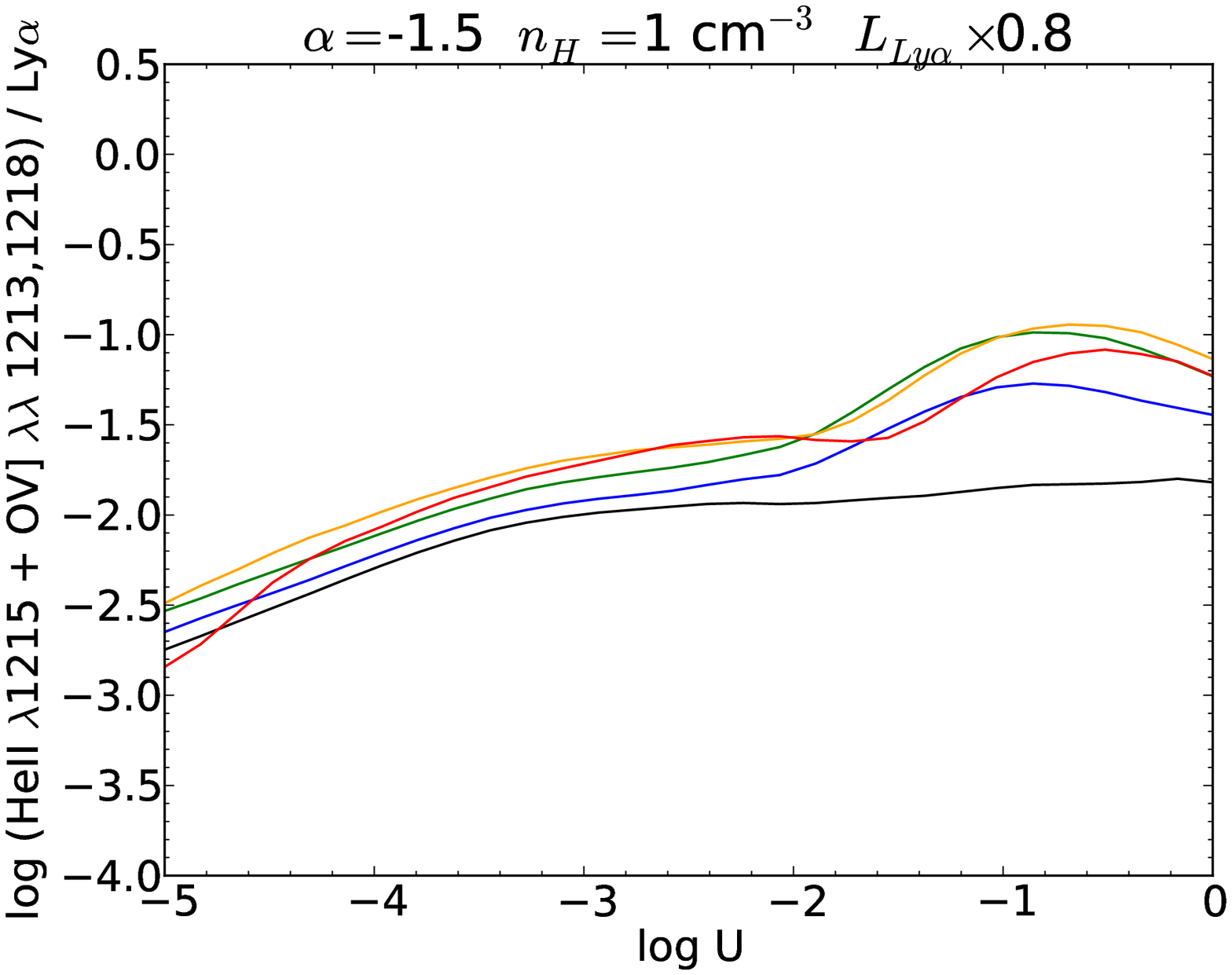}
\includegraphics{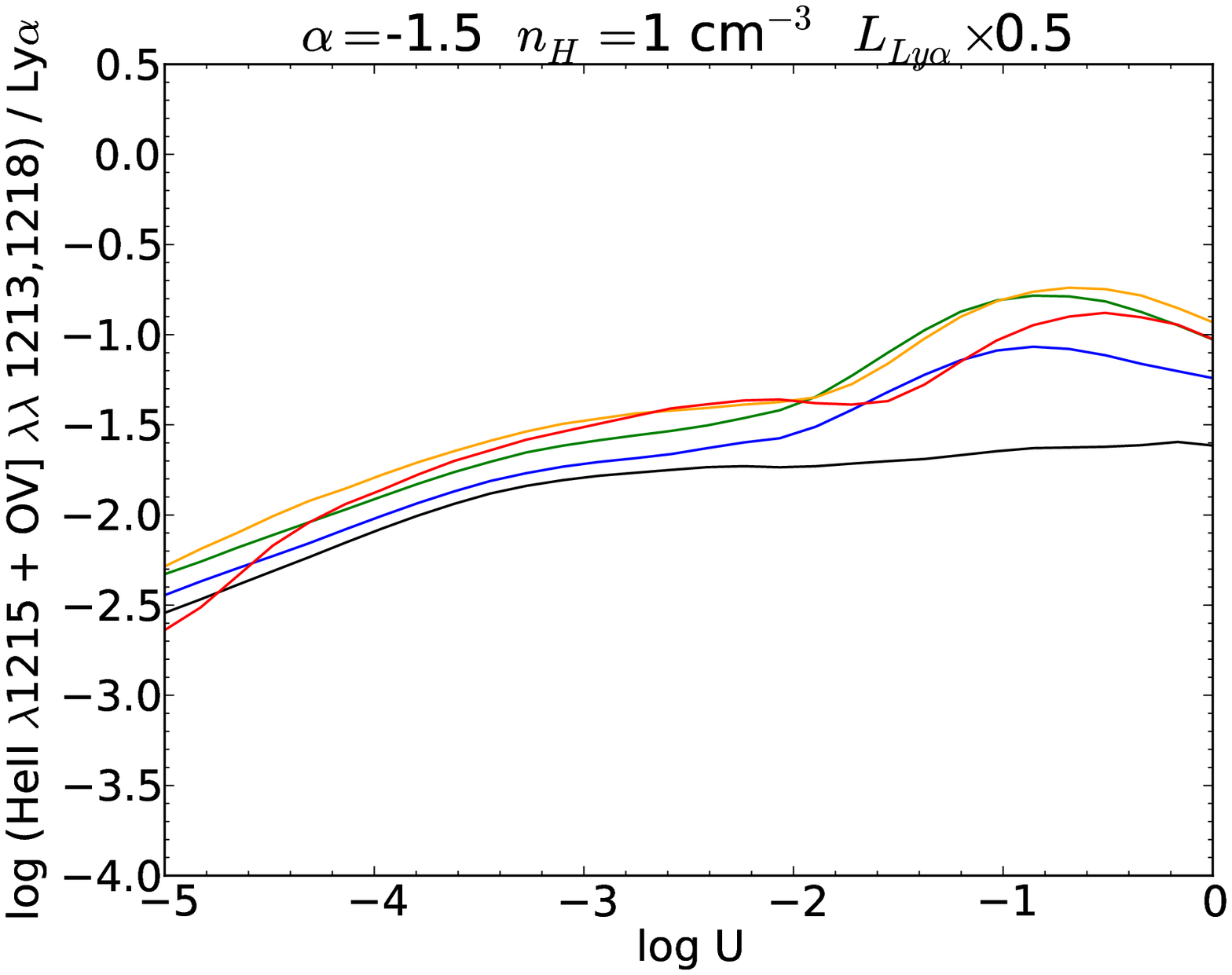}
\includegraphics{./OV_fig8_U.ps}
\includegraphics{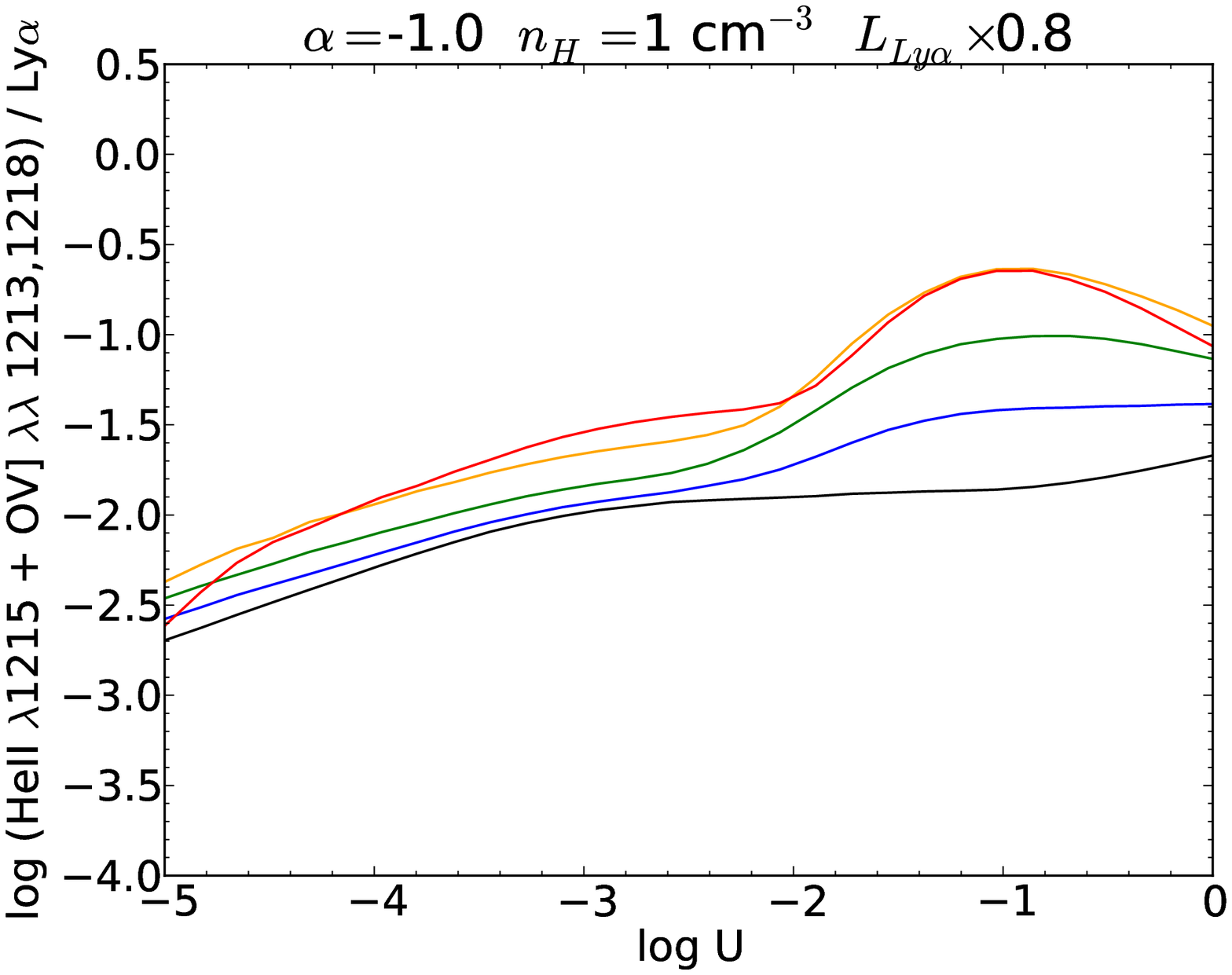}
\includegraphics{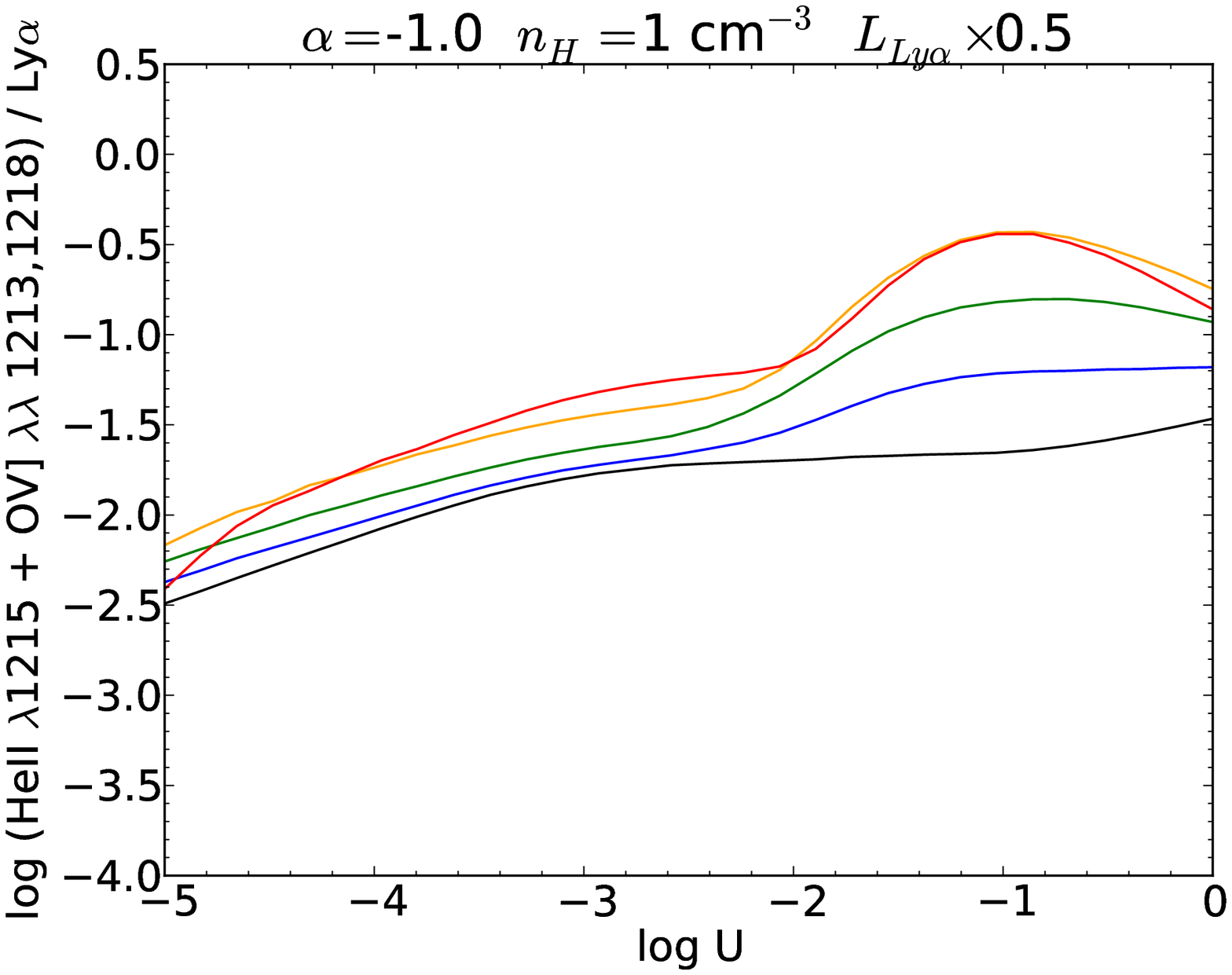}
\includegraphics{./OV_fig6_U.ps}
\includegraphics{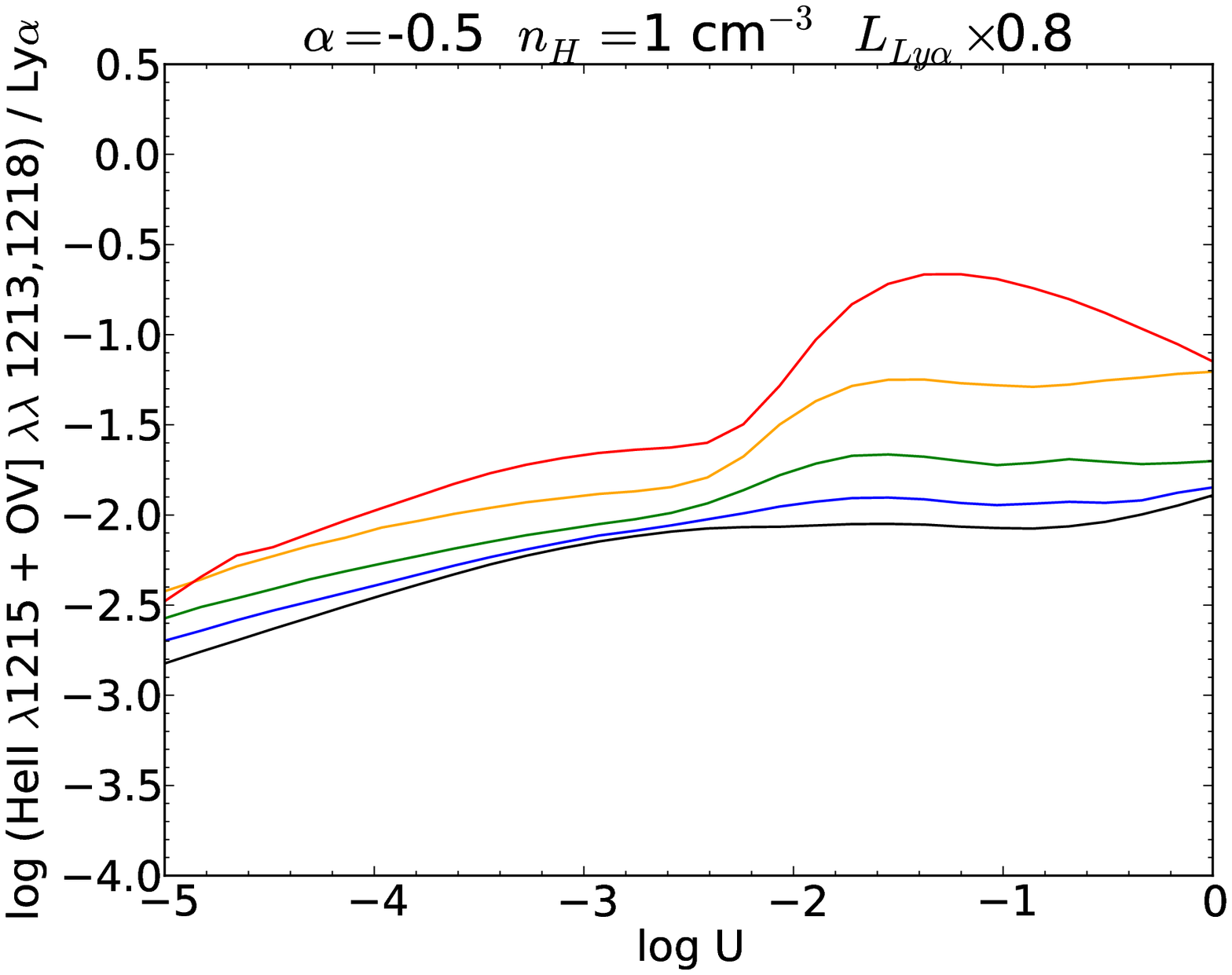}
\includegraphics{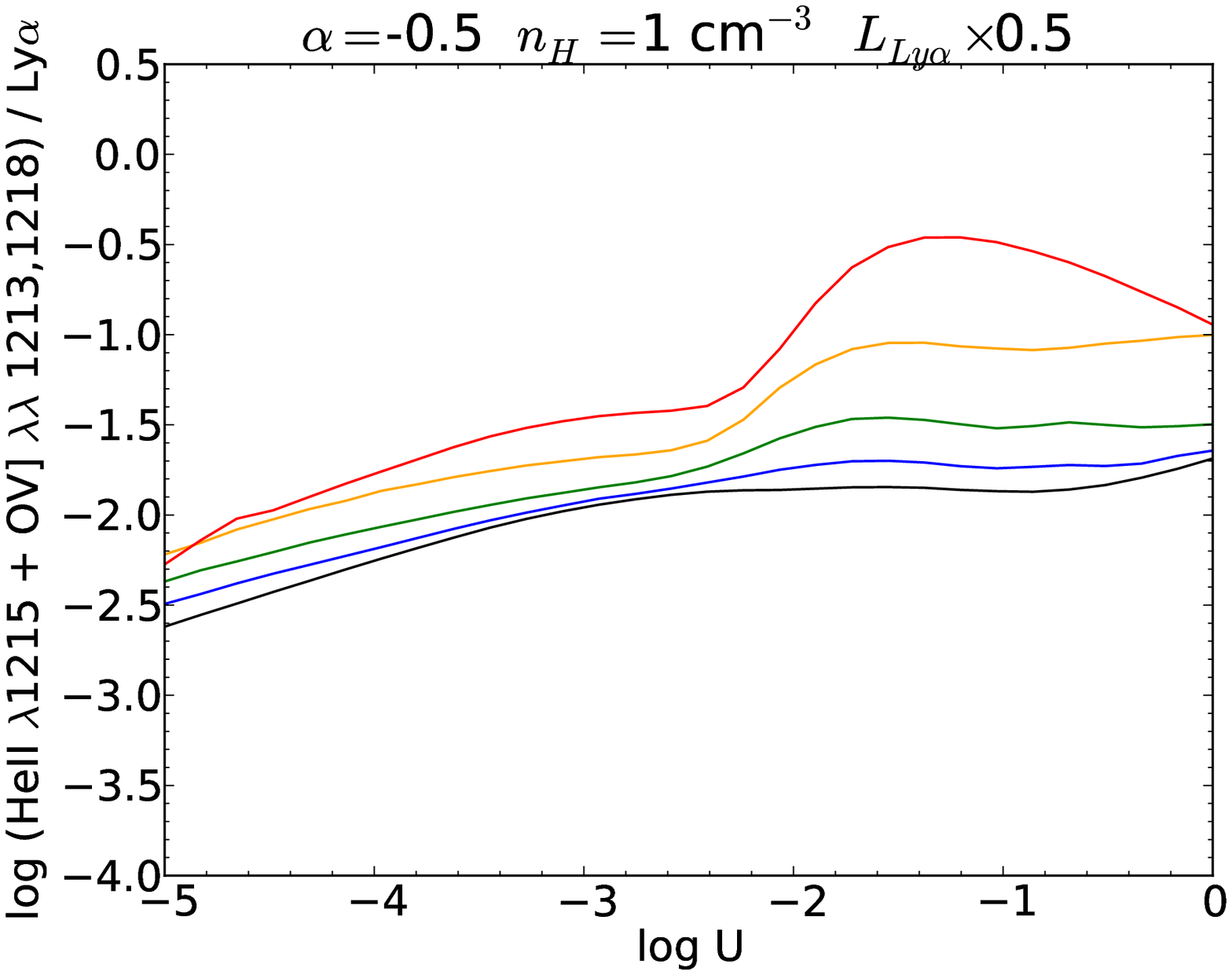}
\vspace{6.05in}
\caption{The impact of Ly$\alpha$ absorption on the HeII+OV] /
  Ly$\alpha$ vs U curves for ionization-bounded (optically-thick)
  photoionization models. The presence of Ly$\alpha$ absorption has
  been simulated by multiplying the Ly$\alpha$ luminosity by 1.0 (no
  absorption, top row), 0.2 (middle row) or 0.5 (bottom row). Due to
  the fact that gas densities of 1, 100 and 10,000 cm$^{-3}$ give
  essentially identical results, here we show only the subset of our
  grid that uses $n_H$=1 cm$^{-3}$.}
\label{fig5}
\end{figure*}

\begin{figure*}
\includegraphics{./OV_fig10_U.ps}
\includegraphics{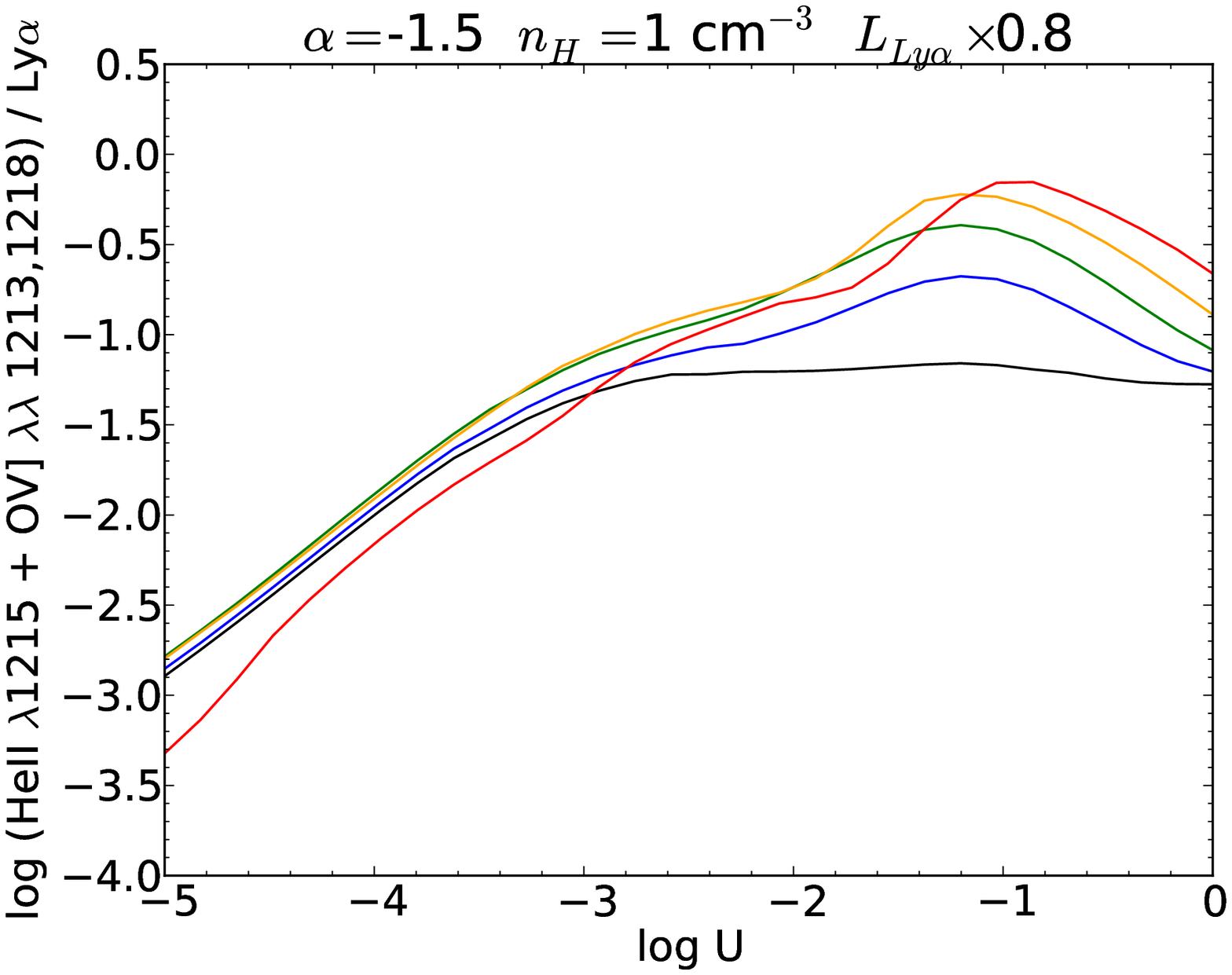}
\includegraphics{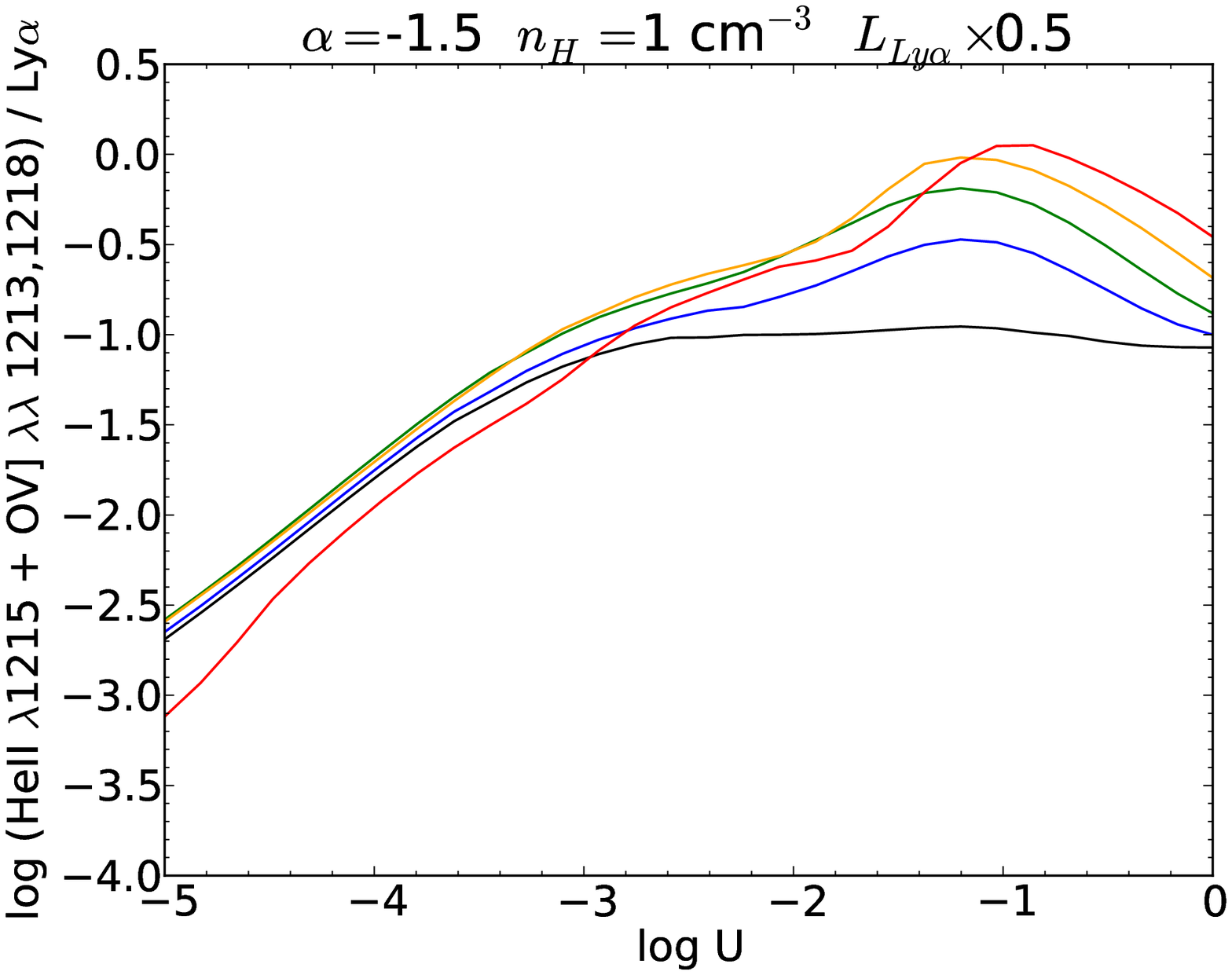}
\includegraphics{./OV_fig11_U.ps}
\includegraphics{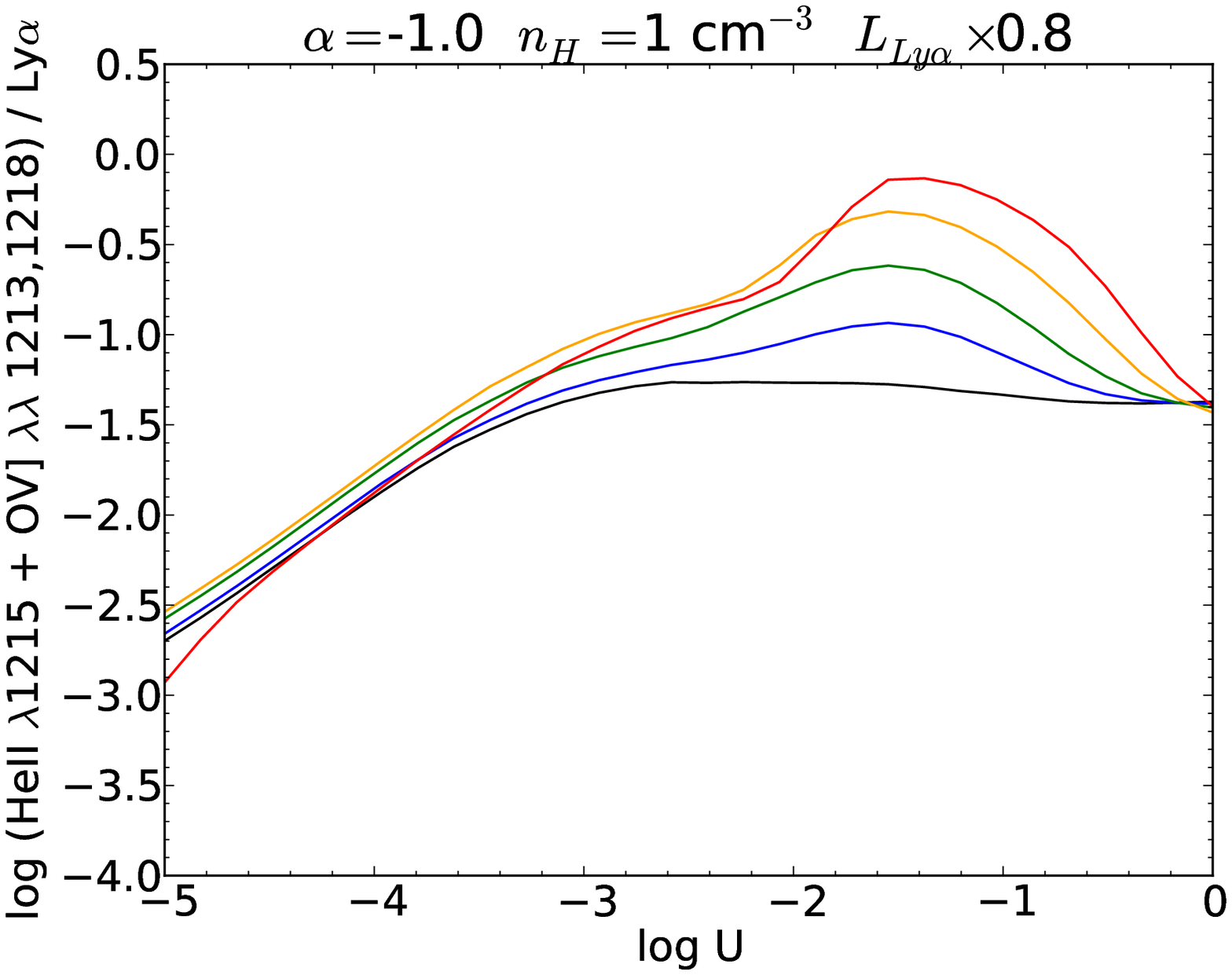}
\includegraphics{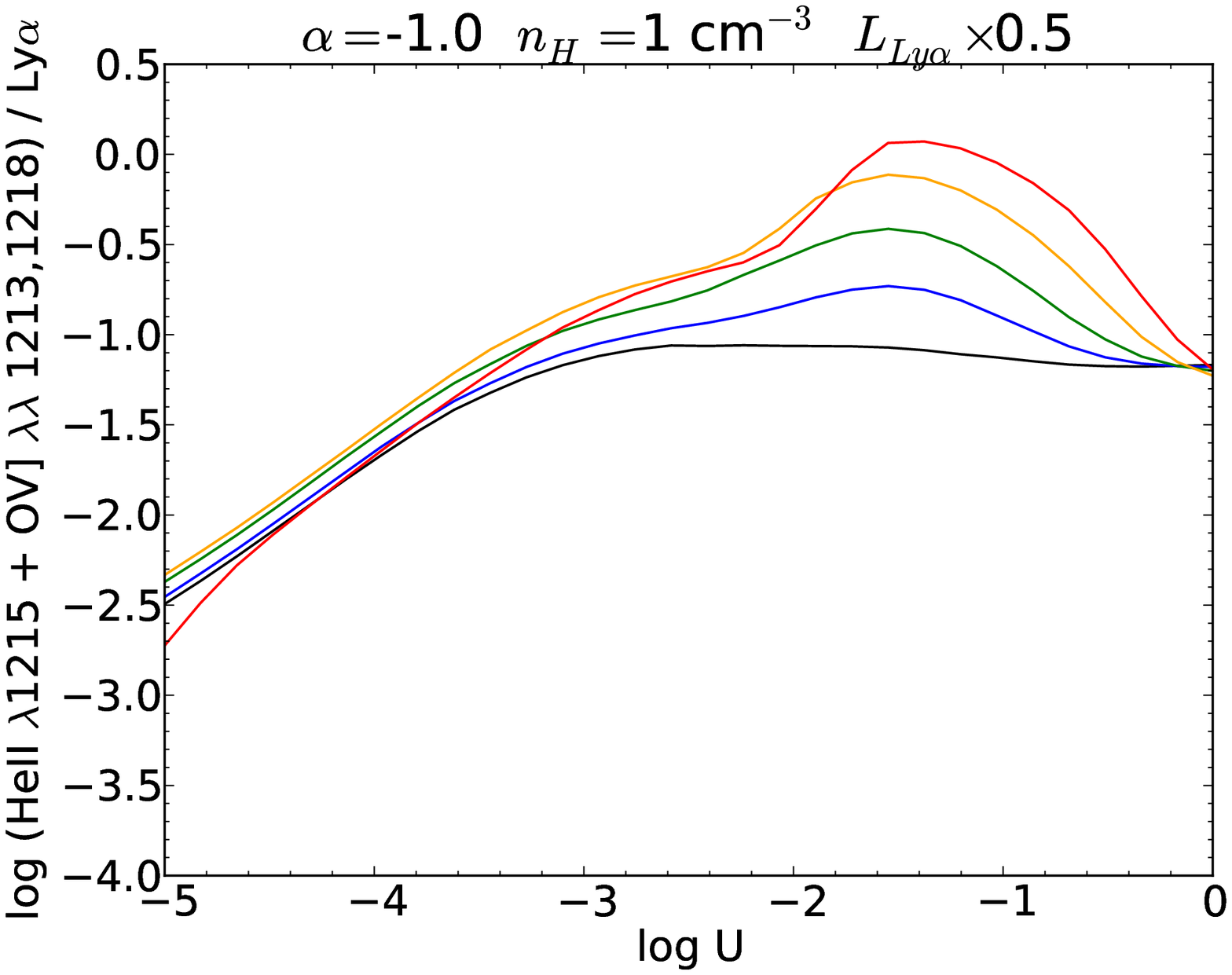}
\includegraphics{./OV_fig12_U.ps}
\includegraphics{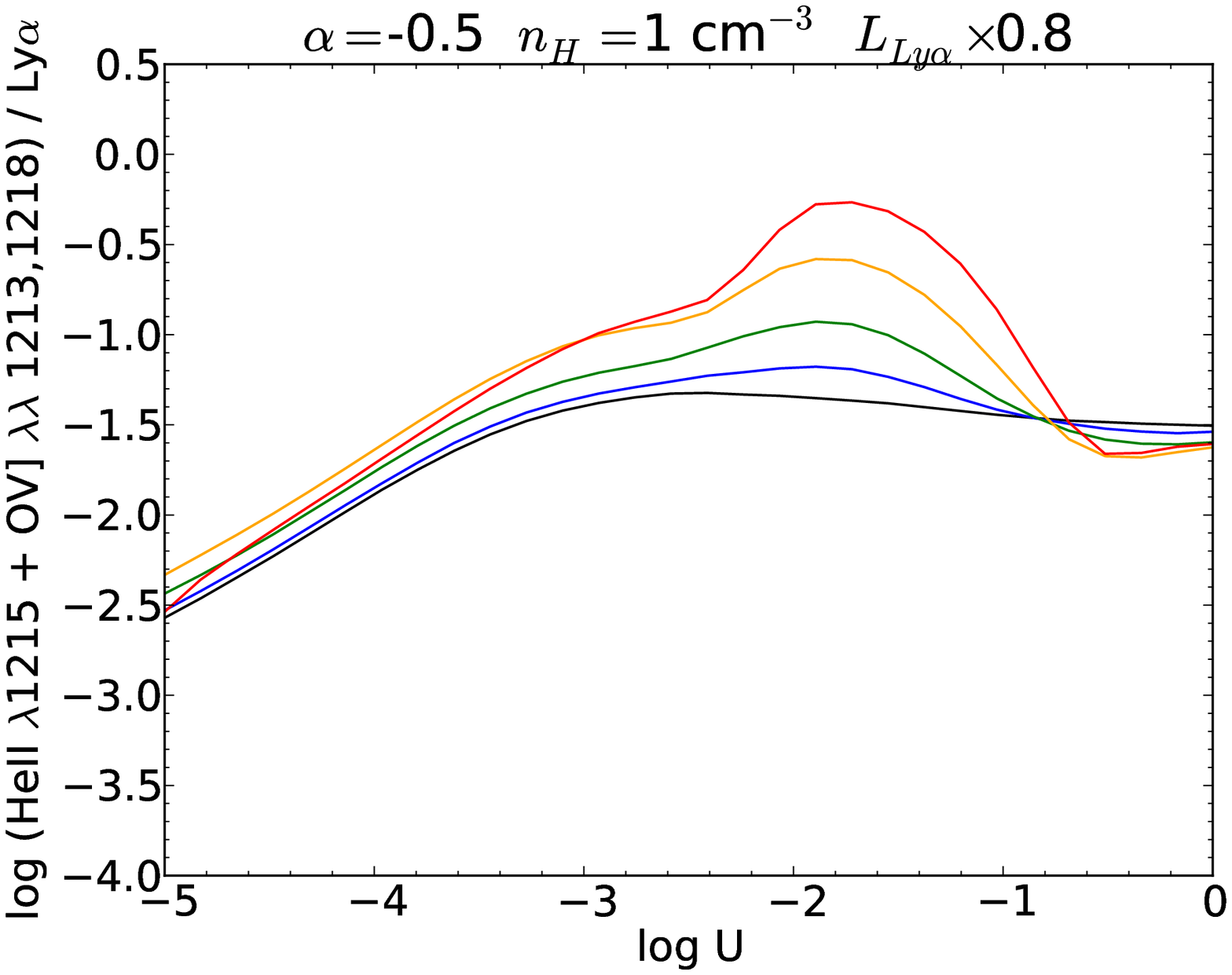}
\includegraphics{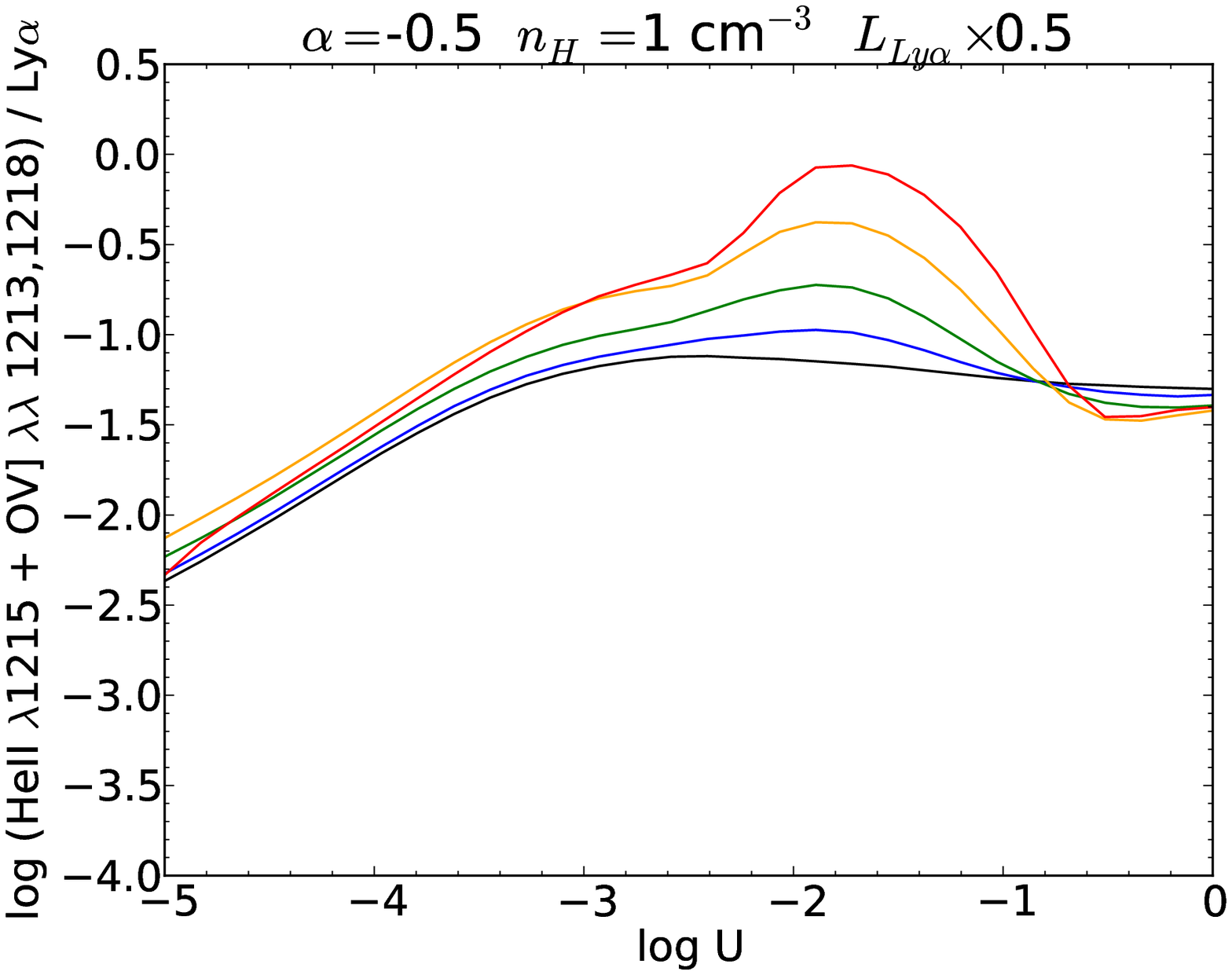}
\vspace{6.15in}
\caption{The impact of our simple Ly$\alpha$ absorption model on the HeII+OV] / Ly$\alpha$ vs U curves for matter-bounded (optically-thin) photoionization models. The presence of Ly$\alpha$ absorption has been simulated by multiplying the Ly$\alpha$ luminosity by 1.0 (no absorption, top row), 0.2 (middle row) or 0.5 (bottom row). Due to the fact that gas densities of 1, 100 and 10,000 cm$^{-3}$ give essentially identical results, here we show only the subset of our grid that uses $n_H$=1 cm$^{-3}$.}
\label{fig6}
\end{figure*}

\begin{figure*}
\includegraphics{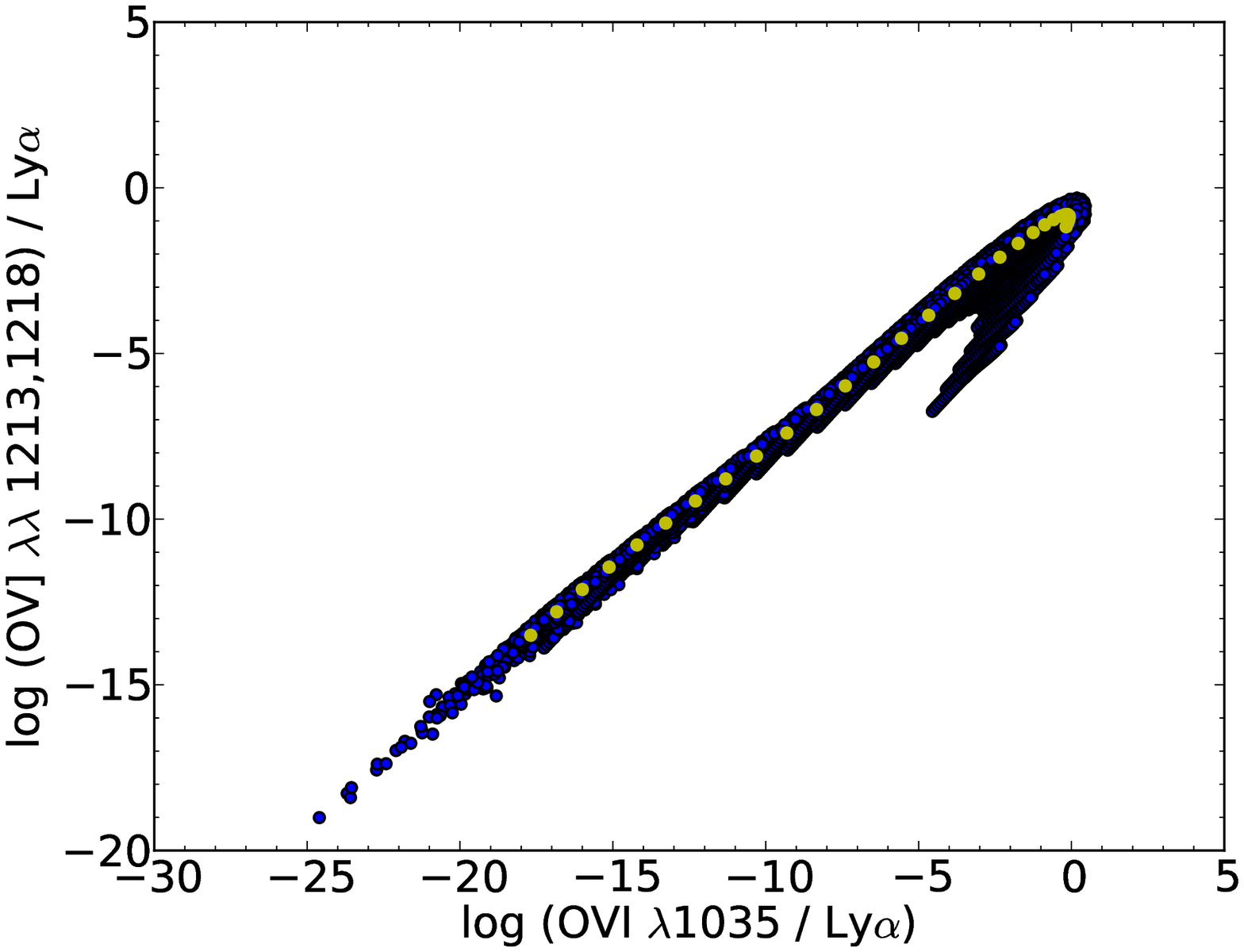}
\includegraphics{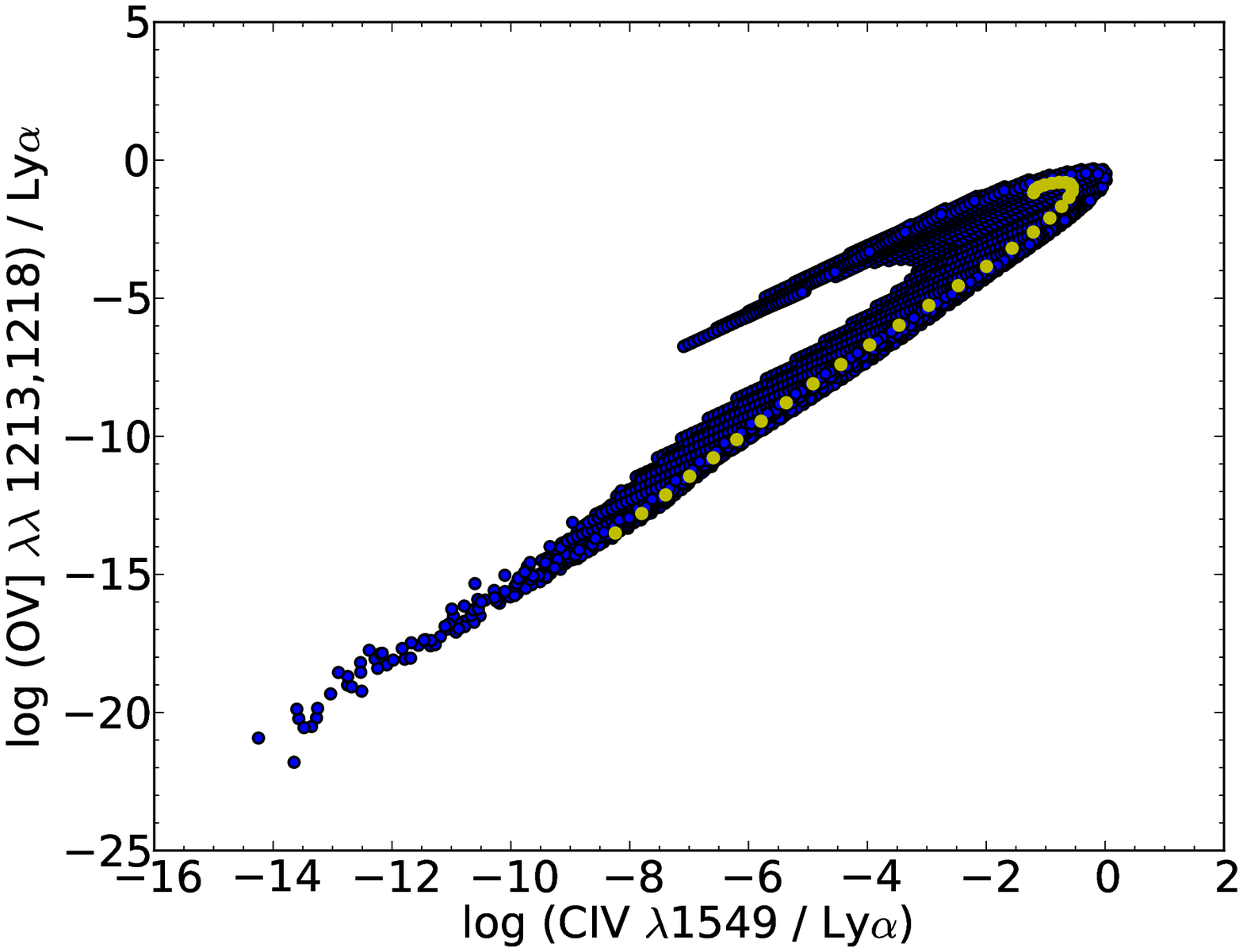}

\includegraphics{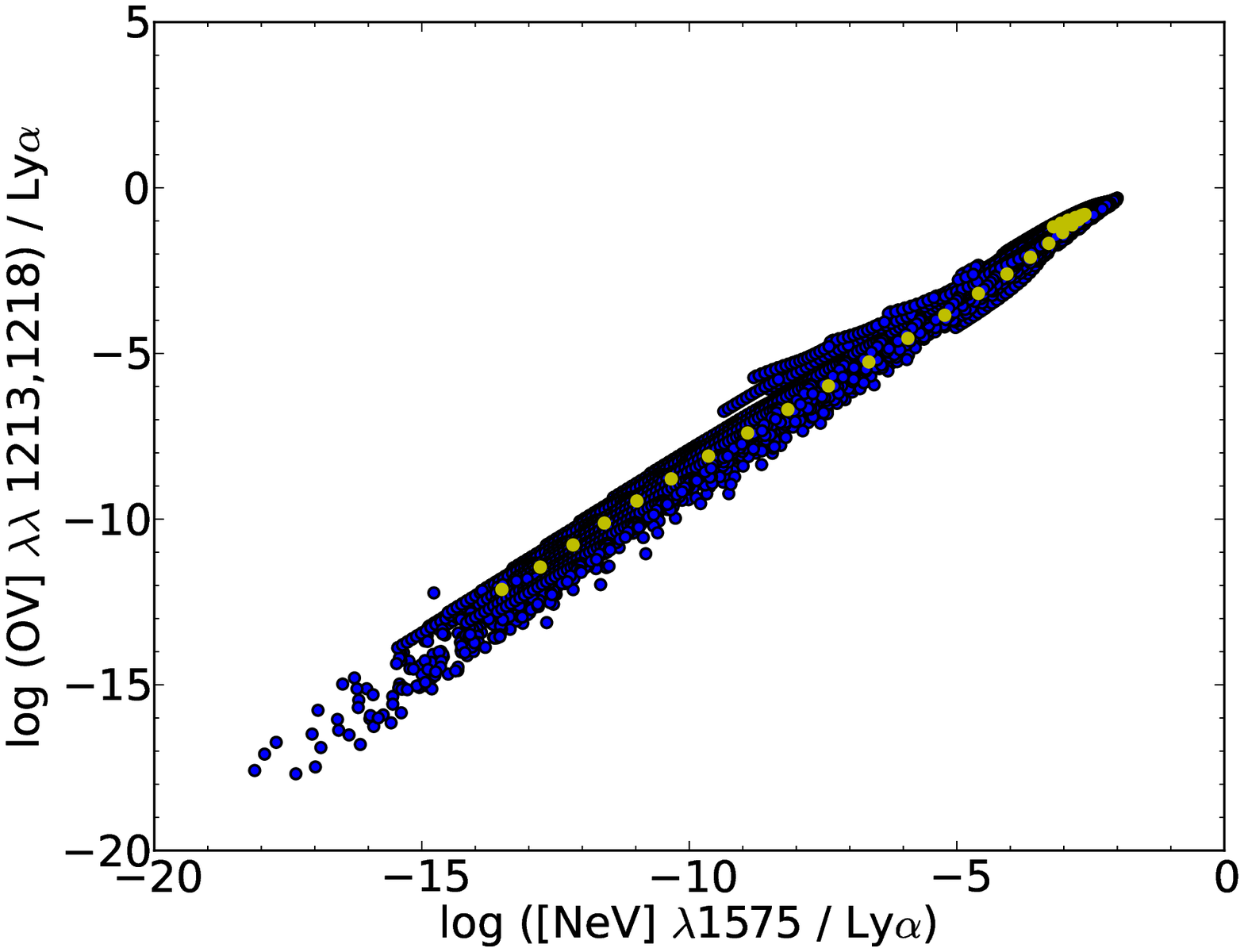}
\includegraphics{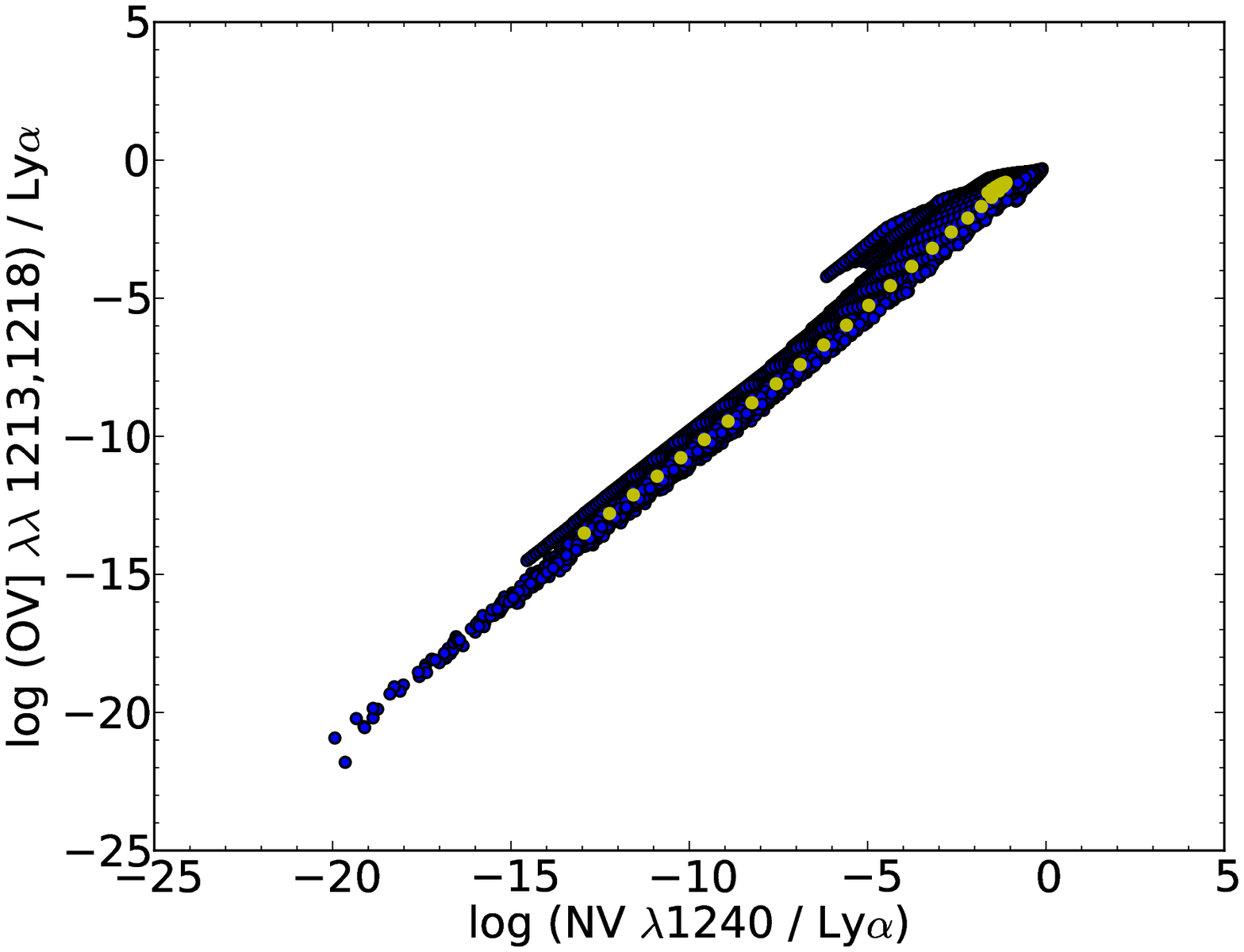}

\vspace{4.1in}
\caption{Log OV]/Ly$\alpha$ vs. log NV/Ly$\alpha$, [NeV]/Ly$\alpha$,
  OVI/Ly$\alpha$, and CIV/Ly$\alpha$. Every photoionization model from
  our grid is plotted (blue circles), with the exception of models
  that include absorption of Ly$\alpha$ by a HI screen, which are not
  shown here. All combinations of gas metallicity, density, U and
  $\alpha$ used in our grid are represened in these plots. Our
  ionization-bounded model sequence with $Z/Z_{\odot}$=1.1,
  $\alpha$=-1.0, and $n_H$=100 cm$^{-3}$ is highlighted using yellow
  circles -- as described in the text, we use this subset of our grid
  to obtain an extrapolation of the expected OV] flux from NV, [NeV],
  OVI and CIV.   } 
\label{fig7}
\end{figure*}

\section{Introduction}
The HI Ly$\alpha$ emission line at 1215.7 \AA~is among the brightest
lines in the ultraviolet spectra of active galaxies, and remains one of the
principal means to detect and study extended gaseous material around
quasars in the distant Universe (e.g. Villar-Mart\'{i}n et
al. 2002; Weidinger, M{\o}ller \& Fynbo 2004;  Christensen et
al. 2006; Borisova et al. 2016). 

The study of the distribution and dynamics of extended envelopes of
gas around active galaxies at high redshift now makes
extensive use of the Ly$\alpha$ emission line, in order to examine
key processes such as feedback activity and accretion of cold gas
(e.g. Vernet et al. 2017: Silva et al. 2018a; Arrigoni Battaia et
al. 2018; Dors et al. 2018). As such, it is crucial to understand the
physics of the production and radiative transfer of this emission
line.

One of the main channels for production of Ly$\alpha$ emission in
quasar nebulae is recombination fluorescence, where ultraviolet photons
emitted by the accretion disc of the quasar photoionize hydrogen,
leading to emission of a line and continuum spectrum upon
recombination (e.g. Heckman et al. 1991). As a resonant line,
Ly$\alpha$ can also be subject to absorption and transfer effects
that, in some circumstances, can significantly alter its observed
kinematics and flux (e.g. Villar-Mart\'{i}n, Binette \& Fosbury 1996;
Dijkstra, Haiman \& Spaans 2006). In addition, collisional excitation
can, under certain circumstances, make a significant contribution to
the production of Ly$\alpha$ photons (e.g. Villar-Mart\'{i}n et
al. 2007a), possibly even becoming the dominant channel of Ly$\alpha$ 
production at low gas metallicities (Humphrey et al. 2018).

A further possible complication is the presence of additional emission
lines very close to the wavelength of Ly$\alpha$, namely, the
non-resonant recombination line HeII $\lambda$1215.1, and the
semi-forbidden doublet OV] $\lambda\lambda$1213.8,1218.4. Clearly, it
is important to understand whether, and to what extent the
measured fluxes of Ly$\alpha$ are enhanced by the presence of HeII
$\lambda$1215.1 and OV] $\lambda\lambda$1213.8,1218.4. This issue was
discussed in the context of the high-density ($n_H\ga$10$^{6}$
cm$^{-3}$) broad-line region (BLR) of quasars by
Shields et al. (1995), who found that the contribution from HeII and
OV] can become non-negligible in the case of highly-ionized,
optically-thin BLR clouds, with flux ratios of up to
OV]/Ly$\alpha$$\sim$1 and HeII $\lambda$1215.1/Ly$\alpha$$\sim$0.2
being reached in their model grid (see also Ferland et
al. 1992). However, we are not aware of any previous photoionization
calculations for the lower-density, narrow-line emitting region (NLR)
or Ly$\alpha$ halo that considers the possible contamination of the
Ly$\alpha$ flux from these lines. Addressing this question for the NLR
and Ly$\alpha$ halo of AGN is the main focus of this work. 

In a previous paper, we applied the modelling code Mappings 1e
(Binette, Dopita \& Tuohy 1985; Ferruit et al. 1997; Binette et
al. 2012) to model the combined effects of several mechanisms to
enhance Ly$\alpha$ emission from AGN-photoionized nebulae at high
redshift (Humphrey et al. 2018). Here we present a follow-up study to
quantify the potential contamination from HeII and OV] emission to
flux measurements of Ly$\alpha$ in low density, narrow-line nebulae
that are photoionized by an AGN.

\section{Photoionization models}
\label{models}
Our objective is to simulate the ionization/excitation conditions in
spatially extended, low density gas that is being photoionized by a
central AGN, such as the NLR (e.g. Pogge 1988) or Ly$\alpha$ halo
(e.g. Reuland et al. 2003). We have used the multi-purpose modeling
code MAPPINGS 1e (Binette, Dopita \& Tuohy 1985; Ferruit et al. 1997;
Binette et al. 2012) to compute our grid of AGN photoionization models. 

Our grid consists of 16200 individual photoionization models, each
one representing an ionized cloud with a specific set of physical and
ionization properties. Our models consist of a plane-parallel slab of
gas which is illuminated by a powerlaw of the form $S_\nu \propto
\nu^{\alpha}$, where $\alpha$ is the spectral index, $\nu$ is emission frequency and $S_\nu$ is
flux density. As a starting point from which to define non-Solar gas 
chemical abundances\footnote{We define gas metallicity $Z$ as the
  oxygen to hydrogen number ratio, normalizing by the Solar 
  value $Z_{\odot}$ for simplicity. For example, O/H = 4.90
  $\times$10$^{-4}$ corresponds to $Z/Z_{\odot}$ = 1.0.}, we adopt the
Solar abundance set 
of Asplund et al. (2006). To vary gas metallicity, we start from
  the Solar abundance set, and we then scale all metals linearly such
  that $Z/Z_{\odot}$ $\propto$ O/H, except in the case of nitrogen, for
  which we take into account secondary production 
  (e.g. Henry et al. 2000) by assuming N/O $\propto$ O/H for
  $Z/Z_{\odot} >$ 0.3 and N/H $\propto$ O/H for $Z/Z_{\odot} <$ 0.3.

For three different values of hydrogen gas
density ($n_H$ = 1, 100, 10$^4$ cm$^{-3}$), three different ionizing
power law indices ($\alpha$=-1.5, -1.0 or -0.5) and two optical depth
regimes (thick or thin), we computed 30$\times$30 sub-grids of 900
models with 30 values of ionization parameter U\footnote{We define U=Q
/ (4$\pi$ r$^2$ c $n_H$), where Q is the isotropic ionizing photon
number luminosity of the ionizing source, r is the distance of the
illuminated gas from the source, c is the speed of light, and $n_H$ is
the number density of hydrogen in the gas.} (10$^{-5}$--1.0) and
30 values of gas metallicity $Z/Z_{\odot}$ (0.01--3.0).

Our optically thick models adopt the {\it ionization-bounded model
termination}, which ends the calculation when the ionization fraction
of hydrogen has fallen below 0.01. This simulates a cloud which
absorbs essentially all the ionizing photons from the incident
spectrum, with the exception of some photons at the high-energy end of
the spectrum and, in some circumstances, UV photons at $hv<$13.6 eV
that ionize some neutral metal species (e.g., C, S, Si, etc.)\footnote{The possible
  presence of singly-ionized C, S, Si, etc. beyond the Str\"omgren
  radius, and our neglect thereof, does not affect our results.}. 

In the case of our optically thin models, the calculation terminates
when the fraction of absorbed H-ionizing photons (by number) reaches
0.05, resulting in a {\it Lyman-continuum-leaking, matter-bounded
  cloud}. This corresponds to a cloud that is insufficiently
  optically thick to absorb all of the ionizing EUV photons, so that
  some of these photons pass through unabsorbed to escape
  from the rear of the cloud.

The Ly$\alpha$ emission line of Type II active galaxies and
  Ly$\alpha$ blobs sometimes show strong, narrow absorption features
  due to associated, intervening HI stuctures (e.g. van Ojik et
  al. 1997;  Wilman et al. 2004, 2005), or unusually low
  Ly$\alpha$/HeII flux ratios suggestive of partially absorbed
  Ly$\alpha$ (e.g. van Ojik et al. 2004). To simulate the effect of
absorption on Ly$\alpha$ by an external HI 
screen, we have multiplied the Ly$\alpha$ luminosity by a transmission
factor of 0.5 (strong absorption), 0.8 (moderate absorption) or 1.0
(no absorption). It is non-trivial to convert these values into
neutral hydrogen column density $N_{HI}$, because the transmission
factor also depends on the kinematic width and covering factor of the
absorbing gas, and the kinematic properties of the Ly$\alpha$ emitting
gas. Nevertheless, for the Ly$\alpha$ flux to be significantly
reduced, an HI column density of $N_{HI} \ga$ 10$^{14}$ cm$^{-2}$
would be required. (See e.g. Silva et al. 2018a for a more detailed
discussion of the degeneracies involved in recovering $N_{HI}$ from
the Ly$\alpha$ velocity profiles of quasars.) In the
interest of simplicity, we assume that the HI absorption (if present)
is positioned near the centre of the Ly$\alpha$ emission line and thus
does not absorb any of the HeII $\lambda$1215.1 or OV]
$\lambda\lambda$1213.8,1218.4 emission\footnote{Even though HeII
$\lambda$1215.1 and OV] $\lambda\lambda$1213.8,1218.4  
are non-resonant lines and thus do not suffer absorption from their
respective ions, they still can potentially be absorbed by neutral
hydrogen given the correct relative velocity shifts between the
emitting O$^{+4}$ or He$^+$ and HI. However, in the interest of
simplicity, such an effect is not considered here.}. 

We emphasize that our treatment of HI absorption is, strictly
  speaking, valid only for the simple case of absorption of Ly$\alpha$
  by an external HI screen that produces a narrow absorption feature
  centred on or near the peak of the Ly$\alpha$ emission. More complex
  HI geometries, such as multiple HI absorption systems offset 
from the centre of the Ly$\alpha$ emission line, or absorption and
destruction of Ly$\alpha$ photons {\it in situ} within the emitting
gas, may produce significantly different results. However, a detailed
treatment of other (more complex) HI absorpion geometries is beyond
the scope of this paper.

\begin{figure*}
\includegraphics{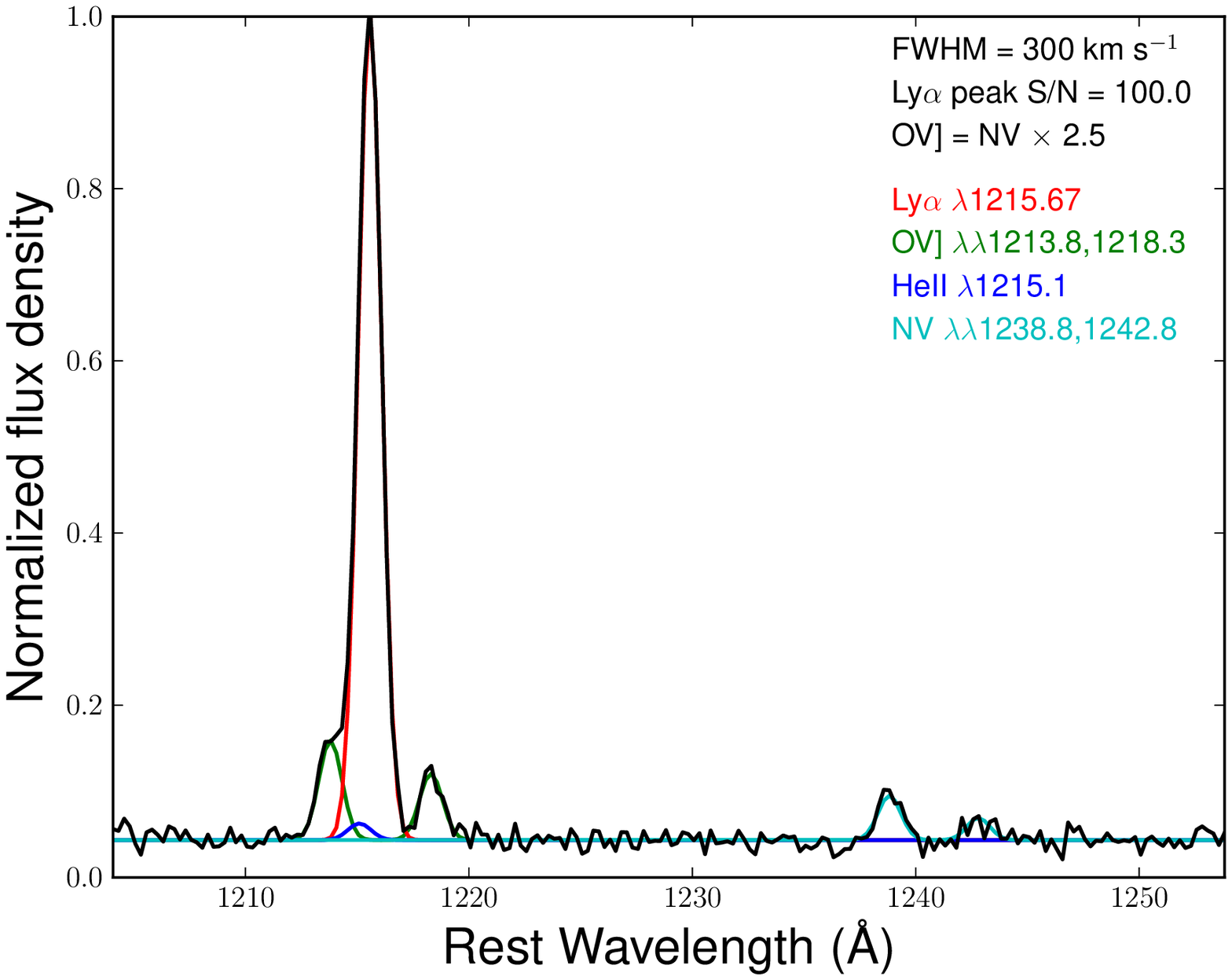}
\includegraphics{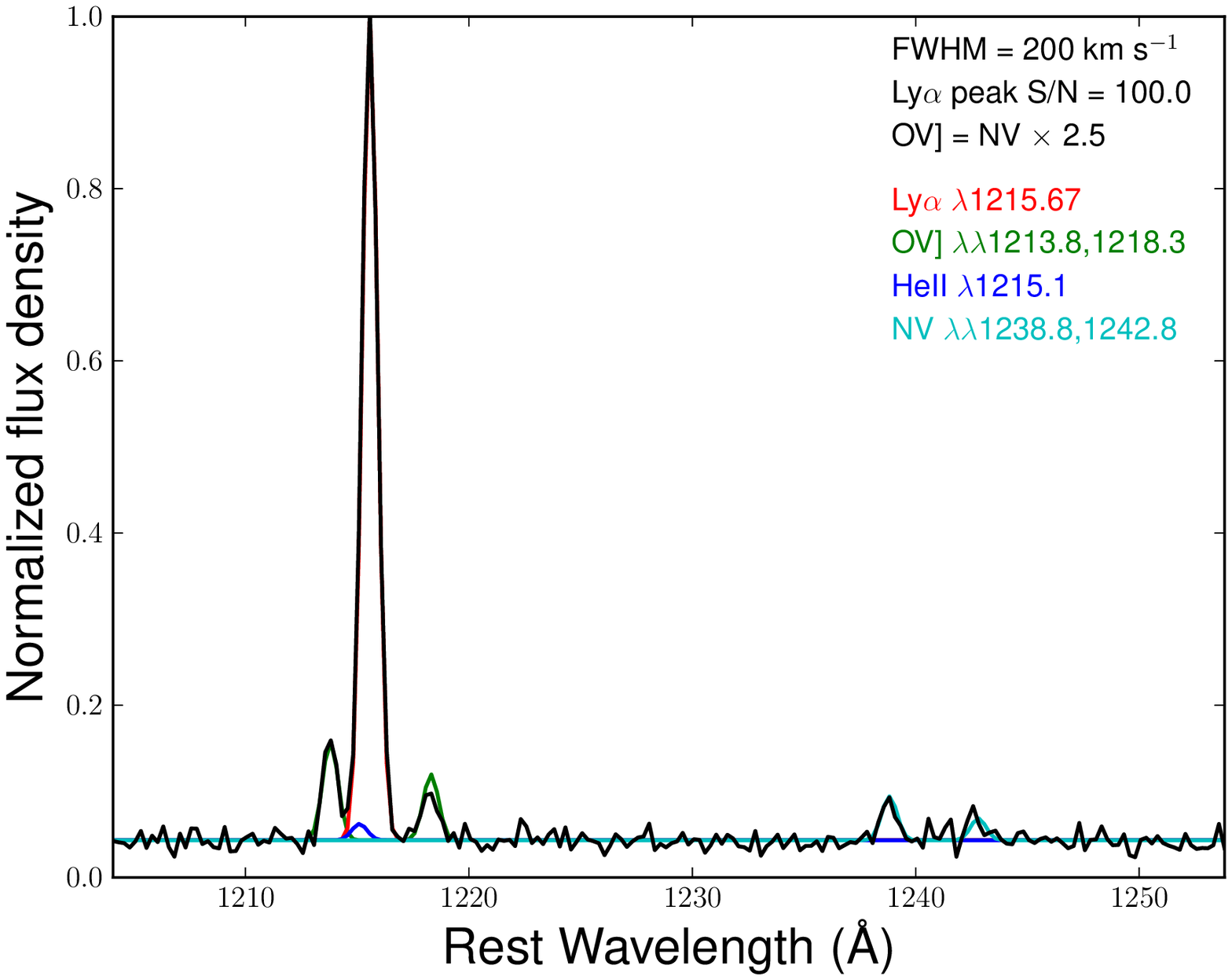}

\includegraphics{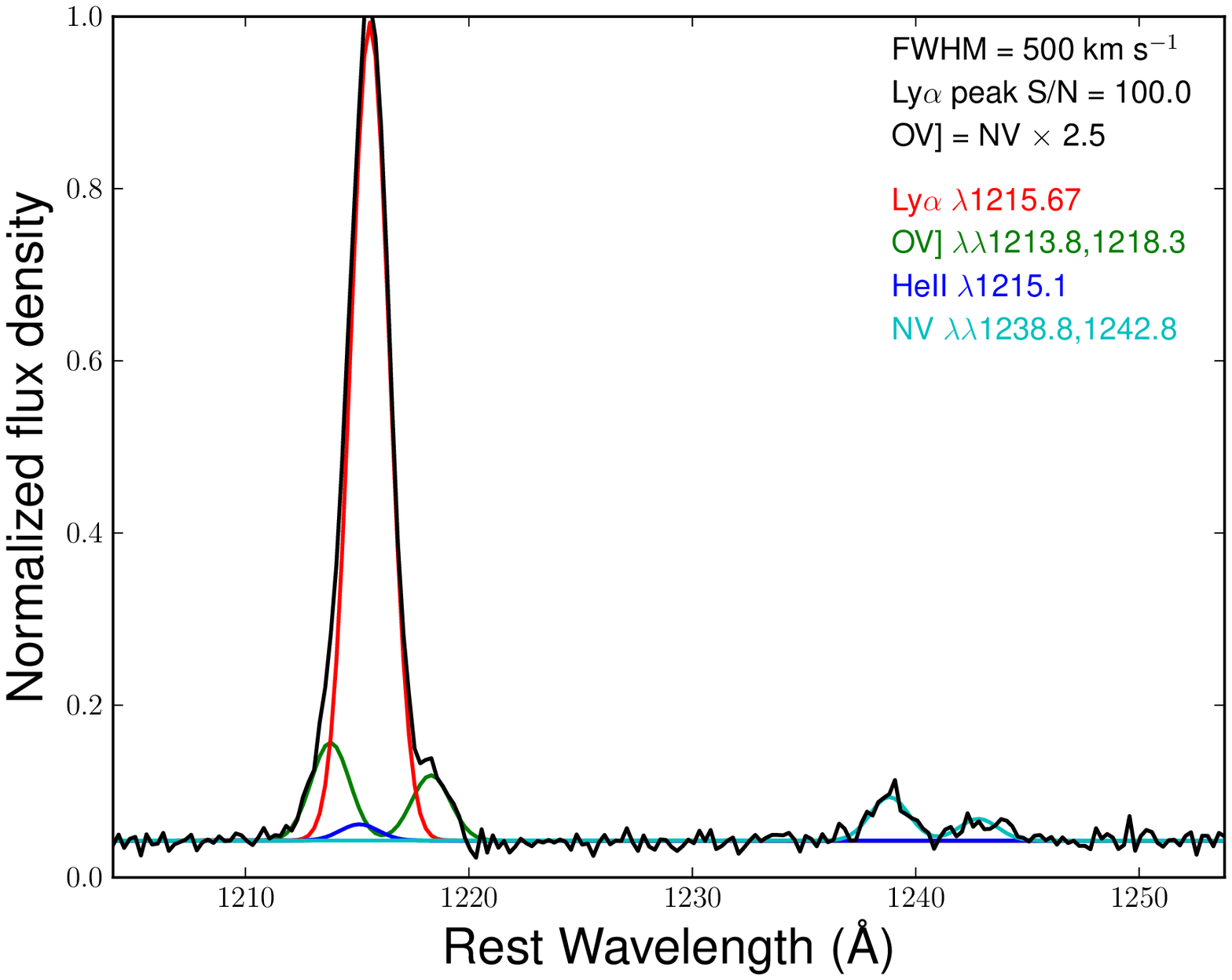}
\includegraphics{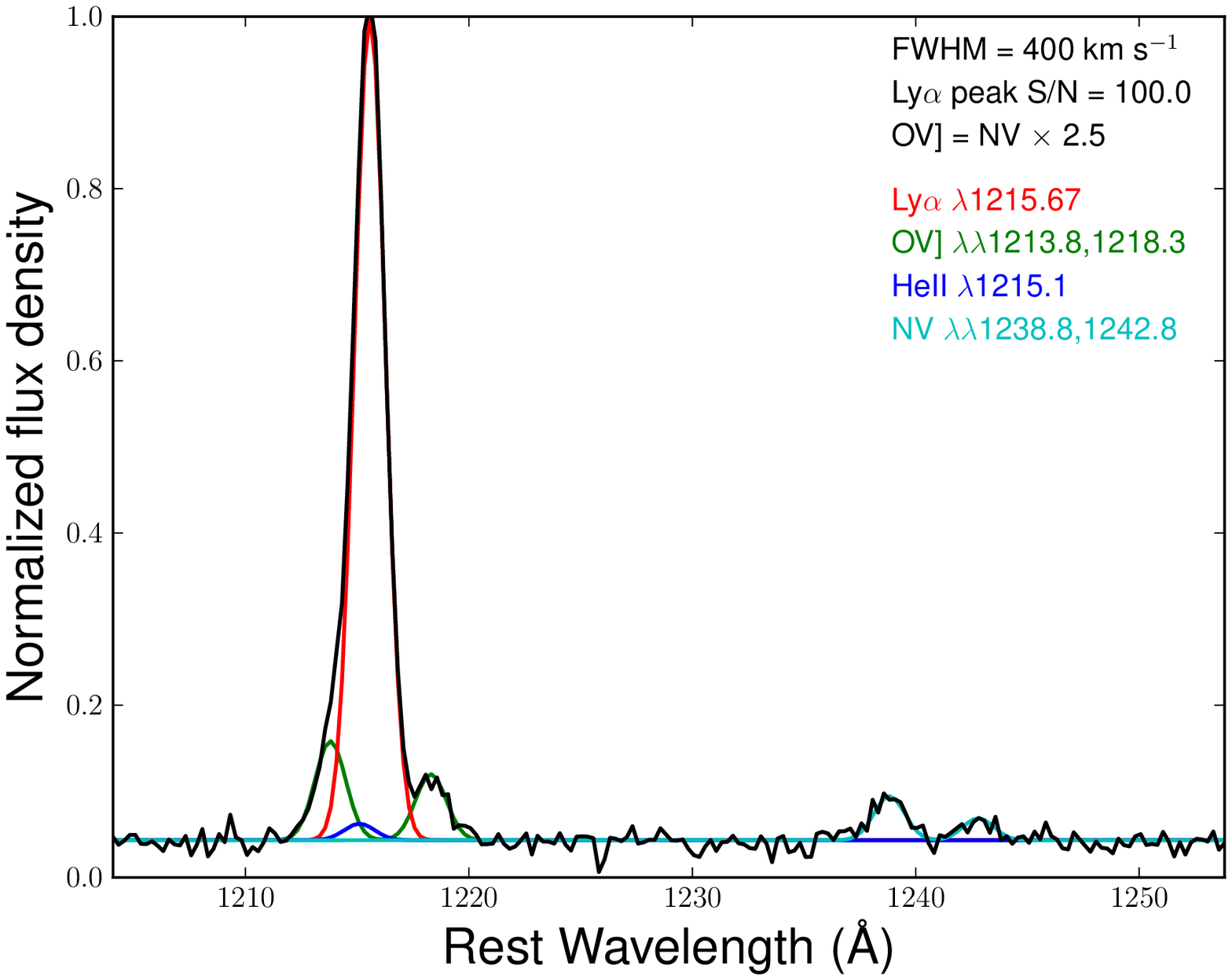}

\includegraphics{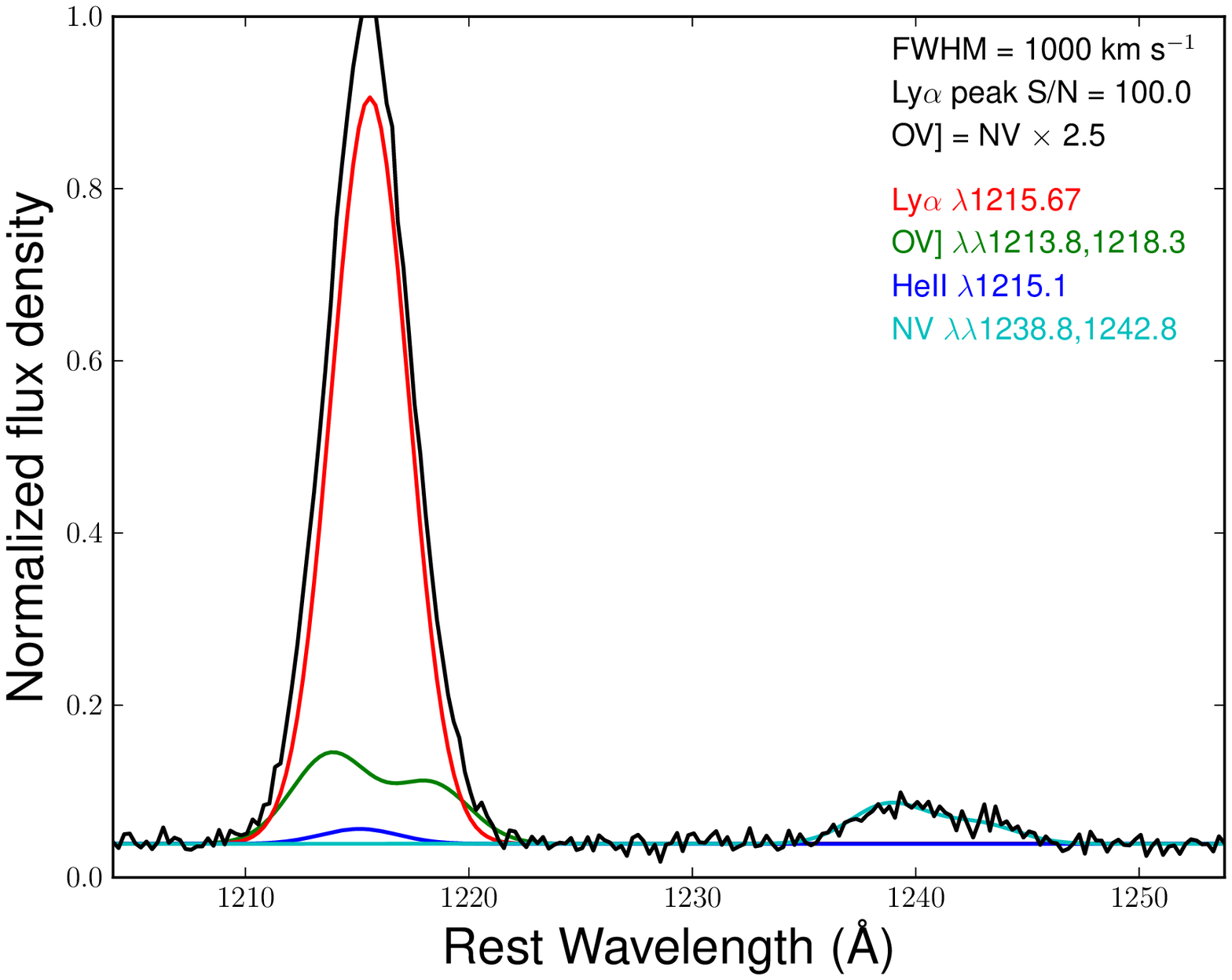}
\includegraphics{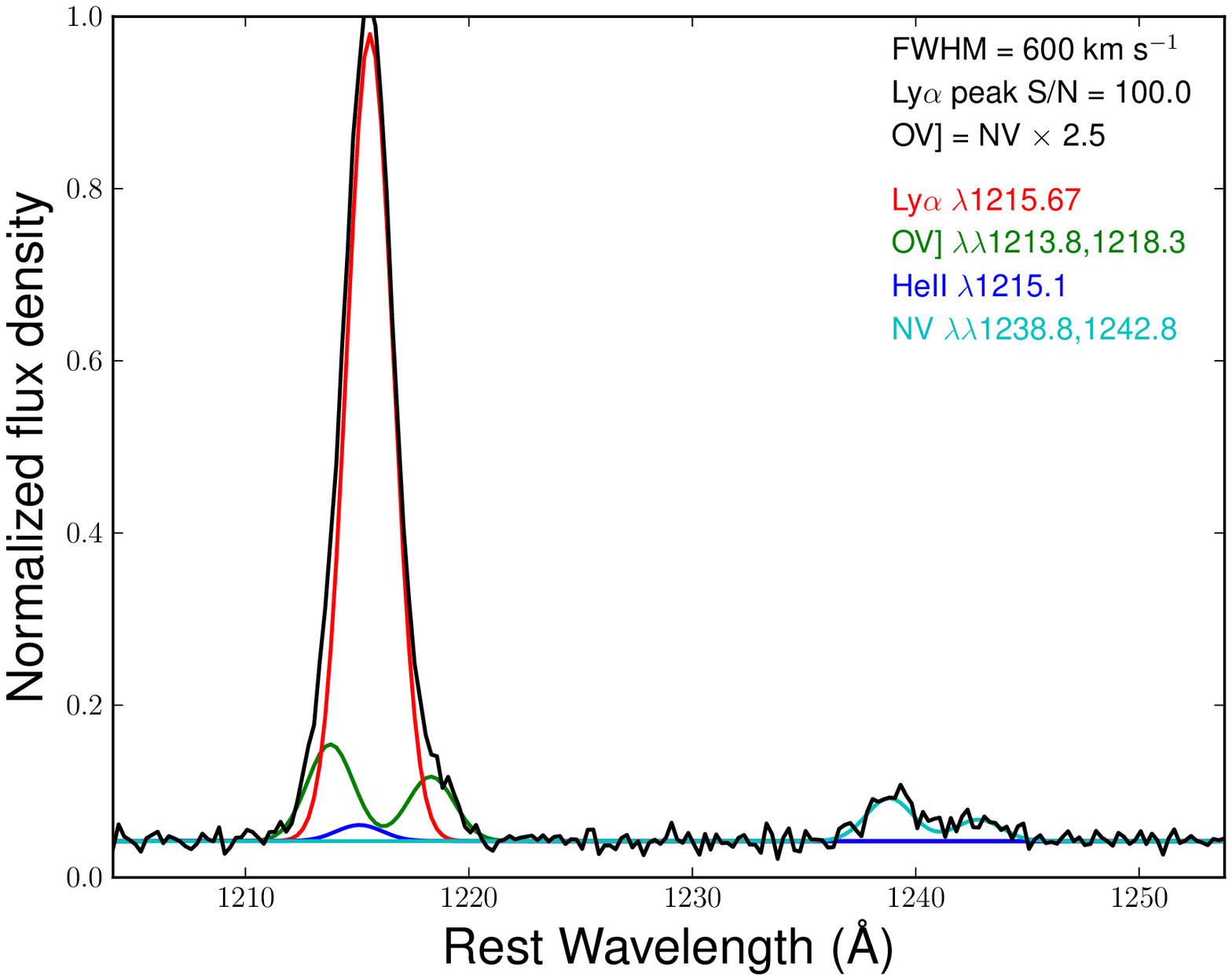}

\vspace{8.3in}
\caption{Model spectra for the NLR or Ly$\alpha$ halo of active
  galaxies, for different values of line FWHM. The emission lines
  shown are Ly$\alpha$ (red), HeII $\lambda$1215.1 (blue), OV]
  $\lambda\lambda$1213.8,1218.3 (green) and NV
  $\lambda\lambda$1238.8,1242.8 (cyan). The black line shows the
  sum of the emission lines, plus a flat continuum component
  containing Gaussian noise. As discussed in the text, the
  detectability of OV] depends strongly on the FWHM of the line
  emission. When the FWHM is large (e.g., 1000 km s$^{-1}$), OV] and
  Ly$\alpha$ are blended to such an extent that the velocity profile
  shows no discernable sign of the presence of the OV] lines (top
  left). At FWHM$\sim$500-600 km s$^{-1}$ (top right and centre left),
  the presence of OV] $\lambda$1218.3 becomes apparent as small
  excess of flux in the red wing of the Ly$\alpha$ profile. When FWHM
  $\la$400 km s$^{-1}$ (centre right and bottom panels), OV]
  $\lambda$1218.3 is resolved from Ly$\alpha$. The blue component of
  the OV] doublet (OV] $\lambda$1218.3) becomes discernable at
  FWHM$\sim$300 km s$^{-1}$ (bottom left), and is fully resolved from
  Ly$\alpha$ at FWHM$\la$200 km s$^{-1}$. Note that HeII
  $\lambda$1215.1 remains blended with Ly$\alpha$ in all the model
  spectra considered here.} 
\label{ov_sim}
\end{figure*}

\begin{figure*}
\includegraphics{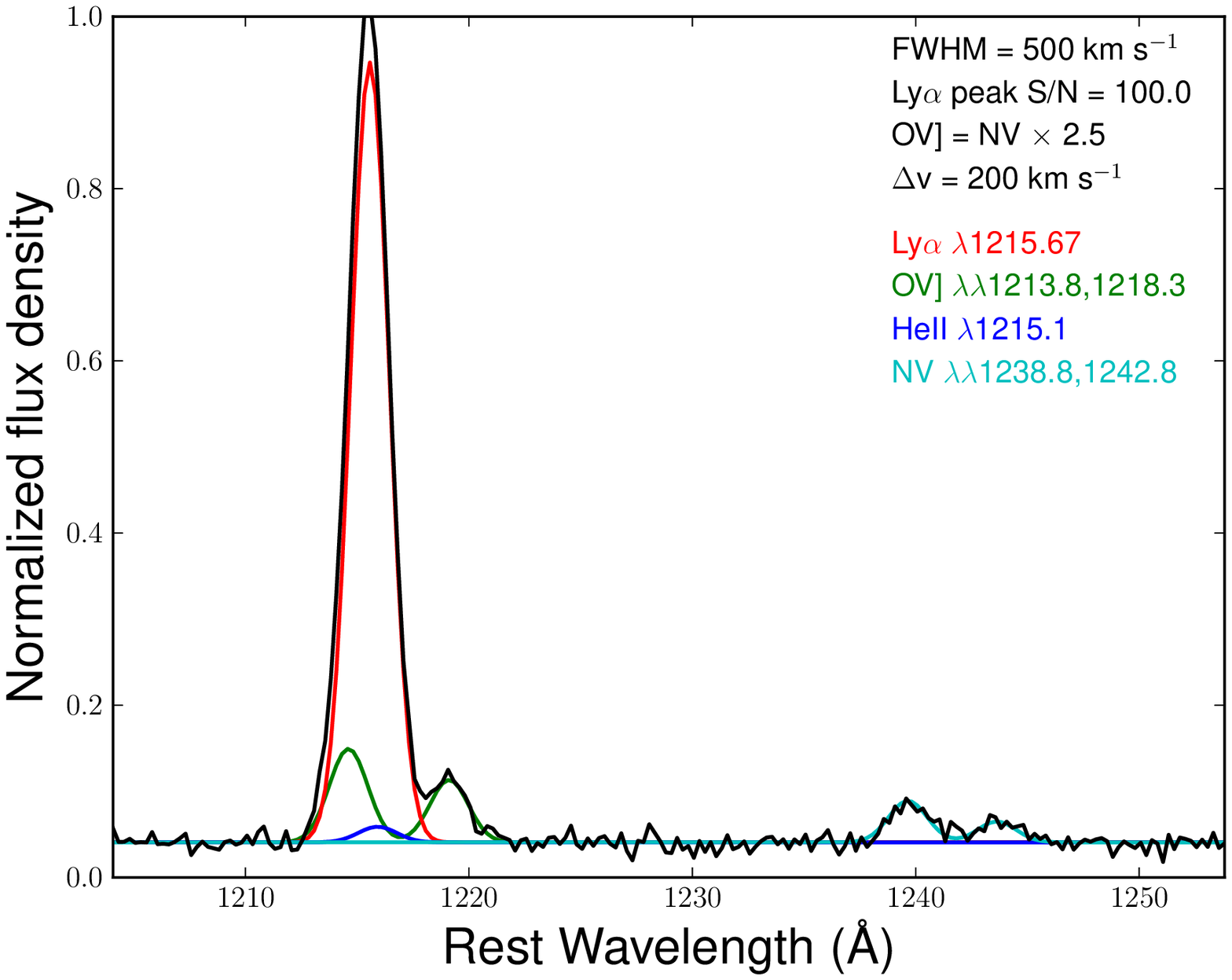}
\includegraphics{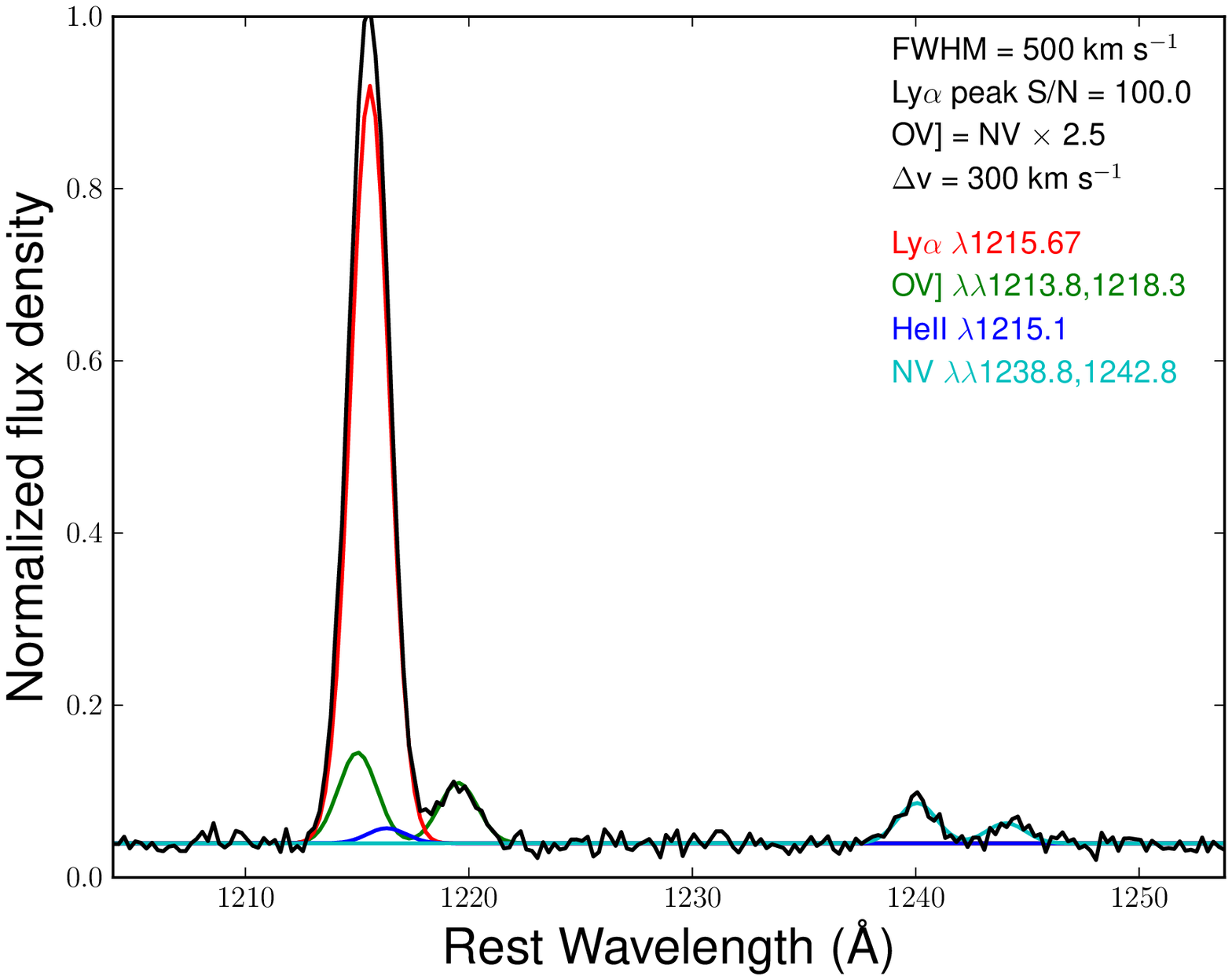}

\includegraphics{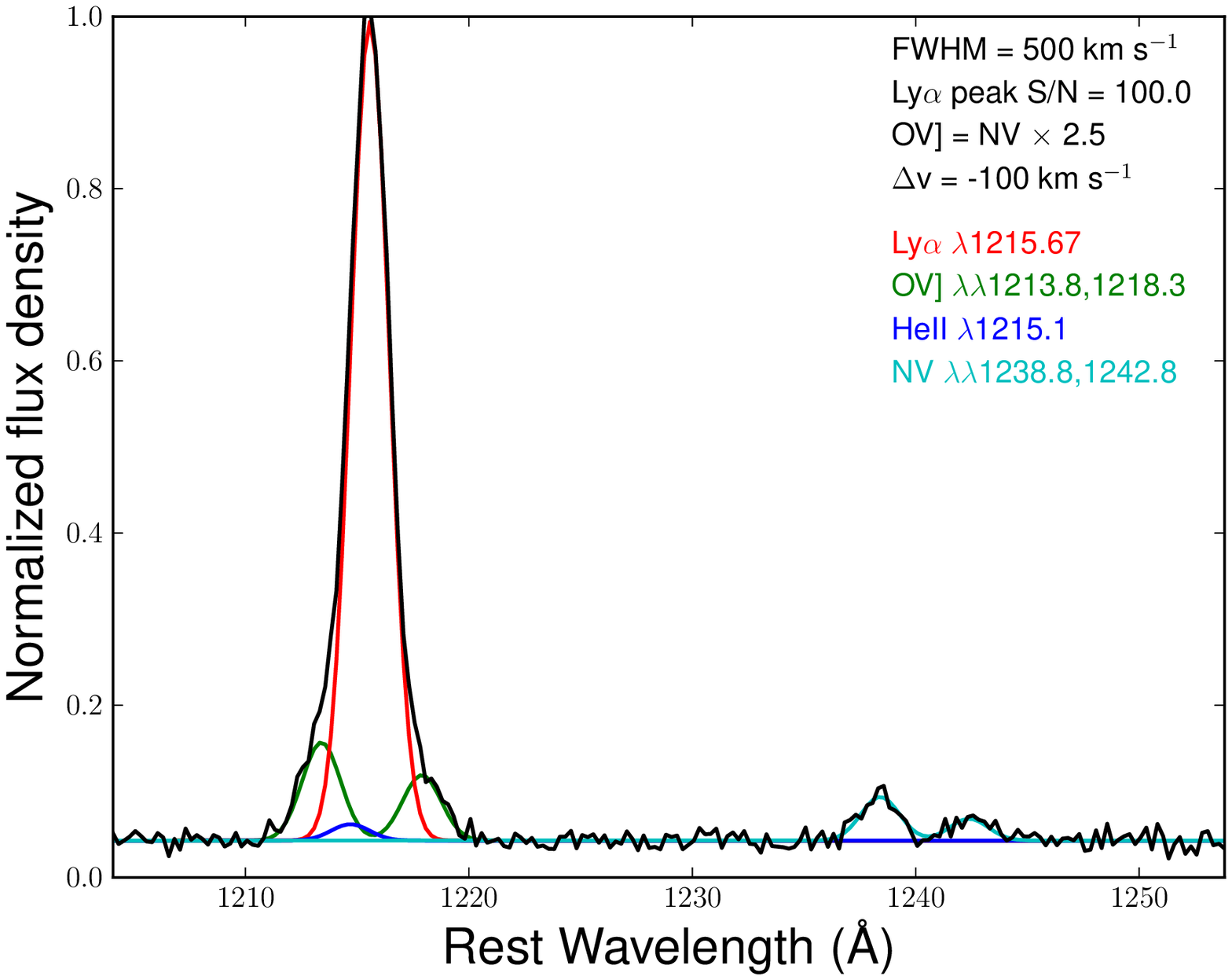}
\includegraphics{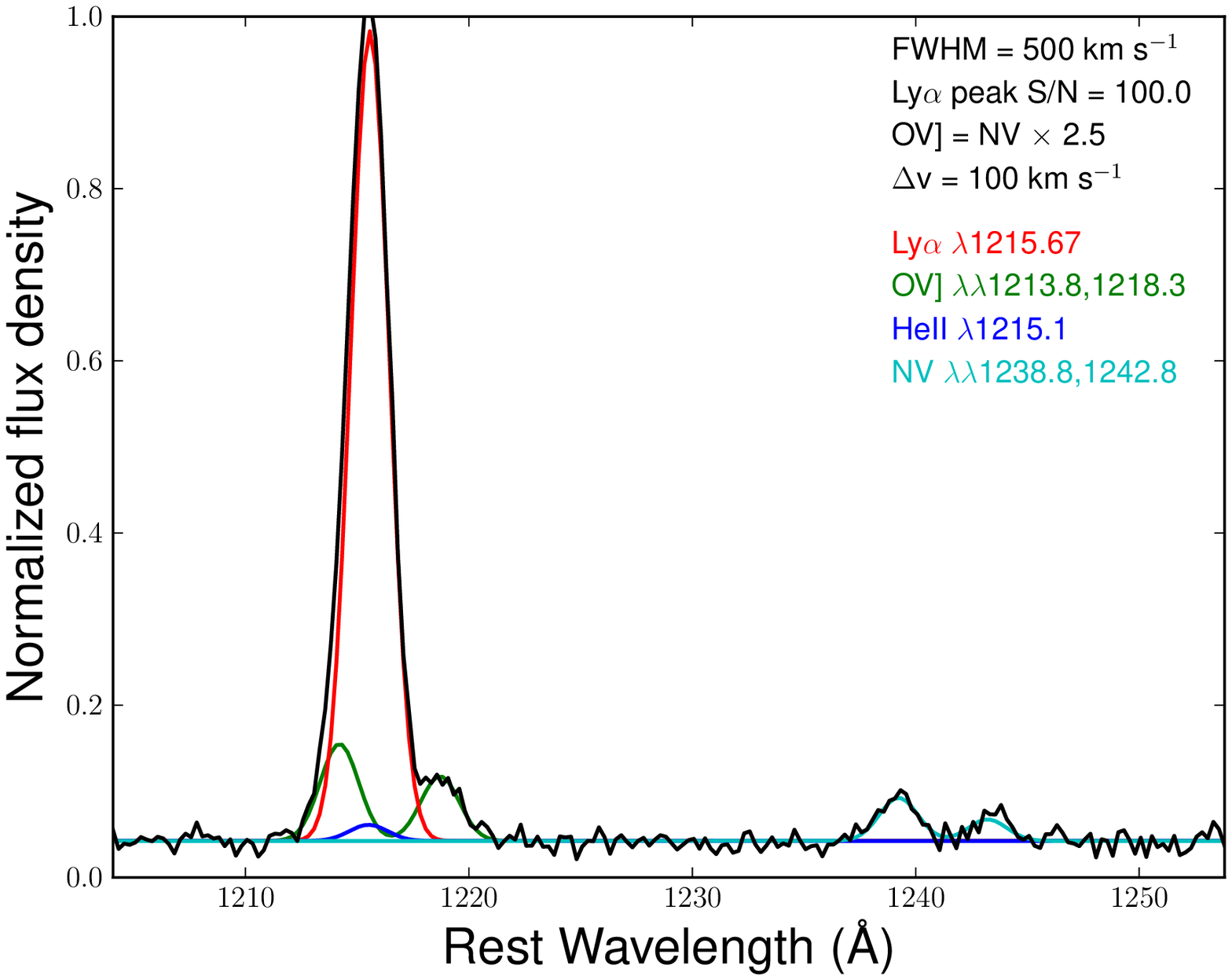}

\includegraphics{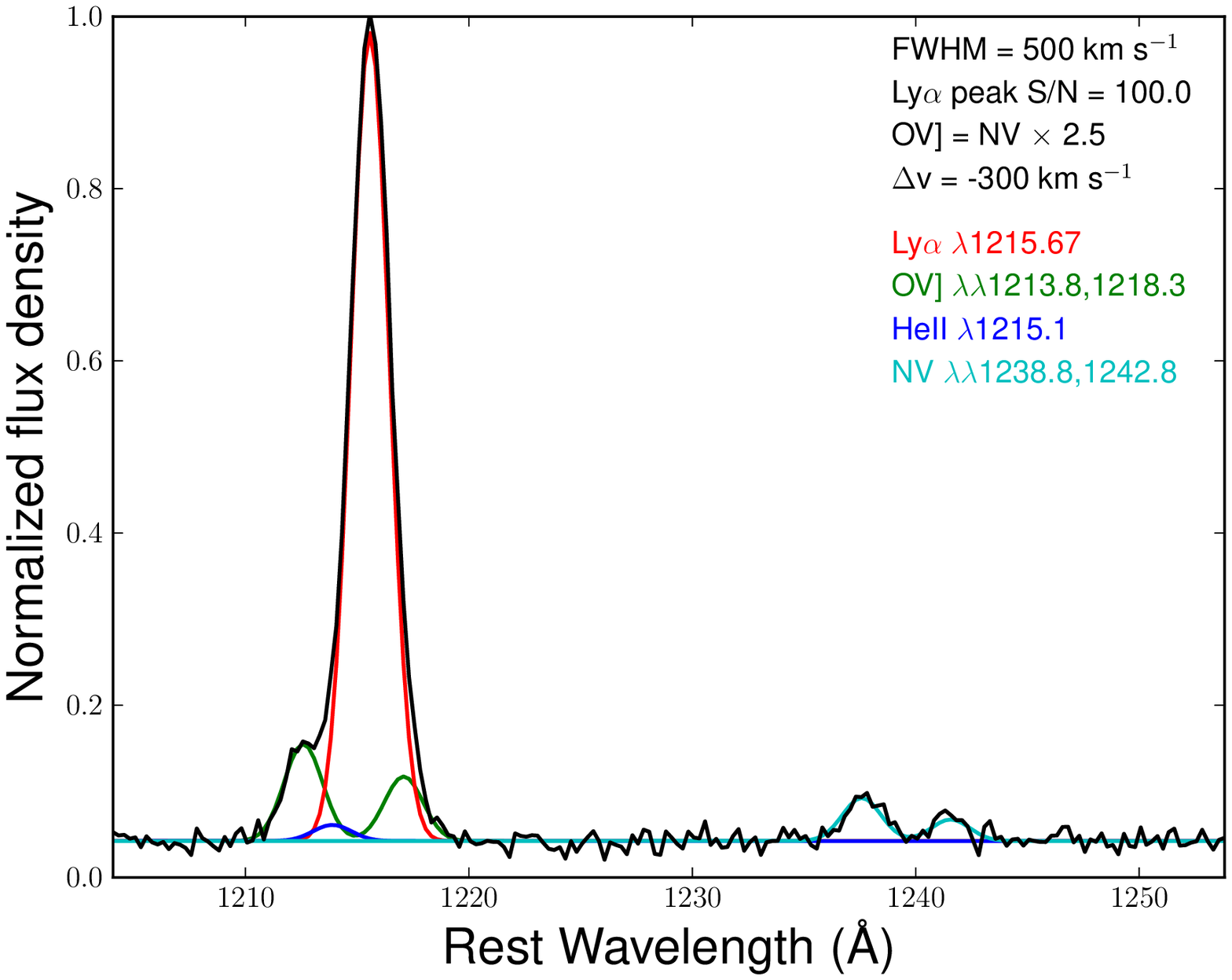}
\includegraphics{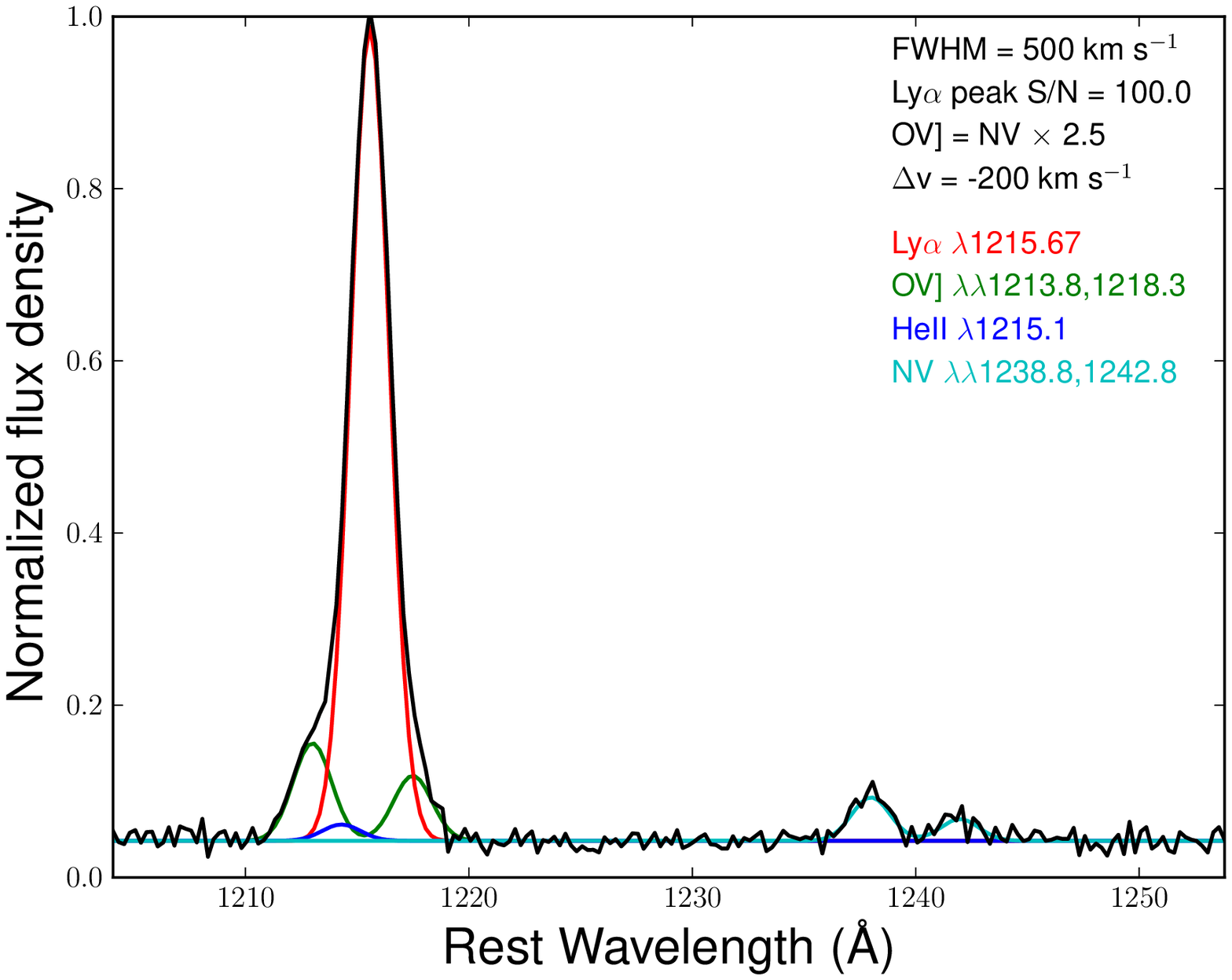}

\vspace{8.3in}
\caption{Model spectra for the NLR or Ly$\alpha$ halo of active
  galaxies, for FWHM=500 km s$^{-1}$, with the high-ionization lines
  shifted from Ly$\alpha$ by -300, -200, -100, 100, 200 and 300 km
  s$^{-1}$. Lines have the same meaning as in Fig. ~\ref{ov_sim}.} 
\label{ov_sim_sh}
\end{figure*}

\section{Model Results}
\label{results}

\subsection{Optically Thick Models}
\label{thick}
In this section we describe the behaviour of the fluxes of HeII and
OV] relative to that of Ly$\alpha$ in our optically-thick
photoionization models. We quantify the contribution of HeII and OV]
to the Ly$\alpha$+HeII+OV] blend using log HeII+OV] / Ly$\alpha$. 
In Figures ~\ref{fig1} and ~\ref{fig2} we illustrate the behaviour of
this ratio for cross-cuts along the $U$ or $Z/Z_{\odot}$ axis of our
grid. The results described in this subsection are principally
  derived from Fig. ~\ref{fig1}, with Fig. ~\ref{fig2} providing an
  additional, supplementary view.

Starting at the low $U$ end of our grid (log $U=$-5), the contribution
from HeII and OV] is negligible, with log HeII+OV] / Ly$\alpha$
$\la$-2.5. As $U$ increases, the contribution from HeII grows until
it reaches a plateau at log $U\ga$-3. However, even at its maximum
flux ratio with Ly$\alpha$, the contribution to the
Ly$\alpha$+HeII+OV] blend is negligible (log HeII / Ly$\alpha$ $\la$
-1.8 or $\la$2\%). 

In the very low metallcity
regime ($Z/Z_{\odot}\le$0.01), OV] does not make a significant
contribution to the HeII+OV] + Ly$\alpha$ blend for any value of U. 
At moderate and high gas metallicity ($Z/Z_{\odot}\ga$0.1), the HeII+OV] /
Ly$\alpha$ curve shows a bump near the high-U end of the grid (log $U
\ga$-2), where OV] becomes much more luminous than HeII due to the
high abundance of O$^{+4}$. When metallicity and U are both
relatively high ($Z/Z_{\odot}\ga$0.3, $U\ga$0.02), the combined
luminosity of HeII and OV] becomes significant\footnote{We consider
  the combined contribution from HeII and OV] to be `significant' when their
  combined luminosity is equal to or greater than one-tenth of the
  luminosity of Ly$\alpha$.} compared to that of
Ly$\alpha$ for some combinations of parameters. For instance, we obtain
log HeII+OV] / Ly$\alpha$ = -0.7 for $Z/Z_{\odot}$=1,
$U$=0.1, $\alpha$=-1.0 and $n_H$=100 cm$^{-3}$. 

We also find that gas density $n_H$ has little or no impact on our
HeII+OV] / Ly$\alpha$ curves, and will not be discussed
further. Nevertheless, we show curves for the full range of density to
illustrate the lack of density dependence of our our results, and to
emphasize that the results are applicable to the wide range of
narrow-line emitting nebulae associated with quasars, from the
classical NLR to the 100-kpc scale Ly$\alpha$ halos or 'blobs'
associated with some distant quasars. 

\subsection{Optically Thin Models}
\label{thin}
In Figs. ~\ref{fig3} and ~\ref{fig4} we show log HeII+OV] / Ly$\alpha$ 
vs. $U$ and $Z/Z_{\odot}$, respectively, for our optically-thin
models. We find a broadly similar behaviour to that seen in the
optically-thick models (see ~\ref{thick} above), but with a steeper
dependence on $U$ and, subsequently, higher values of HeII+OV] /
Ly$\alpha$ in the high-$U$ regime. This difference is due to the
truncated ionization structure of the optically-thin models,
  which lack the relatively lower-ionization zones that are present in
the optically-thick models, and which emit Ly$\alpha$ but not HeII
or OV]. As before, HeII does not make a significant contribution by
itself, but its flux ratio to Ly$\alpha$ is $\sim$0.5 dex higher than
in our optically-thick models.

Compared to the optically-thick models, we find that the contribution
from OV] can be significant for a much wider range in $Z$ and
$U$. However, unlike the optically thick models, high gas metallicity
is not required for the flux of OV] (or HeII+OV]) to become
significant relative to Ly$\alpha$ (compare Figs. ~\ref{fig2} and
~\ref{fig4}). For instance, at $Z/Z_{\odot}$=0.1, 
$U$=0.06 and $\alpha$=-1.5, we obtain HeII+OV] / Ly$\alpha$
$\sim$0.2. Moreover, at high metallicity the sum of HeII and OV]
begins to compete with Ly$\alpha$ itself. For example, at
$Z/Z_{\odot}$=1.0, $U$$\sim$0.03 and $\alpha$=-1.0, we obtain HeII+OV]
/ Ly$\alpha$ $\sim$0.5.

\subsection{HI Absorption}
\label{abs}
The impact of our simple HI absorption model on the log HeII+OV] /
Ly$\alpha$ vs log $U$ diagram is shown for our optically-thick models
(Fig. ~\ref{fig5}) and optically-thin models (Fig. ~\ref{fig6}). In each Figure,
the top row corresponds to no HI absorption ($L_{Ly\alpha}\times1$),
the middle row corresponds to moderate absorption
($L_{Ly\alpha}\times0.8$), and the lower row corresponds to strong
absorption ($L_{Ly\alpha}\times0.5$). In the interest of simplicity,
we show only models with $n_H$=1 cm$^{-3}$; essentially identical
results were obtained at  $n_H$=100 and 10$^4$ cm$^{-3}$.

As expected, the impact of absorption of Ly$\alpha$ is to increase log
HeII+OV] / Ly$\alpha$. The log HeII+OV] / Ly$\alpha$ vs log $U$
curves maintain the same shape as they have without absorption, but
with a systematic shift towards higher values of log HeII+OV] /
Ly$\alpha$. With moderate HI absorption ($L_{Ly\alpha}\times0.8$), log HeII+OV] /
Ly$\alpha$ is increased by 25\% ($\sim$0.1 dex). In the strong absorption case
($L_{Ly\alpha}\times0.5$), the increase is a factor of 2 ($\sim$0.3 dex) above the
no-absorption case. 

In our optically-thick models we find a maximum log HeII+OV] / Ly$\alpha$
= -0.40, using $Z/Z_{\odot}\ga1.0$, $\alpha$=-1.0, $U\sim$0.1 and
strong absorption ($L_{Ly\alpha}\times$0.5). In the case of our
optically-thin models, a maximum log HeII+OV] / Ly$\alpha$ of $\sim$0.11 is
reached, using $Z/Z_{\odot}=3.0$, $\alpha$=-1.0, $U\sim$0.02 and
strong absorption ($L_{Ly\alpha}\times$0.5). 

\section{Corrections for contamination of Ly$\alpha$ flux measurements}

\subsection{Correction for HeII $\lambda$1215.1}
\label{corr_heii}
Despite the expectation that the flux of HeII $\lambda$1215.1 should
be negligible compared to that of Ly$\alpha$ (see \S\ref{thick} and
\S\ref{thin}), there might arise circumstances where it is useful to
estimate the contribution from HeII to the Ly$\alpha$+HeII+OV]
blend. Because HeII $\lambda$1215.1 and HeII
$\lambda$1640 correspond to the $\beta$ and $\alpha$ lines in the Balmer series of
singly ionized helium, their flux ratio is expected to occupy a fairly narrow range, despite its
slight temperature and density sensitivity (see e.g. Osterbrock \&
Ferland 2005). Under the assumption of Case-B conditions, this ratio
is expected to range from 0.28 at $T$=5000 K and $n$=100 cm$^{-3}$, to
0.33 at $T$=20,000 K. Thus, we suggest estimating the HeII
$\lambda$1215.1 flux as 0.3 times the flux of HeII $\lambda$1640.

\subsection{Corrections for OV] $\lambda\lambda$1213.8,1218.3}
\label{corr}
Given the potential for OV] to strongly contaminate Ly$\alpha$ flux
measurements (see \S\ref{thick} and \S\ref{thin}), we propose here a
means to estimate the OV] flux. Extrapolation from other
high-ionization metal lines is likely to 
provide the most reliable estimate of the OV] flux, because such
lines are expected to be emitted from similar locations within an
ionized cloud.

In Fig. ~\ref{fig7} we show how the luminosity of OV] varies in
our grid compared to OVI $\lambda$1035, NV $\lambda$1240, CIV
$\lambda$1549 and [NeV]1575, all normalized to the
luminosity of Ly$\alpha$. These figures show our entire model grid,
with the exception of models that include absorption of Ly$\alpha$ by an
external HI screen, which are not shown. Thus, all the combinations of gas
metallicity, density, U and $\alpha$ are represented. 

We find that the luminosity of OV] is closely correlated with that of
the other high-ionization lines. The 
lines NV $\lambda$1240 and [NeV] $\lambda$1575 show the most linear
correlation with OV] (Fig. ~\ref{fig7}), primarily because they are
also from quadruply-ionized species. Thus, NV $\lambda$1240 and [NeV]
$\lambda$1575 are likely to be among the most reliable lines from
which to extrapolate the OV] flux. 

Ideally, one would use measurements of the observed OV] flux relative to
 the other high-ionization lines to obtain an empirical relation for
 estimating the OV] flux. However, because the OV] doublet has not
 yet been detected from an active galaxy, to the best of our
 knowledge, we have little choice but to use models. For this task, we have  
 used our photoionization models with $Z/Z_{\odot}$=1.1,
 $\alpha$=-1.0 and $n_H$=100 cm$^{-3}$ (yellow points in
 Fig. ~\ref{fig7}), which should be generally appropriate for the NLR
 of powerful, Type 2 active galaxies (e.g. Humphrey et al. 2008). For
 the sake of simplicity we have assumed log U = -1, but
broadly similar values are obtained for other values of log U within
the range -3 $\la$ log U $\la$ 0. Note that at lower values of log U (i.e., log
U $\la$ -3), the flux of OV] is expected to be so low compared to that
of Ly$\alpha$ (i.e., $<<$1\%) that the issue of contamination by OV]
becomes essentially irrelevant. From this photoionization model, we
obtain flux ratios between OV] and several 
other high-ionization lines, to be used as coefficients to extrapolate
the flux of OV] from observed fluxes of those other high-ionization
lines.

\begin{figure}
\includegraphics{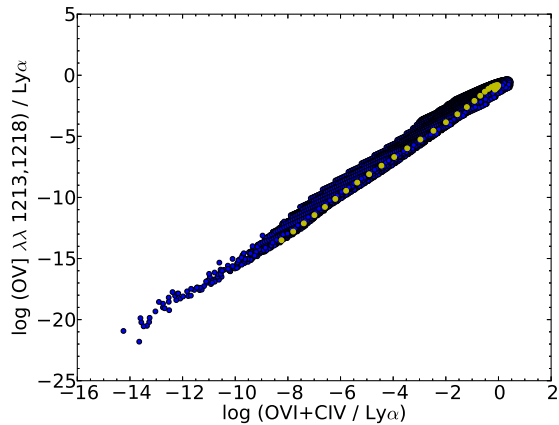}
\vspace{2.2in}
\caption{Log OV]/Ly$\alpha$ vs. log OVI+CIV/Ly$\alpha$. Every
  photoionization model of our grid is plotted (blue circles). Our
  ionization-bounded model sequence with $Z/Z_{\odot}$=1.1,
  $\alpha$=-1.0, and $n_H$=100 cm$^{-3}$ highlighted using yellow
  circles.}  
\label{fig8}
\end{figure}

Thus, we obtain the following relations between the flux of OV] and NV or
[NeV] $\lambda$1575:
\\ 
\\
\noindent OV] $\sim$ 2.5 $\times$ NV\\
\\
\noindent OV] $\sim$ 70 $\times$ [NeV] $\lambda$1575\\
\\
The luminosities of OVI $\lambda$1035 and CIV $\lambda$1549 similarly show a
strong correlation with that of OV], but with a reversal near the
high-ionization parameter end, leading to a larger dispersion and/or
double-values, which reduces their individual usefulness as indicators of
OV] luminosity. This degeneracy can be partially mitigated by
summing the fluxes of OVI and CIV as illustrated in Fig. ~\ref{fig8},
where we show log OV] / Ly$\alpha$ vs log OVI+CIV / Ly$\alpha$. Using
the same model sequence as described above, we obtain the relation:
\\
\\
\noindent OV] $\sim$ 0.07 $\times$ OVI+CIV\\
\\
As a caveat, we emphasize that our corrections for OV] contamination are
approximate and depend, among other parameters, on the gas chemical 
abundances and/or on the ionization parameter U. This is highlighted by
Fig. ~\ref{ov_nv}, where we show OV]/NV vs. $Z/Z_{\odot}$ for model loci with three
different values of U. We stress that observations of the OV] doublet
 will be essential to derive more accurate relations between the flux of this
 and other emission lines.

\section{Discussion}

\subsection{General remarks}
In the previous sections we have shown that for a certain range of
conditions (or model parameters), the OV]
$\lambda\lambda$1213.8,1218.3 doublet should significantly contaminate
the Ly$\alpha$ fluxes of low density AGN-photoionized nebulae, i.e.,
the NLR or the Ly$\alpha$ halo. This
result is in qualitative agreement with Shields et al. (1995), who
obtained a similar result in relation to high-density, broad line
region clouds. However, in contrast to Shields et al. (1995) for
  the high density BLR of Type 1 quasars (i.e., log
$n_H$$\ga$10$^{6}$ cm$^{-3}$), here we find
that HeII $\lambda$1215.1 does not significantly contaminate the
Ly$\alpha$ flux at the lower gas densities of the NLR (i.e., log
$n_H$$\la$10$^{4}$ cm$^{-3}$).

\subsection{OV] contamination in Type 2 quasars}

Thus far we have only considered model flux ratios, but it is
also important to examine whether the observed UV line ratios of
active galaxies suggest any significant contamination of
their Ly$\alpha$ flux measurements by OV] or HeII $\lambda$1215.1. 
For this purpose, we use line flux measurements for 95 SDSS BOSS
Type 2 quasars at z$>$2 from Silva et al. (2019), originally
  selected as candidate Type II quasars by Alexandroff et
  al. (2013). This sample has the advantage of having spectra obtained
  under a relatively homogeneous instrumental configuration (see
  Alewxandroff et al. 2013), and with line parameters determined
  using a single analysis methodology (see Silva et
  al. 2019). Furthermore, the redshift range of this sample places the
  Ly$\alpha$ line within the optical observational window, unlike
  lower-redshift objects (i.e., z$\la$2), thereby allowing comparision
  between the flux of Ly$\alpha$ and those of other UV lines such as
  NV, CIV, HeII, etc. 

To complement this data sample, we also include line flux 
  measurements for 12 radio galaxies at z$>$2 from the Keck II sample
  of Cimatti et al. (1998), Vernet et al. (2001) and Humphrey et
  al. (2008), selecting only those galaxies for which Ly$\alpha$, NV
  and HeII $\lambda$1640 have been detected. Again, this data sample
  has the advantage of being relatively homogenously observed and
  analysed\footnote{See Silva et al. (2019) for detailed ionization
    modeling of the Type II quasar sample, and an intercomparison with
    the high-z radio galaxy sample}.

 The modeling discussed herein applies equally to the extended, 
  narrow line emitting gas of Type 1 quasars. However, the broad (FWHM$\ga$
  2000 km s$^{-1}$) Ly$\alpha$ emission from the BLR usually
  overwhelms any narrow Ly$\alpha$, NV, and CIV emission from the NLR or Ly$\alpha$
  halo, severely complicating the measurement and analysis of this
  narrow emission unless it is very extended (e.g. Borisova et
  al. 2016). For this reason, we do not include Type 1 quasars in our
  sample of objects. In any case, as Type 1 and Type 2 quasars are
  thought to be similar objects merely viewed at different
  orientations (e.g. Antonucci 1993 and references therein),
  conclusions derived from our Type 2 sample ought also to be applicable to
  Type 1 quasars.

\begin{figure}
\includegraphics{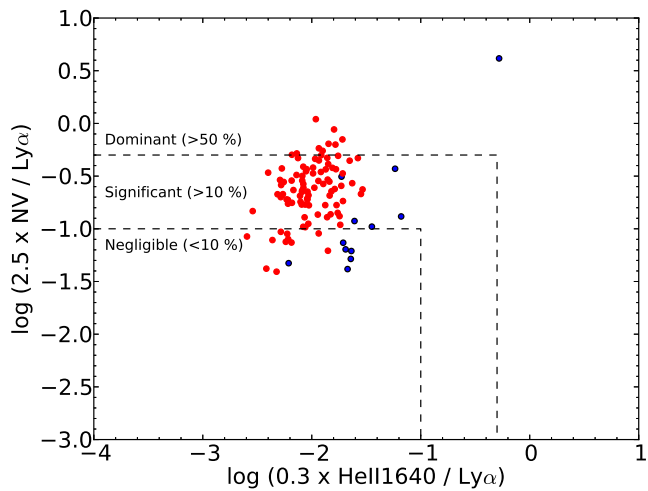}
\vspace{2.5in}
\caption{Plot of Log (2.5$\times$NV / Ly$\alpha$) vs. log (0.3$\times$HeII1640
/ Ly$\alpha$) for z$\sim$2.5 radio galaxies from Vernet et al. (2001:
blue circles), and Type 2 quasars at z$>$2 with detections of
Ly$\alpha$, NV and HeII $\lambda$1640 from Silva et al. (2018: red
circles). The vertical axis, log (2.5$\times$NV / Ly$\alpha$), is
intended as a surrogate for log (OV] / Ly$\alpha$), while the
horizontal axis, log (0.3$\times$HeII1640 / Ly$\alpha$), is a
surrogate for log (HeII $\lambda$1215.1 / Ly$\alpha$).}
\label{obs1}
\end{figure}

The selected sample is shown in Fig. ~\ref{obs1},
where we plot log (2.5 $\times$ NV / Ly$\alpha$) vs. log (0.3
$\times$ HeII $\lambda$1640 / Ly$\alpha$) as estimators of the
strength of OV] $\lambda\lambda$1213.8,1218.3 and HeII $\lambda$1215.1
relative to Ly$\alpha$, respectively (see \S\ref{corr}).

The majority of the objects have 2.5 $\times$ NV / Ly$\alpha$ $\ge$
0.1, implying that the OV] flux is usually significant compared to that
of Ly$\alpha$ (90/107 or 84\% of objects). In addition, we find that a
small but significant fraction of the objects (10/107 or 9\% of cases)
have 2.5 $\times$ NV / Ly$\alpha$ $>$ 0.5, implying that
the flux of OV] exceeds that of Ly$\alpha$.

Only in the case of one object shown in Fig. ~\ref{obs1}, the radio
galaxy TXS 0211-122 (z=2.34), do we 
find a predicted value of 0.3$\times$HeII $\lambda$1640 / Ly$\alpha$ that lies
above 0.5. In fact, its ``Ly$\alpha$'' flux can be more than accounted
for by the combined expected fluxes of OV] and HeII $\lambda$1215.1,
with (2.5 $\times$ NV / Ly$\alpha$) + (0.3$\times$HeII $\lambda$1640 /
Ly$\alpha$) = 4.7. In this case it is clear that we have overestimated
the flux of at least OV], if not also HeII $\lambda$1215.1, since our
predicted OV]/Ly$\alpha$ ratio is 4.1. In other words, the predicted
flux of OV], based on our extrapolation from NV, is 4.1 times higher
than the observed flux of Ly$\alpha$ (see the Ly$\alpha$ and NV fluxes
given in Vernet et al. 2001). We suggest 
that this overestimation might be the result of scattering and
subsequent absorption of OV] and HeII photons by HI and dust, or an N/O
abundance ratio that is several times higher than its Solar value (see
e.g. van Ojik et al. 1994). It seems plausible that this
`overestimation effect' might also affect other objects in the sample,
though it would be difficult to definitively verify without direct
detections of OV].

As a further
caveat, we stress that the selection criteria we have used to build
our sample of quasars and radio galaxies specifically requires the detection of NV,
and it is possible that this has introduced a bias favouring objects
whose narrow line regions are highly-ionized and have a high abundance
of nitrogen. Both of these conditions are expected to favour significant
contamination of Ly$\alpha$ fluxes from OV] (see \S\ref{results}).

\subsection{Prospects for direct detection of OV]}
Despite the fact that photoionization models for AGN predict their presence
(see also Ferland et al. 1992; Shields et al. 1995),
HeII $\lambda$1215.1 or OV] $\lambda\lambda$1213.8,1218.3 have never,
to the best of our knowledge, been directly detected in
AGN-photoionized gas. Is there any prospect of directly detecting
and deblending these lines from Ly$\alpha$ in active galaxies?
We argue that under certain conditions, it should indeed be
possible, at least in the case of OV]. If there is no velocity shift
between OV] and Ly$\alpha$, then OV] $\lambda$1213.8 and OV]
$\lambda$1218.3 would be offset with respect to Ly$\alpha$ by -1.9 \AA~
(-461 km s$^{-1}$) and +2.6 \AA~ (+649 km s$^{-1}$), respectively. This
means that if Ly$\alpha$ is sufficiently narrow and is observed at
sufficient spectral resolution and signal to noise ratio, then it
should be possible to kinematically resolve OV] from Ly$\alpha$. 

To illustrate this, we have created model spectra of the Ly$\alpha$
and NV spectral region as shown in Fig. ~\ref{ov_sim}. For the sake of
simplicity, each line is represented by a single Gaussian emission
profile, with all lines having the same full width at half maximum (FWHM)
and, unless otherwise stated, we have not introduced any
relative velocity shifts between lines. We have assumed the
following flux ratios: OV] $\lambda$1213.8 / OV] $\lambda$1218.3 = 1.5
corresponding to the low density limit of 
this ratio (n$_e$ $\le$10$^{4}$ cm$^{-3}$; McKenna et al. 1997); NV
$\lambda$1238.8 / NV $\lambda$1242.8 = 2.0, corresponding to the
optically thin case; OV] $\lambda\lambda$1213.8,1218.3 / NV
$\lambda\lambda$1238.8,1242.8 = 2.5; OV] $\lambda\lambda$1213.8,1218.3
/ Ly$\alpha$ = 0.2; HeII $\lambda$1215.1 / Ly$\alpha$
= 0.02. The continuum level has been set at 0.05 times the
peak flux density of Ly$\alpha$. This value is consistent with
  observations of Type 2 active galaxies, although there can be large
  variation between objects (e.g., Vernet et al. 2001). In addition, we have added random
(Gaussian) noise to our model spectra such that the $\sigma$ of the noise
spectrum is 0.01 the peak flux density of the Ly$\alpha$
line. This value is somewhat arbitrary, given that the
  signal to noise ratio of the continuum depends on a variety of
  parameters, such as the UV continuum flux density of the target, the
  exposure time and conditions of the observations, the instrumental
  configuration, etc.

We have adopted FWHM values in the range 200 $<$ FWHM $<$ 1000 km
s$^{-1}$, based on the observed kinematic properties of the NLR and
extended emission halo of active galaxies, which are thought to be
driven by a combination of gravitational motion and feedback activity
(e.g., van Ojik et al. 1997; Baum \& McCarthy 2000; Villar-Mart\'{i}n
et al. 2003, 2007b; Das et al. 2005; Humphrey et al. 2006). In
comparison, the extended Ly$\alpha$ emission associated with z$\ga$2
star forming galaxies typically lies in the range FWHM$\sim$100--500
km s$^{-1}$ (e.g. Leclercq et al. 2017).

We find that when the FWHM is large (e.g., $\ge$1000 km s$^{-1}$), OV] and
Ly$\alpha$ are blended to such an extent that the velocity profile
shows no discernable sign of the presence of the OV] lines
(Fig. ~\ref{ov_sim}, top left). However, at FWHM$\sim$500-600 km
s$^{-1}$ (Fig. ~\ref{ov_sim}, top right and centre left), the presence
of OV] $\lambda$1218.3 becomes apparent as a small excess of flux in
the red wing of the Ly$\alpha$ profile. At even lower values of FWHM
(i.e., FWHM $\la$400 km s$^{-1}$: Fig. ~\ref{ov_sim}, centre right and bottom panels), OV]
$\lambda$1218.3 is now resolved from Ly$\alpha$, rendering it
detectable. The short wavelength component of the OV] doublet (OV]
$\lambda$1213.8) becomes discernable at FWHM$\sim$300 km s$^{-1}$
(Fig. ~\ref{ov_sim}, bottom left), and is fully resolved from
Ly$\alpha$ at FWHM$\la$200 km s$^{-1}$ (Fig. ~\ref{ov_sim}, bottom
right), at which point it too should be readily detectable.

In Fig. ~\ref{ov_sim_sh} we show the impact of introducing a velocity
shift between the high-ionization lines and Ly$\alpha$. We use a line
FWHM of 500 km s$^{-1}$, and apply a velocity shift of -300,-200,
-100, +100, +200 or +300 to the high-ionization lines, relative to the
line of sight velocity of Ly$\alpha$. From the
selection of model spectra shown in this Figure, it can be seen that
even a small velocity shift (i.e., $\sim$100 km s$^{-1}$) between OV]
and Ly$\alpha$ can alter the detectability of OV], and
significantly change the total velocity profile of the
blend. Generally speaking, a larger velocity shift improves the
visibility of OV]. In the case where the OV] emission has a relative
blueshift, we find the the flux asymmetry of the blend moves from its
red wing to its blue wing. In the case of HeII $\lambda$1215.1, its -0.6 \AA~ offset from
Ly$\alpha$ (equivalent to -141 km s$^{-1}$) and low expected relative
flux should make this line highly challenging to directly
detect (see Fig. ~\ref{ov_sim}).

If neglected, the presence of the OV doublet may complicate kinematic
analyses that rely on the Ly$\alpha$ line. For instance,
Fig. ~\ref{ov_sim} reveals that when FWHM is in the range
$\sim$400-600 km s$^{-1}$, the presence of strong OV] emission mimics
the presence   
of a broader underlying kinematic component of Ly$\alpha$ emission,
qualitatively similar to what is sometimes seen in high-z radio
galaxies (e.g. Villar-Mart\'{i}n et al. 2003), potentially leading to 
the false detection of a gas outflow if a similar kinematic pattern 
cannot be confirmed in other emission lines (e.g., CIV, HeII $\lambda$1640, etc.). 
In addition, when one of the OV] lines is resolved from Ly$\alpha$,
the dip in flux between the two lines could potentally be misinterpreted as a
Ly$\alpha$ absorption feature, instead of as two resolved emission
lines. 

Thus, there is clearly some potential for ambiguity in the
interpretation of the kinematic properties of the Ly$\alpha$+OV]
blend, where a high velocity component of Ly$\alpha$ may be
misinterpreted as OV] emission, and vice-versa. To overcome this
degeneracy, we suggest extrapolating the expected wavelengths of the
OV] lines from observations of other high-ionization lines, such as NV or CIV, to
determine whether an observed feature in the profile of Ly$\alpha$ is
likely to be an OV] line.

\subsection{OV] as a diagnostic for AGN in Ly$\alpha$-emitters}
We suggest that OV] $\lambda\lambda$1213.8,1218.3 should be
useful as a means to confirm the presence of AGN activity in
Ly$\alpha$-emitters at high redshift, if OV] is at least partly
resolved from Ly$\alpha$ (e.g., at FWHM $\la$500 km s$^{-1}$; see
Fig. ~\ref{ov_sim}). This is because ionizing O$^{+3}$ to O$^{+4}$,
the species responsible for OV emission, requires a photon energy of
$hv$$\ge$77.4 eV, which in turn requires an ionizing spectrum that is much
harder than produced by young stellar populations, but which can be
produced by an AGN. Thus, like NV $\lambda\lambda$1238,1242
(e.g. Villar-Mart\'{i}n et al. 1999), a detection of OV] emission ought to
be considered a `smoking gun' of AGN activity in a Ly$\alpha$-emitter\footnote{For comparison,
  the presence of N$^{+4}$ (to make NV 
  emission possible) requires $hv$$\ge$77.5 eV; He$^+$ (for HeII
  emission) requires $hv$$\ge$54.4 eV; and C$^{+3}$ (for CIV emission) requires
  $hv$$\ge$47.9 eV.}. 

\begin{figure}
\includegraphics{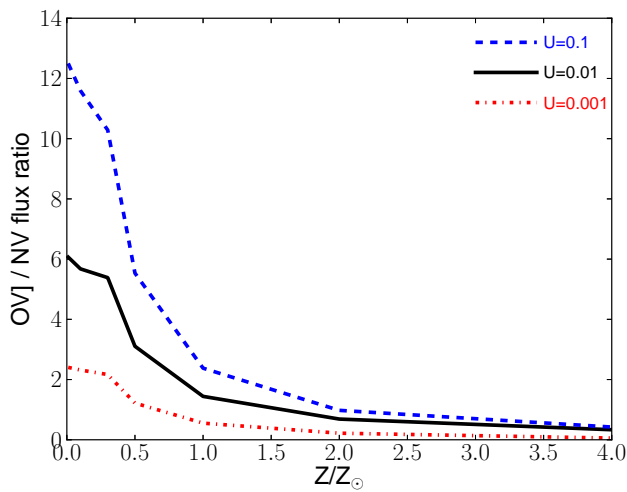}
\vspace{2.5in}
\caption{OV] over NV versus gas metallicity $Z/Z_{\odot}$ for
  $\alpha$=-1.0 and $n_H$ = 100 cm$^{-3}$, for three different values of
ionization parameter U. The shown models are ionization-bounded
(optically-thick). As discussed in the main text, the OV]/NV ratio
increases with decreasing gas metallicity $Z/Z_{\odot}$, such that OV]
may become more easily detectable than NV when
0.1$\la Z/Z_{\odot} \la$1.0, provided OV] and Ly$\alpha$ can be deblended.}
\label{ov_nv}
\end{figure}

A further point of interest is that under some circumstances, OV]
should be easier to detect than NV. For instance, given the secondary
enrichment of the N abundance, the
OV]/NV flux ratio is expected to increase significantly towards lower
gas metallicity as shown in Fig. ~\ref{ov_nv}, such that OV] should
become more easily detectable than NV for sub-Solar gas metallicities,
provided OV] and Ly$\alpha$ can be deblended. The precise
range of parameters where OV] should become easier to detect than NV
is model dependent, but based on our modelling we expect this to be
the case when FWHM $\la$500 km s$^{-1}$ and
0.1$\la$$Z/Z_{\odot}$$\la$1.0 -- properties broadly corresponding to
those expected for low to intermediate mass
Ly$\alpha$-emitters at high redshift (e.g. McGreer et al. 2018; Sobral
et al. 2018a,b; Mainali et al. 2018; Leclercq et al. 2017; Jiang et al. 2013; Fosbury et al. 2003).

Thus, as advances in instrumentation and telescope collecting area
provide access to ever more distant and ever fainter galaxies
(e.g. Hashimoto et al. 2017; Ouchi et al. 2018), we expect OV] to become a useful diagnostic
of the presence of AGN activity in high-z Ly$\alpha$-emitting systems,
particularly in intermediate to low mass galaxies whose gas metallicity is sub-Solar.

\section{Summary}
We have used a grid of photoionization models to examine the potential
impact of OV] $\lambda\lambda$1213.8,1218.3 and HeII $\lambda$1215.1
emission on measurements of the Ly$\alpha$ flux from the NLR and
Ly$\alpha$ halos of active galaxies. We find that the HeII flux is
essentially always negligible, but OV] can contribute significantly
($\ga$10\%) when the ionization parameter and the gas metallicity are
high (log U $\ga$-2; $Z/Z_{\odot}$$\ga$0.3). We also find that using
optically-thin clouds can increase the relative contributions from OV]
and HeII. 

In addition, we have provided means to estimate the fluxes of
HeII $\lambda$1215.1 and OV] $\lambda\lambda$1213.8,1218.3 by
extrapolating from other UV emission lines, and have estimated the
contribution from these lines in a sample of 107 Type 2 active
galaxies (QSO2s and HzRGs) at z$>$2, finding evidence for significant 
contamination of Ly$\alpha$ fluxes ($\ge10\%$) in 84\% of cases. This
suggests that Ly$\alpha$ flux measurements of type 2 active galaxies
are often contaminated at the $\ga$10\% level by these other lines. 

We have also found that the presence of
OV] emission can impact the apparent kinematics of Ly$\alpha$,
potentially mimicking the presence of high-velocity outflows.

Additionally, we have shown that, where its flux is 
significant, OV] ought to be detectable when the FWHM of Ly$\alpha$ is
less than $\sim$500 km s$^{-1}$, and we have proposed using detection
of OV] as a new diagnostic of AGN activity in high-z Ly$\alpha$
emitters.

\section*{Acknowledgments}
AH thanks the anonymous referee for their helpful comments and
suggestions. AH also thanks Montse Villar-Mart\'{i}n, Luc Binette and
Jarle Brinchmann for useful discussions, and Marckelson Silva for making
available the emission line measurements of Type 2 quasars. AH
acknowledges FCT Fellowship SFRH/BPD/107919/2015; Support from
European Community Programme (FP7/2007-2013) under grant agreement
No. PIRSES-GA-2013-612701 
(SELGIFS); Support from FCT through national funds
(PTDC/FIS-AST/3214/2012 and UID/FIS/04434/2013), and by FEDER through
COMPETE (FCOMP-01-0124-FEDER-029170) and COMPETE2020 
(POCI-01-0145-FEDER-007672). In addition, AH acknowledges support from
the FCT-CAPES Transnational Cooperation Project "Parceria
Estrat\'egica em Astrof\'{i}sica Portugal-Brasil".

\end{document}